\documentclass[11pt]{article}

\usepackage{amsmath}
\usepackage{amssymb}
\usepackage{amsthm}
\usepackage{amsfonts}
\usepackage{graphicx}
\usepackage{textcomp}
\usepackage{color}
\usepackage[hidelinks]{hyperref}
\usepackage{stmaryrd}
\usepackage{setspace}
\usepackage[utf8]{inputenc}
\usepackage[english]{babel}
\usepackage{mathrsfs}
\usepackage[a4paper, total={5.7in, 9.5in}]{geometry}
\usepackage{cite}

\newcommand{\hh}{{\hspace{.3mm}}}

\let\OLDitemize\itemize
\renewcommand\itemize{\OLDitemize\addtolength{\itemsep}{-8pt}}

\numberwithin{equation}{section}

\def\sideremark#1{\ifvmode\leavevmode\fi\vadjust{\vbox to0pt{\vss
			\hbox to 0pt{\hskip\hsize\hskip1em
				\vbox{\hsize3cm\tiny\raggedright\pretolerance10000
					\noindent #1\hfill}\hss}\vbox to8pt{\vfil}\vss}}}%

\def\be{\begin{equation}}
\def\ee{\end{equation}}
\def\bea{\begin{eqnarray}}
\def\eea{\end{eqnarray}}
\def\ba{\begin{array}}
\def\ea{\end{array}}

\begin{document}

\include{preamble}

\begin{titlepage}

\title{Drinfel'd Double of Bialgebroids for String and M Theories: \\ Dual Calculus Framework}

\author{Aybike Çatal-Özer, Keremcan Doğan$^{\dagger}$, Cem Yetişmişoğlu\footnote{E-mails: ozerayb[at].itu.edu.tr, dogankerem[at]itu.edu.tr, yetismisoglu[at]itu.edu.tr} \\ 
\small{Department of Mathematics, İstanbul Technical University, İstanbul, Türkiye} \\
\small{$^{\dagger}$Corresponding author}}

\date{}

\maketitle

\begin{abstract}
\noindent We extend the notion of Lie bialgebroids for more general bracket structures used in string and M theories. We formalize the notions of calculus and dual calculi on algebroids. We achieve this by reinterpreting the main results of the matched pairs of Leibniz algebroids. By examining a rather general set of fundamental algebroid axioms, we present the compatibility conditions between two calculi on vector bundles which are not dual in the usual sense. Given two algebroids equipped with calculi satisfying the compatibility conditions, we construct its double on their direct sum. This generalizes the Drinfel'd double of Lie bialgebroids. We discuss several examples from the literature including exceptional Courant brackets. Using Nambu-Poisson structures, we construct an explicit example, which is important both from physical and mathematical point of views. This example can be considered as  the extension of triangular Lie bialgebroids in the realm of higher Courant algebroids, that automatically satisfy the compatibility conditions. We extend the Poisson generalized geometry by defining Nambu-Poisson exceptional generalized geometry and prove some preliminary results in this framework. We also comment on the global picture in the framework of formal rackoids and we slightly extend the notion for vector bundle valued metrics. 
\end{abstract}

\vskip 2cm



\thispagestyle{empty}

\end{titlepage}

\maketitle


\section{Introduction and Motivation}
\label{s1}

\noindent String and M theories are strong candidates for a coherent theory of quantum gravity, which unifies all the known forces in nature. These theories  require extra dimensions for their consistency, and consequently they depend on a compactification procedure. Toroidal compactification of Type~II string theories yields an effective action in lower dimensions, which is known to have a duality symmetry containing the group $O(d, d)$, where $d$ is the dimension of the internal torus \cite{giveon1989duality, giveon1994target}. Double field theory (DFT) provides a reformulation of the effective action, where this $O(d, d)$ symmetry group becomes manifest \cite{hull2009double, hull2009gauge,  Hohm:2010pp, aldazabal2011effective, zwiebach2012doubled, Hohm:2011dv, Hohm:2013bwa, geissbuhler2013exploring}. This is achieved by introducing dual coordinates associated with the winding modes of strings, in addition to the usual spacetime coordinates. The action for DFT of the NS-NS sector of Type II string theory is written in terms of the generalized metric and the generalized dilaton field, whose gauge transformations are governed by the C-bracket.  The theory is consistent only when a certain type of constraint is imposed. When the fields and the gauge parameters do not depend on the dual coordinates, the constraint is trivially satisfied and the theory is said to be in the supergravity frame. In this case, the action for DFT reduces to the Type II supergravity action and the gauge transformation of the generalized metric yields for the metric $g$ and the Kalb-Ramond 2-form $B$-field the following usual transformation rules:
\begin{align}
    \delta_{U + \omega} g &= \mathcal{L}_U g \, , \nonumber\\
    \delta_{U + \omega} B &= \mathcal{L}_U B + d \omega \, ,
\end{align}
which closes as \cite{hull2009gauge}
\begin{equation}
    [\delta_{U + \omega}, \delta_{V + \eta}] = \delta_{[U + \omega. V + \eta]_{\text{Cour}}} \, .
\end{equation}
Here, the Courant bracket is defined as
\begin{equation}
    [U + \omega, V + \eta]_{\text{Cour}} := [U, V]_{\text{Lie}} \oplus \mathcal{L}_U \eta - \mathcal{L}_V \omega + \frac{1}{2} d \left( \iota_U \eta - \iota_V \omega \right) \, .
\label{standard}
\end{equation}
This bracket first appeared in \cite{courant1990dirac}, and it is a fundamental ingredient for the generalized geometry~\cite{hitchin2010lectures}. It is a natural bracket on the sections of the generalized tangent bundle
\begin{equation}
    TM \oplus T^*M \, .
\label{gtb}
\end{equation}
Equipped with the bracket in \eqref{standard}, the vector bundle \eqref{gtb} has the structure of a Courant algebroid. In the supergravity frame, the gauge parameters are sections of the generalized tangent bundle and the generalized metric encoding the metric and the $B$-field is a tensor on the generalized tangent bundle \cite{hull2009gauge}.

The Courant algebroid structure is also convenient to describe the geometric and non-geometric fluxes appearing in string theory compactifications. It has been long understood that under T-duality, a compactification on a $d$-dimensional torus with non-trivial $H$-flux $H_{abc}$ threading the cycles of the torus is dual to compactification on a twisted torus with no $H$-flux \cite{kachru2003new, shelton2005nongeometric, Grana:2006is, grana2009t}, where the twist is measured by the so-called geometric flux $f^{a}{}_{ b c}$. Further T-dualities lead to the following chain involving the \textit{non-geometric} $Q$-flux and $R$-flux \cite{shelton2005nongeometric}:
\begin{equation}
    H_{a b c} \longleftrightarrow f^{a}{}_{ b c} \longleftrightarrow Q^{a b}{}_{c} \longleftrightarrow R_{a b c} \, .
\end{equation}
These fluxes are the structure functions of the Kaloper-Myers algebra \cite{kaloper1999dd}:
\begin{align}
    [e_a, e_b] &= f^c{}_{a b} e_c + H_{a b c} e^c \, , \nonumber\\
    [e_a, e^b] &= Q^{b c}{}_{a} e_c + f^b{}_{ a c} e^c \, , \nonumber\\
    [e^a, e^b] &= R^{a b c} e_c + Q^{a b}{}_{c} e^c \, ,
\label{km}
\end{align}
giving the form of the most general gauge algebra arising from a generalized duality twisted reduction \cite{aldazabal2011effective}. 

There is a flux reformulation of DFT, where the geometric and non-geometric fluxes become `dynamical' and field dependent \cite{aldazabal2011effective, geissbuhler2013exploring}. If $(e_a, e^a)$ is a frame for the generalized tangent bundle underlying the geometry of DFT, the above algebra \eqref{km} gives the most general form of the bracket in this frame, sometimes referred to as the Roytenberg bracket \cite{roytenberg2002quasi}, governing the gauge algebra of DFT under the strong constraint. Consistency of the action following from the closure of the gauge algebra imposes some constraints on the dynamical fluxes, usually referred to as  \textit{Bianchi identities} \cite{geissbuhler2013exploring, blumenhagen2012bianchi}. These identities can be identified with the Courant algebroid axioms written in local coordinates \cite{ikeda2003chern, chatzistavrakidis2018double}. On the other hand, given the data of a Courant algebroid, one can write down a unique membrane sigma model  \cite{roytenberg2007aksz}:
\begin{equation} 
    S[X, A, F] = \int_{\Sigma_3} \left( F_i \wedge d X^i + \frac{1}{2} \eta_{I J} A^I \wedge d A^J - \rho^i_{\ I} A^I \wedge F_i + \frac{1}{6} T_{I J K} A^I \wedge A^J \wedge A^K \right) \, ,
\end{equation} 
where $T_{IJK}$ are the fluxes. Gauge invariance of this Courant sigma model is also guaranteed by Bianchi identities on the structure functions of the Courant algebroid in local coordinates (the dynamical fluxes) imposed by the Courant algebroid axioms \cite{roytenberg2007aksz, ikeda2003chern}. This triple point of view is investigated and generalized in \cite{chatzistavrakidis2019fluxes} to M-theory fluxes for $SL(5)$ case.

One of the most fundamental results on Courant algebroids is the \v{S}evera classification theorem for the exact ones \cite{vsevera2017letters}, which gives a one-to-one correspondence between isomorphism classes of exact Courant algebroids and the third de Rham cohomology class of the base manifold. According to this result, the bracket on any exact Courant algebroid $E$ can be written as
\begin{equation}
    [U + \omega, V + \eta]_E = [U, V]_{\text{Lie}} \oplus \mathcal{L}_U \eta - \mathcal{L}_V \omega + d \iota_V \omega + H(U, V) \, ,
\label{severaintro}
\end{equation}
where the $H$-flux is given by a closed 3-form. Here, the first four terms on the right-hand side constitute the standard Dorfman bracket whose anti-symmetrization is the standard Courant bracket. The first appearance \cite{liu1997manin} of Courant algebroids was as the Drinfel'd doubles of Lie bialgebroids, where the latter notion is defined in terms of two Lie algebroid structures on two dual vector bundles $A$ and $A^*$ satisfying the following compatibility condition \cite{mackenzie1994lie}:
\begin{equation}
    d^*[U, V]_{\text{Lie}} = \mathcal{L}_U d^* V - \mathcal{L}_V d^* U \, .
\label{compintro}
\end{equation}
Given a Lie bialgebroid $(A, A^*)$, the following ``doubled'' bracket induces a Courant algebroid structure on the Drinfel'd double $A \oplus A^*$:
\begin{equation}
    [U + \omega, V + \eta]_{A \oplus A^*} = [U, V]_A + \mathcal{L}^*_{\omega} V - \mathcal{L}^*_{\eta} U + d^* \iota_U \eta \oplus [\omega, \eta]_{A^*} + \mathcal{L}_U \eta - \mathcal{L}_V \omega + d \iota_V \omega \, ,  
\label{bracketintro}
\end{equation}
which has the same form as the D-bracket of DFT \cite{hull2009gauge}. Such Courant algebroids are a natural generalization of the Drinfel'd double of Lie bialgebras, and play an important role in T-duality, non-abelian T-duality and more generally Poisson Lie T-duality. If a given supergravity solution is Poisson T-dualizable \cite{klimvcik1995dual, Klimcik:1995dy}, a frame $(e_a)$ for the tangent bundle can be constructed in such a way that together with the dual coframe for the cotangent bundle, they form an algebra, which is just the algebra in \eqref{km} with $R = H = 0$, and $f$ and $Q$ constants \cite{Hassler:2017yza, Demulder:2018lmj, Demulder:2019bha, Sakatani:2019jgu, Thompson:2019ipl, Catal-Ozer:2019tmm}. This means that the commutation relations for $(e_a)$ is that of a Lie algebra $\mathfrak{g}$, whereas the Lie algebra for the dual frames can be identified with $\mathfrak{g}^*$ with a bracket compatible with the bracket on $\mathfrak{g}$, so that $(\mathfrak{g}, \mathfrak{g}^*)$ is a Lie bialgebra. The Drinfel'd double is then the unique Lie algebra on $\mathfrak{d} = \mathfrak{g} \oplus \mathfrak{g}^*$ satisfying certain conditions. It is worth noting that this algebraic structure on the double is an example of the Courant algebroid where the base manifold of the underlying vector bundle is taken to be a point and the fiber is the Lie algebra $\mathfrak{g}$. The triple $(\mathfrak{d}, \mathfrak{g}, \mathfrak{g}^*)$ is an example of a Manin triple. Poisson-Lie T-duality relies on the idea that a given Drinfel'd double (up to isomorphism) can be decomposed into different Manin triples. If a different Manin triple gives the same Drinfel'd double, one can construct a set of vielbeins associated with this new Manin triple and it can be shown that the corresponding background is also a solution of supergravity \cite{klimvcik1995dual}. 

It is a well-known fact that Lie bialgebras and Manin triples are in one-to-one correspondence. One can also define the notion of matched pairs of Lie algebras, and they also have a one-to-one correspondence with Lie bialgebras \cite{mokri1997matched}. This triple correspondence can be extended to the realm of Leibniz algebras \cite{tang2022leibniz}. However, the theory presented in \cite{tang2022leibniz} is not an extension of Lie bialgebras. One of the aims of this paper is to take the first steps towards this correspondence in more general algebroid structures, even though we leave the full analysis of this possible generalization in a future study. We expect that this would be useful for non-abelian generalizations of U-duality analogously to the Poisson-Lie T-duality and Manin triples relation.

Mimicking the ideas coming from Poisson-Lie T-duality, exceptional Drinfel'd algebras have been introduced recently \cite{sakatani2020u, malek2020poisson}, custom-tailored for $SL(5)$ U-duality. Here, the relevant bundle is~$TM \oplus \Lambda^2 T^*M $ where $M$ is a 4-dimensional manifold equipped with the action of a Lie group $G$ with Lie algebra $\mathfrak{g}$. The exceptional Drinfel'd algebra $\mathcal{E}$ is an extension of $\mathfrak{g}$ and was constructed in \cite{malek2020poisson} led by the guiding principle that it should be the same as the gauge algebra of the 7-dimensional gauged supergravity resulting from a compactification of 11-dimensional supergravity on the 4-dimensional base $M$. In \cite{sakatani2020u}, the same algebra is obtained in a somewhat reverse way. They start by constructing the `generalized vielbeins' by utilizing a trivector constructed in a way analogous to the construction of the bivector associated with the Poisson-Lie group structure required for Poisson-Lie T-duality. Then they deduce the form of the algebra $\mathfrak{d}$ by looking at what happens at a certain point on the base manifold where the trivector is assumed to vanish. The second approach is more aligned with the purposes of this paper. Indeed, our main focus will be on the algebroid structure: a direct sum vector bundle equipped with a bracket which determines a generalized Lie derivative acting on its sections. For this reason, in prelude we outline exceptional Drinfel'd algebras following \cite{sakatani2020u}. The $SL(5)$ exceptional Drinfel'd algebra $\mathcal{E}$ turns out to be a 10-dimensional Leibniz algebra. For certain manifolds $M$, it is possible to construct generalized vielbeins realizing this exceptional Drinfel'd algebra, which are sections of the vector bundle $TM \oplus \Lambda^2 T^*M$. The topic of exceptional Drinfel'd algebras quickly got attention and their generalizations for larger duality groups were constructed \cite{malek2021e6, blair2022generalised, kumar202310}. The $O(d, d)$ T-duality arising from the toroidal compactification of 11-dimensional supergravity on a $d$-dimensional torus extends to the U-duality group $E_{d(d)}$ (with $E_{4(4)} = SL(5)$). Exceptional field theories make these symmetries manifest, in a similar way to how DFT makes the $O(d, d)$ T-duality group manifest, by extending the spacetime with the introduction of dual coordinates associated with string and membrane charges   \cite{Berman:2011jh,Hohm:2013pua,Hohm:2013vpa,Berman:2020tqn, Musaev:2015ces}.

Depending on the dimension, the field content of the theory and the relevant vector bundle that underlies the necessary geometry change \cite{hull2007generalised}. Here is a list of some direct sum bundles, each of which is relevant for a different duality group: 
\begin{align}
    & TM \oplus T^*M\, , \nonumber\\
    & TM \oplus \Lambda^2 T^*M\, , \nonumber\\
    & TM \oplus \Lambda^2 T^*M \oplus \Lambda^5 T^*M\, , \nonumber\\
    & TM \oplus \Lambda^2 T^*M \oplus \Lambda^5 T^*M \oplus \left( T^*M \otimes \Lambda^7 T^*M \right) \, . 
\label{physicsbundles}
\end{align} 
In order to find the analogues of Poisson-Lie T-dualizable backgrounds for U-duality, one should be able to construct vielbeins, which are sections of these extended vector bundles and which realize the exceptional Drinfel'd algebra.

Extending the ideas of exceptional Drinfel'd algebras, one would expect the individual summands in the direct sums (\ref{physicsbundles}) to be equipped with distinct algebra structures satisfying certain compatibility conditions so that their ``doubles'' on the full vector bundle can be equipped with a bracket inducing a desired algebra structure. This would necessitate the \textit{addition} of two algebra or more generally algebroid structures, leading one to the realm of bialgebroids \cite{mackenzie1994lie} and matched pairs of algebroids \cite{ibanez2001matched}. Considering the form of the bracket (\ref{bracketintro}) on the Drinfel'd double of a Lie bialgebroid and the compatibility condition (\ref{compintro}) given in terms of Lie derivatives, interior product and exterior derivatives, one would expect a generalization of the notion of Cartan calculus to be relevant for more general bialgebroid structures. Moreover, as the \v{S}evera classification theorem for exact Courant algebroids forces the bracket to have the form (\ref{severaintro}), and its proof heavily relies on Cartan calculus relations, it is clear that there is a relation between the properties of the Cartan calculus and the Courant algebroid axioms. As the terms in the bracket is defined in terms of Cartan calculus elements, relaxing the axioms defining the bracket amounts to introducing new analogous operators which satisfy a relaxed set of properties compared to the usual Cartan calculus. This observation was the main motivation of our previous work \cite{ccatal2022pre} in which we extended \v{S}evera's fundamental result for a broader class of ``exact'' algebroids. On the other hand, the exact cases are too restrictive as two distinct non-trivial subalgebra structures in exceptional Drinfel'd algebras indicate. Furthermore, the physically motivated examples as in Equation (\ref{physicsbundles}) are seemingly arbitrary from a mathematical point of view, and the summands in the direct sums are not \textit{dual} in the usual sense. Hence, it is of importance to extend the ideas of Lie bialgebroids for vector bundles that are non-dual.

Our main aim is to extend the relation between Cartan calculus, Courant algebroid axioms and Drinfel'd doubles of Lie bialgebroids to vector bundles of the form
\begin{equation}
    E = A \oplus Z \, ,
\end{equation}
where $A$ and $Z$ are now not dual. Therefore, we first formalize the notion of \textit{calculus} on algebroids, and introduce a notion of \textit{duality} between two calculi. In order to achieve this, we reinterpret and improve the results from matched pair of algebroids literature \cite{mokri1997matched, ibanez2001matched}. Our strategy is to analyze the frequently used algebroid axioms in their rather general forms, and find out \textit{compatibility conditions} for each. We start with an exposition of the results from matched pair of Leibniz algebroids literature \cite{ibanez2001matched} in a form more familiar to physicists. As Leibniz algebroids only satisfy the right-Leibniz rule and the Jacobi identity, we aim to extend these results for other important algebroid axioms; the symmetric part of the bracket, metric invariance property, left-Leibniz rule and certain bracket morphisms. The symmetric part is especially crucial for our formalism, since a certain form of the symmetric part makes it possible to interpret the algebroid representations as generalizations of Cartan calculus. We consider bracket structures where the symmetric part can be decomposed as a bundle valued metric and a first-order differential operator acting on it \cite{bugden2021g, ccatal2022pre}. In light of such decompositions, we define the notion of calculus, and present the compatibility conditions coming from each algebroid axiom separately. These compatibility conditions between two calculi on $A$ and $Z$ make the pair $(A, Z)$ a \textit{bialgebroid} with desired properties. Moreover, these conditions are necessary for inducing an algebroid structure on their \textit{Drinfel'd double} $A \oplus Z$ with analogous properties. This notion of calculus is based on our earlier work \cite{ccatal2022pre}, where we also introduced metric-Bourbaki algebroids which satisfy each of the axioms we analyzed. We extend this work by introducing the notion of metric-Bourbaki bialgebroids, and show that many physically and mathematically motivated examples fit into this framework.

We then focus on the well-known relation between Poisson structures and triangular Lie bilgebroids \cite{mackenzie1994lie}, which satisfy the compatibility condition (\ref{compintro}) automatically. We present another construction of triangular Lie bialgebroids from the physics literature \cite{halmagyi2009non}. This construction is based on a twist of the Dorfman bracket given in terms of a bivector. This approach was used to construct the Roytenberg algebra \cite{roytenberg2002quasi} related to generalized actions involving Wess-Zumino terms based on their earlier work about current algebras \cite{halmagyi2008non}. By building on these ideas, we construct concrete examples of dual calculi by using Nambu-Poisson structures \cite{nambu1973generalized, takhtajan1994foundation}, which we expect to be relevant for U-duality. These structures can be considered as the generalization of \textit{triangularity} in the realm of higher Courant algebroids \cite{bi2011higher}. The calculus elements that we construct by using Nambu-Poisson structures are natural extension of the ones used in Poisson generalized geometry \cite{asakawa2015poisson}. In this sense we extend this framework to Nambu-Poisson exceptional generalized geometry and present some analogous results. As the algebroid structures that we construct by using Nambu-Poisson structures are based on the vector bundle
\begin{equation}
    TM \oplus \Lambda^p T^*M \, ,
\end{equation}
they are directly related to the fluxes of $SL(5)$ exceptional Drinfel'd algebras. We plan to further investigate this relation in the near future. We also present some observations about the exceptional Courant brackets \cite{pacheco2008m}, and their relations to our framework. Lastly, we touch upon the coquecigrue problem \cite{loday1993version} about the global aspects of group-like objects that are yielded by the \textit{integration} of algebroids. We slightly extend the notion of metric bundle rackoids introduced in \cite{ikeda2021global} for vector bundle valued metrics.

The organization of the paper is as follows: In Section \ref{s2}, we set the notation of the paper, point out our conventions, and summarize the basics of anchored vector bundles. In Section \ref{s3}, we explain the basics of Lie bialgebroids and Courant algebroids, and briefly mention exceptional Drinfel'd algebras. Section \ref{s4} starts with a summary of matched pairs of Leibniz algebroids in our own notation, and the rest of the section consists of a detailed analysis of algebroid axioms. Here, we present our main definitions about the notions of calculus and dual calculi, and we present compatibility conditions required for algebroid properties. Particularly in Subsection \ref{s4d}, we collect all of our previous observations and state our main theorems. In Section \ref{s5}, we present some examples from the mathematics literature. Section \ref{s6} consists of a crucial construction that generalizes the triangular Lie bialgebroids, where we make use of higher Courant algebroids and Nambu-Poisson structures. We continue with the discussion of another example central for physical motivations, namely the exceptional Courant brackets in Section \ref{s7}. In the first part of Section \ref{s8} we extend the Poisson generalized geometry to Nambu-Poisson exceptional generalized geometry, and in the second part we focus on metric bundle rackoids related to global picture. We finish the paper with some concluding remarks and possible research directions in Section \ref{s9}. We leave most of the proofs to the Appendix, where we also summarize Cartan calculus in the beginning.


\section{Notation and Conventions}
\label{s2}

Now we set our conventions that will be used and introduce the notation of the paper. Moreover, we recall some of the basics about anchored vector bundles and generalizations of fundamental differential geometric objects like tensors on arbitrary vector bundles.

We always consider an arbitrary connected, paracompact, Hausdorff, orientable, smooth manifold $M$. Moreover, every construction in this paper is assumed to be smooth. All vector bundles are assumed to be real and finite-rank, with a specified projection map over the same base manifold~$M$. In particular, the tangent bundle is denoted by $TM$, and the cotangent bundle, $T^*M$. The ring of real-valued smooth functions is denoted by $C^{\infty}M$, which can be considered as sections of~$\Lambda^0 T^*M$, where $\Lambda^p$ denotes the $p$th exterior power of a vector bundle, and in particular the sections of $\Lambda^p T^*M$ are (exterior differential) $p$-forms. Any vector field $V \in \Gamma TM$ acts as a derivation on a smooth function $f$, which is denoted by $V(f)$.  

We only consider vector bundle morphisms that are over the identity map. That is, given two vector bundles $E_1$, $E_2$, the vector bundle morphisms we consider are pairs of smooth maps of the form $(\Phi: E_1 \to E_2,\text{id}_M:M\to M)$ satisfying $\pi_2 \Phi = \text{id}_M \pi_1$ where $\pi_i: E_i \to M$ is the projection of the vector bundle. Therefore we ignore the identity map, $\text{id}_M$, from the notation and write only~$\Phi: E_1 \to E_2$. The crucial part of the definition of the vector bundle morphism is that the induced maps between the fibers are $C^{\infty}M$-linear. Note that we do not use any symbol for the composition of two maps; it is just juxtaposition.

For an arbitrary vector bundle, one can easily generalize some of the usual notions, including tensors, vector fields, $p$-forms, interior product, (local) frames and coframes (see for a detailed exposition \cite{dereli2021metric}). For example, a $(q, r)$-type tensor on a vector bundle $E$ is just a section of the vector bundle 
\begin{equation}
    \bigotimes_{i = 1}^q E \otimes \bigotimes_{j = 1}^r E^* \, .
\end{equation}
One crucial difference is that the sections of an arbitrary vector bundle do not act as derivations on smooth functions. Hence, one usually needs to introduce an additional structure, the \textit{anchor}, in order to be able to discuss Leibniz rules in different contexts, including connections and brackets. Any vector bundle $E$ in this paper is assumed to carry an anchor $\rho_E: E \to T M$, which is a vector bundle morphism, so that it induces a map which will be denoted by the same letter, $\rho_E: \Gamma E \to \Gamma TM$. The anchor also induces a map to the dual bundle,
\begin{equation}
    D_E := \rho_E^t d: C^{\infty} M \to \Gamma E^* \, ,
    \label{zazaza}
\end{equation}
where the $t$ superscript denotes the transpose of a vector bundle morphism, and $d$ is the usual exterior derivative. This map satisfies $(D_E f)(u) = \rho_E(u)(f) \, $, for all $f \in C^{\infty} M, u \in \Gamma E \, $. Given an anchored vector bundle $(E, \rho_E)$, one can define an $E$-connection on a vector bundle $R$ as an~$\mathbb{R}$-bilinear map $\nabla: \Gamma E \times \Gamma R \to \Gamma R$ satisfying \cite{fernandes2002lie}
\begin{align}
    \nabla_u (f r) &= f \nabla_u r + \rho_E(u)(f) r \, , \nonumber\\
    \nabla_{f u} r &= f \nabla_u r \, ,
\end{align}
for all $u \in \Gamma E, r \in \Gamma R, f \in C^{\infty}M$.

We will be interested in the vector bundles of the form of a direct sum $E = A \oplus Z$, and we use~$\oplus$ notation for the vector bundle sections when we want to make it clear the $A$ and~$Z$ parts. When we have such a direct sum decomposition, we use the notation $\text{pr}_A$ and $\text{pr}_Z$ for projection maps to $A$ and $Z$, respectively. Sections of $E$ will be denoted by small Latin letters $u, v, w$, whereas the sections of $A$ and $Z$ will be denoted by capital Latin $U, V, W$, and small Greek letters $\omega, \eta, \mu$, respectively since they are in some sense generalizations of $TM$ and $T^*M$ in light of exact Courant algebroids of the form $TM \oplus T^*M$.

In the following parts of the paper, we will drop the section notation $\Gamma$, so when we write~$u \in E$, it actually means $u \in \Gamma E$. Moreover, every map between vector bundles will be understood as section-wise. We will also shamelessly behave like $C^{\infty}M$ is a fiber bundle and write for example~$\mathcal{A} = C^{\infty}M$ for a vector bundle $\mathcal{A}$. We will use Einstein's summation convention of repeated indices. The anti-symmetrization of a set of indices is indicated by $[\ldots]$-brackets on that set.

A caution is in order. We use the notation for the maps $\mathcal{L}, \iota, d$ in various contexts. For example, they are sometimes the usual Cartan calculus operators that are summarized in Appendix A, but they frequently denote their abstractions which we introduce in Section \ref{s4}.


\section{Prelude: Lie Bialgebras to Exceptional Drinfel'd Algebras}
\label{s3}

We now give a brief review of the notions of Lie bialgebroids \cite{mackenzie1994lie}, Drinfel'd doubles \cite{drinfeld1986quantum} and  Courant algebroids \cite{liu1997manin}, which are central to Poisson-Lie T-duality \cite{klimvcik1995dual}. In the rest of the paper, these structures will be used as prototypical examples, and we will be interested in certain generalizations of them. We close the section by making some comments on their extensions at the algebra level, that is, to exceptional Drinfel'd algebras \cite{sakatani2020u, malek2020poisson, malek2021e6} required for U-duality. 

A Lie bialgebra is a doublet $(\mathfrak{g}, \mathfrak{g}^*)$ where both $\mathfrak{g}$ and its dual $\mathfrak{g}^*$ carry Lie algebra structures compatible in a certain sense \cite{drinfeld1986quantum}. Given a Lie bialgebra structure $(\mathfrak{g}, \mathfrak{g}^*)$, there is a unique Lie algebra structure on $\mathfrak{d} = \mathfrak{g} \oplus \mathfrak{g}^*$ such that both $\mathfrak{g}$ and $\mathfrak{g}^*$ are Lie subalgebras and the canonical pairing is invariant under the adjoint action on $\mathfrak{d}$. One can extend the notion of Lie bialgebra to the algebroid framework \cite{mackenzie1994lie}. A Lie algebroid is a triplet $(A, \rho_A, [\cdot,\cdot]_A)$, where $\rho_A$ is the anchor and~$[\cdot,\cdot]_A$ is an anti-symmetric bracket satisfying the Jacobi identity, \textit{i.e.}, $(A, [\cdot,\cdot]_A)$ is a Lie algebra over $\mathbb{R}$, further satisfying the right-Leibniz rule
\begin{equation}
    [U, f V]_A = f [U, V]_A + \rho_A(U)(f) V \, ,
\end{equation}
for all $U, V \in A, f \in C^{\infty} M$\footnote{The triplet $(TM, \text{id}_{TM}, [\cdot,\cdot]_{\text{Lie}})$ is a Lie algebroid, where $\text{id}_{TM}$ is the identity map on $TM$ and $[\cdot,\cdot]_{\text{Lie}}$ is the usual Lie bracket of vector fields.}. A Lie bialgebroid is then a pair $(A, A^*)$ such that both $A$ and $A^*$ are Lie algebroids which are compatible in the sense that they satisfy \cite{mackenzie1994lie}
\begin{equation}
    d^*[U, V]_A = \mathcal{L}_U d^* V - \mathcal{L}_V d^* U \, ,
\label{compatibilityLie}
\end{equation}
or equivalently  \cite{kosmann1995exact}
\begin{equation}
   d^*[U, V]_A = [U, d^*V]_{\text{SN},A} + [d^*U, V]_{\text{SN},A} \, .
\end{equation}
Here, the Schouten-Nijenhuis bracket $[\cdot,\cdot]_{\text{SN},A}: \Lambda^p A \times \Lambda^q A \to \Lambda^{p + q - 1} A$  is the unique extension of the bracket $[\cdot,\cdot]_A$ defined by 
\begin{equation}
    [U_1 \wedge \ldots \wedge U_p, V_1 \wedge \ldots \wedge V_q]_{\text{SN},A} := \sum_{i = 1}^p \sum_{j = 1}^q  (-1)^{i + j} [U_i, V_j]_A \wedge U_1 \wedge \ldots \wedge \check{U_i} \wedge \ldots \wedge U_p \wedge V_1 \wedge \ldots \wedge \check{V_j} \wedge \ldots \wedge V_q \, ,
\label{schouten}
\end{equation}
that makes the multivector fields a Gerstenhaber algebra. Here, $\check{U}$ indicates that $U$ is excluded. On a frame $(e_a)$ of $A$ with dual coframe $(e^a)$ on $A^*$, the compatibility condition (\ref{compatibilityLie}) reads 
\begin{equation}
    \rho_{A^*} \left( e^{\left[ m \right.} \right) \left( f_{a b}{}^{\left. n \right]} \right) - \frac{1}{2} f_{a b}{}^k \tilde{f}^{m n}{}_k = - \rho_A \left( e_{\left[ a \right.} \right) \left( \tilde{f}^{m n}{}_{\left. b \right]} \right) + 2 \tilde{f}^{k \left[ n \right.}{}_{\left[ a \right.} f_{\left. b \right] k}{}^{\left. m \right]} \, ,
\label{complieframe}
\end{equation}
where $f$ and $\tilde{f}$ are the structure functions of algebroids $A$ and $A^*$, respectively. The maps $\mathcal{L}, d$ and~$\mathcal{L}^*, d^*$ are the Lie derivative and exterior derivative induced by the Lie algebra structures $[\cdot,\cdot]_A$ and $[\cdot,\cdot]_{A^*}$, respectively. The pair of maps $\mathcal{L}$ and $d$ can be defined as follows \cite{mackenzie1987lie}: For a $p$-form $\omega$ on $A$ with anchor $\rho_A$
\begin{equation} 
    (\mathcal{L}_V \omega)(V_1, \ldots, V_p) := \rho_A(V)(\omega(V_1, \ldots, V_p)) - \sum_{i = 1}^p \omega(V_1, \ldots, [V, V_i]_A, \ldots V_p) \, ,
\label{liederivative}
\end{equation}
and 
\begin{align}
    (d \omega)(V_1, \ldots, V_{p+1}) &:= \sum_{1 \leq i \leq p+1} (-1)^{i+1} \rho_A(V_i)(\omega(V_1, \ldots, \check{V_i}, \ldots, V_{p+1})) \nonumber\\
    & \quad \ + \sum_{1 \leq i < j \leq {p+1}} (-1)^{i + j} \omega([V_i, V_j]_A, V_1, \ldots, \check{V_i}, \ldots \check{V_j}, \ldots, V_{p+1}) \, ,
\label{exteriorderivative}
\end{align}
for all $V_i \in A$. With the interior product $\iota$, they enjoy the same properties as their usual Cartan calculus versions \cite{mackenzie1987lie}. Moreover, since these are valid for any arbitrary Lie algebroid,~$\mathcal{L}^*$ and $d^*$ can be defined analogously for the bracket $[\cdot,\cdot]_{A^*}$. Similarly to the algebra case, when $(A, A^*)$ is a Lie bialgebroid, so is $(A^*, A)$. Interestingly, the double $E = A \oplus A^*$ is not a Lie algebroid anymore, but instead a Courant algebroid \cite{liu1997manin} which is defined as a quadruplet $(E, \rho_E, [\cdot,\cdot]_E, g_E)$ such that the bracket satisfies the Jacobi identity, and the metric invariance property with respect to the metric $g_E: E \times E \to C^{\infty} M$, \textit{i.e.},
\begin{equation}
    \rho_E(u)(g_E(v, w)) = g_E([u, v]_E, w) + g_E(v, [u, w]_E) \, ,
\end{equation}
for all $u, v, w \in E$. Moreover, the symmetric part of the bracket should be given by
\begin{equation}
    [u, v]_E + [v, u]_E = g_E^{-1} D_E g_E(u, v) \, .
\end{equation}
In \cite{liu1997manin}, the following anti-symmetric bracket is introduced in order to construct the double of a Lie bialgebroid
\begin{align}
    [U + \omega, V + \eta]_{A \oplus A^*} &= [U, V]_A + \mathcal{L}^*_{\omega} V - \mathcal{L}^*_{\eta} U - d^*(U + \omega, V + \eta)_- \nonumber\\
    & \quad \ \oplus [\omega, \eta]_{A^*} + \mathcal{L}_U \eta - \mathcal{L}_V \omega + d (U + \omega, V + \eta)_- \, ,
    \label{maninbracket}
\end{align}
where the pairing $(\cdot, \cdot)_-$ is defined by
\begin{equation}
    (U + \omega, V + \eta)_- := \frac{1}{2} \left( \iota_V \omega - \iota_U \omega \right) \, .
\end{equation}
It is possible to choose a frame $(e_a)$ of $A$ with dual coframe $(e^a)$ on $A^*$ such that the bracket on the Drinfel'd double $A \oplus A^*$ takes the form
\begin{align} 
    [e_a, e_b]_{A \oplus A^*} &= f_{a b}{}^c e_c \, , \nonumber\\
    [e_a, e^b]_{A \oplus A^*} &= \tilde{f}^{b c}{}_a e_c -f_{a c}{}^b e^c \, , \nonumber\\
    [e^a, e^b]_{A \oplus A^*} &= \tilde{f}^{a b}{}_c e^c \, .
\end{align}
It should be noted that without the compatibility assumption (\ref{compatibilityLie}), the above bracket (\ref{maninbracket}) induces a metric algebroid structure \cite{vaisman2012geometry}, which can be thought as a Courant algebroid with the Jacobi identity relaxed, as has been proven in \cite{mori2020doubled}. Relaxation of different properties of algebroid structures is often useful for physical purposes. For example, in this case these metric algebroids are directly related to DFT algebroid structures \cite{chatzistavrakidis2018double}, which underlie the geometric setup for double field theory framework.

Courant algebroids were originally defined with respect to an anti-symmetric (Courant \cite{courant1990dirac} instead of Dorfman \cite{dorfman1987dirac}) bracket satisfying a modified version of the Jacobi identity \cite{liu1997manin}, and the original definition has been modified several times \cite{uchino2002remarks,roytenberg1999courant}. Most of the proofs are easier in the non-anti-symmetric language, and the symmetric parts of the brackets play a crucial role in this paper. Hence, we stick with the Dorfman bracket, which is defined on the generalized tangent bundle $TM \oplus T^*M$ by
\begin{equation}
    [U + \omega, V + \eta]_{\text{Dorf}} := [U, V]_{\text{Lie}} \oplus \mathcal{L}_U \eta - \mathcal{L}_V \omega + d \iota_V \omega \, ,
\end{equation}
for $U, V \in TM, \omega, \eta \in T^*M$. This bracket induces a Courant algebroid structure on the generalized tangent bundle with the metric given by the canonical pairing. 

Lie bialgebroids are defined on dual vector bundles $A$ and $A^*$. As we have mentioned, our main aim is to extend the notion of bialgebroids in a more general setting where the vector bundles are not dual as in (\ref{physicsbundles}), so we are interested in the direct sums of the forms
\begin{equation}
    E = A \oplus Z \, ,
\end{equation}
where $A$ and $Z$ are arbitrary. For example in Section \ref{s6}, a prominent example will be
\begin{equation}
    TM \oplus \Lambda^p T^*M \, ,
\end{equation}
where we make use of higher Courant algebroids and Nambu-Poisson structures. Equipping $A$ and $Z$ with algebroid structures with desired properties, we will construct the full bracket on the \textit{Drinfel'd double} $A \oplus Z$ in the framework of our dual calculus. We will present compatibility conditions for various algebroid axioms. In a sense, we will take the first steps towards extending the correspondence between matched pairs of Lie algebroids and Lie bialgebroids, where our primary source will be \cite{ibanez2001matched} where matched pair of Leibniz algebroids are studied. The third component of this correspondence, namely Manin triples, is expected to be of great importance for Nambu-Poisson U-duality since Manin triples of Lie algebras are directly related to Poisson-Lie T-duality~\cite{klimvcik1995dual}. This triple correspondence is recently extended to Leibniz algebra level in \cite{tang2022leibniz}, where some relevant earlier work also appeared in \cite{barreiro2016new}. The full analysis of this correspondence is out of scope of this paper.

Exceptional Drinfel'd algebras have been introduced almost simultaneously by Sakatani \cite{sakatani2020u} and Malek, Thompson \cite{malek2020poisson}, as generalizations of Drinfel'd double of Lie bialgebras custom-tailored for  U-duality. They can be defined as certain subalgebras of the Lie algebra of the exceptional Lie group $E_{n(n)}$, admitting a maximally isotropic subalgebra $\mathfrak g$ corresponding to symmetries of the ``physical spacetime'' in the extended spacetime of exceptional field theories. The isotropy condition is given by the following constraint \cite{malek2021e6} 
\begin{equation}
    \mathfrak{g} \otimes \mathfrak{g} |_{R_2} = 0 \, ,
\end{equation}
where $R_2$ is the representation appearing in the tensor hierarchy of exceptional field theory, associated with the adjoint bundle \cite{palmkvist2014tensor, malek2020poisson}. In order to apply U-duality, one needs to construct generalized vielbeins, which realize the algebra defined by the Dorfman-like bracket on the sections of an appropriate vector bundle. One of the simplest exceptional Drinfel'd algebras is related to the Lie group $SL(5)$ \cite{sakatani2020u}, and in this case the relevant vector bundle is
\begin{equation}
    E = TM \oplus \Lambda^2 T^*M \, ,
\label{vb}
\end{equation}
where $M$ is 4-dimensional.  When $M$ is a Nambu-Poisson Lie group equipped with a Nambu-Poisson trivector, the explicit construction of the generalized vielbeins is given in \cite{malek2020poisson, sakatani2020u}.

Choosing an appropriate basis for the exceptional Drinfel'd algebra $\mathcal{E}$, one has the following commutation relations for the generators:
\begin{align}
    [T_a, T_b]_{\mathcal{E}} &= f_{a b}{ }^c T_c \, , \nonumber\\
    [T_a, T^{b_1 b_2}]_{\mathcal{E}} &= \tilde{f}_a{}^{b_1 b_2 c} T_c + 2 f_{a c}{ }^{\left[ b_1 \right.} T^{\left. b_2 \right] c} \, , \nonumber\\
    [T^{a_1 a_2}, T_b]_{\mathcal{E}} &= - \tilde{f}_b{ }^{a_1 a_2 c} T_c + 3 f_{\left[ c_1 c_2 \right.}{ }^{\left[ a_1 \right.} \delta_{b]}{}^{\left. a_2 \right]} T^{c_1 c_2} \, , \nonumber\\
    [T^{a_1 a_2}, T^{b_1 b_2}]_{\mathcal{E}} & = - 2 \tilde{f}_d{ }^{a_1 a_2 \left[ b_1 \right.} T^{\left. b_2 \right] d}\, , 
\label{SL5EDA}
\end{align}
where $f_{a b}{}^c = f_{[a b]}{ }^c$ and $\tilde{f}_a{ }^{b_1 b_2 b_3} = \tilde{f}_a{ }^{\left[ b_1 b_2 b_3 \right]}$. Moreover, these structure constants should satisfy the following quadratic constraints
\begin{align}
    & 0 = f_{[a b}{ }^e f_{c] e}{ }^d \, , \nonumber\\
    & 0 = f_{b c}{ }^e \tilde{f}_e{}^{a_1 a_2 d} + 6 f_{e \left[ b \right.}{}^{\left[ d \right.} \tilde{f}_{c]}{}^{a_1a_2]c}
      \, , \nonumber\\
    & 0 = f_{d_1 d_2}{ }^{\left[ a_1 \right.} \delta_b{}^{\left. a_2 \right]} \tilde{f}_c{}^{d_1 d_2 e} \, , \nonumber\\
    & 0 = \tilde{f}_c{ }^{e a_1 a_2} \tilde{f}_e{ }^{d b_1 b_2} - 3 \tilde{f}_c{}^{e \left[ b_1 b_2 \right.} \tilde{f}_e{}^{ d] a_1 a_2} \, ,
\label{quadratic}
\end{align}
following from requirement of Jacobi identity. The first and last equations can be considered as Jacobi identities themselves, so that there are two Leibniz algebras (due to anti-symmetry, the one associated with structure constants $f$ is a Lie algebra), while the middle two equations have the role of compatibility conditions between these two algebras. Moreover, on the \textit{double} of these two algebras, we have another Leibniz algebra which is the exceptional Drinfel'd algebra $\mathcal{E}$. 

There are more complicated exceptional Drinfel'd algebras \cite{malek2021e6, blair2022generalised, kumar202310}, and a closer look on these algebras is important to have a better understanding of dualities. It is clear that a more rigorous approach on these structures is worth studying. We believe our formalism which we will present in the consecutive sections might be useful to answer some of the problems in these directions. We will have various preliminary comments on exceptional Drinfel'd algebras relating them to our framework, but the full analysis of them will be out of scope of this paper. We plan to study exceptional Drinfel'd algebras in our dual calculus formalism in the near future.


\section{Matched Pairs, Dual Calculi and Compatibility Conditions}
\label{s4}

\noindent In this section, we will first review matched pairs of Leibniz algebroids following \cite{ibanez2001matched}. After a detailed summary of the general framework in the first subsection, we will focus on a specific subclass and introduce the notion of calculus on algebroids in the second subsection. These will naturally come up after an analysis of algebroid axioms used for generalizations of Courant algebroids, in particular the symmetric part of the bracket, metric invariance property, left-Leibniz rule and the effect of certain bracket morphisms. We will use these axioms as a guiding principle for introducing the notion of dual calculus together with explicit compatibility conditions. Since Courant algebroid axioms, Bianchi identities \cite{blumenhagen2012bianchi} and gauge closure conditions for sigma models \cite{ikeda2003chern} create a \textit{triple point}, which can be extended to higher Courant algebroid case for $SL(5)$ M theory \cite{chatzistavrakidis2019fluxes}, we expect these generalizations on arbitrary vector bundles to be useful for a better understanding of a \textit{larger} class of physically motivated examples. In particular, matched pairs of Leibniz algebras is the natural framework for exceptional Drinfel'd algebras \cite{sakatani2020u, malek2020poisson}, since latter is of the form of the direct sum of two Leibniz algebra structures.


\subsection{Matched Pairs of Leibniz Algebroids}
\label{s4a}

\noindent Here we summarize the notion of matched pairs of Leibniz algebroids, where the relevant algebroid axioms are the right-Leibniz rule and Jacobi identity. We will closely follow the fundamental work \cite{ibanez2001matched}, and explain their main results from a physicist-friendly perspective and in our own notation.

Matched pairs of algebraic structures are abundant in the mathematics literature. Drinfel'd's work on Poisson-Lie groups unveiled the notion of a Lie bialgebra, where a Lie algebra structure on a direct sum of a Lie algebra with its dual has been obtained \cite{drinfel1990hamiltonian}. This work has been expanded to so called twilled extension of Lie algebras where one asks for a Lie algebra structure on a direct sum of two arbitrary Lie algebras, not necessarily dual to each other \cite{kosmann1988poisson}. These works are extended further to the matched pairs of groups on Poisson geometries. In particular in \cite{lu1990poisson, kosmann2004lie}, doubles of Poisson-Lie groups and their dressing transformations are studied, and in \cite{majid1990matched}, matched pairs of Lie groups that come from solutions to Yang-Baxter equation are studied. In the latter, the term ``matched pair'' was first used for a pair of Lie algebras. Another generalization of these works come from generalizing the notions of algebras/groups to algebroids/groupoids. Mackenzie extended the notion of double of a Lie group to Lie groupoids \cite{mackenzie1992double}. Integration of Lie bialgebroids and their relation to Poisson groupoids are investigated in \cite{mackenzie1994lie, mackenzie2000integration}. In \cite{mokri1997matched}, matched pairs of Lie algebroids are introduced and their correspondence with matched pairs of Lie groupoids are studied. Similarly in \cite{ibanez2001matched}, matched pairs of Leibniz algebroids are studied, which will be the main reference for this subsection.

A pair of Leibniz algebras is called a matched pair if their direct sum carries a Leibniz algebra structure such that both of the initial algebras are subalgebras \cite{agore2013unified}. This can be elevated to the algebroid level \cite{ibanez2001matched}. A Leibniz algebroid is defined as a triplet $(E, \rho_E, [\cdot,\cdot]_E)$, where $E$ is a vector bundle with the anchor $\rho_E$ whose sections carry a Leibniz algebra structure over $\mathbb{R}$, \textit{i.e.}, an $\mathbb{R}$-bilinear bracket $[\cdot,\cdot]_E: E \times E \to E$ satisfying the Jacobi identity
\begin{equation} 
    [u, [v, w]_E]_E = [[u, v]_E, w]_E + [v, [u, w]_E]_E \, ,
\label{jacobiloday}
\end{equation}
and the bracket additionally satisfies the right-Leibniz rule
\begin{equation}
    [u, f v]_E = f [u, v]_E + \rho_E(u)(f) v \, ,
\label{leibnizright}
\end{equation}
for all $u, v, w \in E, f \in C^{\infty}M$. When $E$ is a Lie algebroid, that is a Leibniz algebroid with an anti-symmetric bracket, the above Jacobi identity takes the more familiar, cyclic form. In the literature, the definition of Leibniz algebroids often includes also the anchor being a morphism of Leibniz algebroids, that is 
\begin{equation}
    \rho_E ([u, v]_E) = [\rho_E(u), \rho_E(v)]_{\text{Lie}} \, ,
\label{anchormorphism}
\end{equation}
where the tangent Lie algebroid is just the tangent bundle equipped with the Lie bracket, and the identity map as its anchor.
However, Condition (\ref{anchormorphism}) is already implied by the Jacobi identity~(\ref{jacobiloday}) and right-Leibniz rule (\ref{leibnizright}), therefore it is redundant. The proof of this small identity can be found in the Appendix B. 

There is a direct relation between matched pairs of Leibniz algebroids and algebroid representations \cite{ibanez2001matched}. A Leibniz (algebroid) representation of a Leibniz algebroid $A$ on a vector bundle $Z$ is defined as a pair of maps $\mathcal{L}: A \times Z \to Z$ and $\mathcal{K}: A \times Z \to Z$ satisfying:
\begin{align}
     \mathcal{L}_{[U,V]_A} \omega &= \mathcal{L}_U \mathcal{L}_V \omega - \mathcal{L}_V \mathcal{L}_U \omega \, , \nonumber\\
     \mathcal{K}_{[U, V]_A} \omega &= \mathcal{K}_V \mathcal{K}_U \omega + \mathcal{L}_U \mathcal{K}_V \omega \, , \nonumber\\
     \mathcal{K}_U \mathcal{K}_V \omega &= - \mathcal{K}_U \mathcal{L}_V \omega \, ,
\label{leibnizrepresentation}
\end{align}
together with the $C^{\infty}M$-linearity properties
\begin{align}
    \mathcal{L}_U(f \omega) &= f \mathcal{L}_U \omega + \rho_A(U)(f) \omega \, , \nonumber\\
    \mathcal{K}_{f U} \omega &= f \mathcal{K}_U \omega \, ,
\label{leibnizlinearity}
\end{align}
for all $U, V \in A, \omega \in Z, f \in C^{\infty}M$. The first three conditions (\ref{leibnizrepresentation}) mean that the pair of maps~$\mathcal{L}, \mathcal{K}$ form a Leibniz algebra representation \cite{loday1993version}. These maps correspond to $\varphi_1$ and $\varphi_2$ of \cite{ibanez2001matched}, but the domain of $\varphi_2$ is usually taken as $Z \times A$ as it corresponds to a right-action. However this is just a matter of taste, and our choice is better suited for the purposes of this paper, which will become clear later when we choose particular representations. Conditions (\ref{leibnizrepresentation}) are in order to be able to satisfy the Jacobi identity. The remaining conditions (\ref{leibnizlinearity}) are for the right-Leibniz rule, and they make a Leibniz algebra representation to a Leibniz algebroid representation. In particular, Lie algebroid representations are defined with only one map $\mathcal{L}$ \cite{mokri1997matched}, which is given by an $A$-connection on $Z$, and it induces a Leibniz representation with the choice $\mathcal{K} = - \mathcal{L}$. 

With this machinery, the main theorem of \cite{ibanez2001matched} can be stated as follows. Two Leibniz algebroids~$A$ and $Z$ form a matched pair of Leibniz algebroids $(A, Z)$ if and only if there exist two sets of Leibniz representations, one of $A$ on $Z$ given by a pair $\mathcal{L}, \mathcal{K}$, and another one of $Z$ on $A$ given by a pair $\tilde{\mathcal{L}}, \tilde{\mathcal{K}}$, with $\tilde{\mathcal{L}}, \tilde{\mathcal{K}}: Z \times A \to A$ satisfying 
\begin{align}
    &[\mathcal{L}_U \omega, \eta]_Z + [\omega, \mathcal{L}_U \eta]_Z = \mathcal{L}_U [\omega, \eta]_Z - \mathcal{K}_{\tilde{\mathcal{K}}_\eta U} \omega - \mathcal{L}_{\tilde{\mathcal{K}}_\omega U} \eta \, , \nonumber\\
    &[\omega, \mathcal{K}_U \eta]_Z - [\eta, \mathcal{K}_U \omega]_Z = \mathcal{K}_U  [\omega, \eta]_Z - \mathcal{K}_{\tilde{\mathcal{L}}_\eta U} \omega + \mathcal{K}_{\tilde{\mathcal{L}}_\omega U} \eta \, , \nonumber\\
    &[\omega, \mathcal{L}_U \eta]_Z - [\mathcal{K}_U \omega, \eta]_Z = \mathcal{L}_U [\omega, \eta]_Z - \mathcal{K}_{\tilde{\mathcal{K}}_\eta U} \omega + \mathcal{L}_{\tilde{\mathcal{L}}_\omega U} \eta \, , \nonumber\\
    &\rho_Z (\mathcal{K}_U \omega) + \rho_A (\tilde{\mathcal{L}}_\omega U) = [\rho_Z(\omega), \rho_A(U)]_{\text{Lie}} \, ,
\label{compatibility1old}
\end{align}
and their ``dual'' conditions 
\begin{align}
    &[\tilde{\mathcal{L}}_\omega U, V]_A + [U, \tilde{\mathcal{L}}_\omega V]_A = \tilde{\mathcal{L}}_\omega [U, V]_A - \tilde{\mathcal{K}}_{\mathcal{K}_V \omega} U - \tilde{\mathcal{L}}_{\mathcal{K}_U \omega} V \, , \nonumber\\
    &[U, \tilde{\mathcal{K}}_\omega V]_A - [V, \tilde{\mathcal{K}}_\omega U]_A = \tilde{\mathcal{K}}_\omega [U, V]_A - \tilde{\mathcal{K}}_{\mathcal{L}_V \omega} U + \tilde{\mathcal{K}}_{\mathcal{L}_U \omega} V \, , \nonumber\\
    &[U, \tilde{\mathcal{L}}_\omega V]_A - [\tilde{\mathcal{K}}_\omega U, V]_A = \tilde{\mathcal{L}}_\omega [U, V]_A - \tilde{\mathcal{K}}_{\mathcal{K}_V \omega} U + \tilde{\mathcal{L}}_{\mathcal{L}_U \omega} V \, , \nonumber\\
    &\rho_A (\tilde{\mathcal{K}}_\omega U) + \rho_Z (\mathcal{L}_U \omega) = [\rho_A(U), \rho_Z(\omega)]_{\text{Lie}}\, , 
\label{compatibility2old}
\end{align}
for all $U, V \in A, \omega, \eta \in Z$. These are presented as Conditions L1-L4 in \cite{ibanez2001matched} in a way unrecognizable at first glance.
The last of both sets of conditions are due to anchor being a morphism of Leibniz algebroids, and the rest of them are due to Jacobi identity. Yet, as we have already mentioned, the right-Leibniz rule (\ref{leibnizright}) together with the Jacobi identity (\ref{jacobiloday}) imply that the anchor is a morphism of brackets. Therefore, the last conditions in both (\ref{compatibility1old}) and (\ref{compatibility2old}) are superfluous. Note that this main theorem also reproduces the one for matched pair of Lie algebroids given by Theorems 4.2 and 4.3 of \cite{mokri1997matched}. 

Given a matched pair of Leibniz algebroids $(A, Z)$, one can induce a Leibniz algebroid structure on the direct sum $E = A \oplus Z$ equipped with the anchor $\rho_E = \rho_A \oplus \rho_Z$ and the bracket
\begin{equation}
    [U + \omega, V + \eta]_E = [U, V]_A + \tilde{\mathcal{L}}_{\omega} V + \tilde{\mathcal{K}}_{\eta} U \oplus [\omega, \eta]_Z + \mathcal{L}_U \eta + \mathcal{K}_V \omega \, .    
\label{thebracket}
\end{equation}
Since both $A$ and $Z$ are subalgebroids, there are no ``twists'' in the sense that there are no maps of the form
\begin{equation}
    H: A \times A \to Z \, , \qquad \qquad \qquad \qquad \qquad R: Z \times Z \to A \, , \label{twists}
\end{equation}
which we call $H$- and $R$-twists following the physics nomenclature. Note that this bracket can be written as the sum of two Dorfman-like brackets:
\begin{align}
    [U + \omega, V + \eta]_{\mathcal{D}} &:= [U, V]_A \oplus \mathcal{L}_U \eta + \mathcal{K}_V \omega \, , 
\label{dbracket} \\ 
    [U + \omega, V + \eta]_{\tilde{\mathcal{D}}} &:= \tilde{\mathcal{L}}_\omega V + \tilde{\mathcal{K}}_\eta U \oplus [\omega, \eta]_Z \, .
\label{dtildebracket}
\end{align} 

In this subsection, we observed that the matched pair results only deal with the right-Leibniz rule and the Jacobi identity. Yet, Courant algebroids \cite{liu1997manin}, their natural generalizations including higher Courant algebroids \cite{bi2011higher}, and many others (see for instance \cite{Chen_2010, bugden2021g, ccatal2022pre}) satisfy also the metric invariance property and their bracket has a specific symmetric part. Hence, we will extend the results of matched pair literature by considering these additional axioms in a rather general form, further including the left-Leibniz rule and certain bracket morphism compatibility conditions.


\subsection{Calculus and Dual Calculi on Algebroids}
\label{s4b}

\noindent In this subsection, we further investigate the matched pairs for various algebroid axioms. Most importantly, we focus on the symmetric part of the bracket and assume that it can be decomposed in terms of a vector bundle valued metric and a first-order differential operator acting on it. In light of this decomposition, we focus on certain algebroid representations from the previous subsection, and reinterpret the results of matched pairs from a \textit{calculus} perspective. We also analyze the metric invariance property, left-Leibniz rule, and the effect of certain bracket morphisms. Metric invariance property is relevant for both physical motivations including exceptional Drinfel'd algebras \cite{sakatani2020u, malek2020poisson}, and mathematical structures like Manin triples \cite{liu1997manin}. By analyzing the algebroid axioms in their rather general forms, we first introduce the notion of calculus on algebroids, and then we explicitly write down \textit{compatibility conditions} for individual axioms. The compatibility conditions for the Jacobi identity are interpreted as the notion of duality between two calculi, and we construct Drinfel'd doubles for algebroids equipped with such dual calculi. Our results in this section can be considered as reinterpretations/improvements for both Lie bialgebroid \cite{mackenzie1994lie} and matched pair \cite{mokri1997matched, ibanez2001matched} literatures.

We start with the symmetric part of the bracket, where we observe that for many physically and mathematically motivated examples \cite{Chen_2010, bugden2021g, ccatal2022pre}, the symmetric part of the bracket takes the from that can be expressed in terms of a differential operator acting on a metric taking values in an arbitrary vector bundle. Hence, we consider brackets whose symmetric part can be decomposed as
\begin{equation}
    [u, v]_E + [v, u]_E = \mathbb{D}_E g_E(u, v) \, ,
\label{symmetricpart}
\end{equation}
for $u, v \in E$, where $g_E$ now takes values in an arbitrary vector bundle $\mathbb{E}$, and $\mathbb{D}_E: \mathbb{E} \to E$ is a first-order differential operator \cite{mackenzie1987lie}. For instance, for our prototypical example, Courant algebroids, the symmetric part is given by $g_E^{-1} D_E g_E$ for a metric $g_E $ taking values in $C^{\infty}M$, and $g_E^{-1} D_E$ is a first-order differential operator.  A similar decomposition is also considered in the definition of~$G$-algebroids \cite{bugden2021g}, where exceptional generalized geometries associated with $E_{n(n)} \times \mathbb{R}^+$ with $n < 7$ M theories are studied. We also note that, although this form of the symmetric part is quite general and cover many physically interesting examples, there are other constructions such as $Y$-algebroids which include the cases that cannot be put in the form of $(\ref{symmetricpart})$, in particular for $n = 7$ case \cite{hulik2023algebroids}. These algebroid structures are special cases of anti-commutable Leibniz algebroids \cite{dereli2021anti}; see Section \ref{s7} for further remarks. 

If we require for the bracket (\ref{thebracket}) to have such a symmetric part, we see that we should focus on certain Leibniz representations that admit a similar decomposition. The symmetric part contributes with $\mathcal{L} + \mathcal{K}$ and $\tilde{\mathcal{L}} + \tilde{\mathcal{K}}$ in addition to the symmetric parts of the brackets on $A$ and $Z$. Hence, forcing that the full bracket (\ref{thebracket}) to have a similar symmetric part decomposition of the form (\ref{symmetricpart}), we assume that the sum $\mathcal{L} + \mathcal{K}$ can be decomposed into $C^{\infty}M$-linear and first-order differential operator parts as well, \textit{i.e.},
\begin{equation}
    \mathcal{L}_V \omega + \mathcal{K}_V \omega = d \iota_V \omega  \, ,
\label{decomp}
\end{equation}
where the maps are defined as
\begin{equation}
    \iota: A \times Z \to \mathcal{Z} \, , \qquad \qquad \qquad \qquad \qquad d: \mathcal{Z} \to Z \, , 
\end{equation}
for some arbitrary vector bundle $\mathcal{Z}$. Here, we explicitly assume that $\iota$ is $C^{\infty}M$-bilinear, and~$d$ is a first-order differential operator as for the usual interior product and exterior derivative. Analogously to $\mathcal{L}$ and $\mathcal{K}$, we also assume that the sum of $\tilde{\mathcal{L}} + \tilde{\mathcal{K}}$ can be decomposed as 
\begin{equation} \label{calculusdecompose}
    \tilde{\mathcal{L}}_{\omega} V + \tilde{\mathcal{K}}_{\omega} V = \tilde{d} \tilde{\iota}_{\omega} V \, ,
\end{equation}
for a $C^{\infty}M$-bilinear map $\tilde{\iota}: Z \times A \to \mathcal{A}$ and a first-order differential operator $\tilde{d}: \mathcal{A} \to A$, for some arbitrary vector bundle $\mathcal{A}$. 

With these maps, the bracket (\ref{thebracket}) can be reexpressed as
\begin{equation}
[U + \omega, V + \eta]_E = [U, V]_A + \tilde{\mathcal{L}}_{\omega} V - \tilde{\mathcal{L}}_{\eta} U + \tilde{d} \tilde{\iota}_{\eta} U \oplus [\omega, \eta]_Z + \mathcal{L}_U \eta - \mathcal{L}_V \omega + d \iota_V \omega \, ,  
\label{thebracketold}
\end{equation}
for $U, V \in A, \omega, \eta \in Z$. We will interpret the maps $(\mathcal{L}, \iota, d)$ as generalizations of the usual Cartan calculus elements motivated from the decomposition (\ref{maninbracket}) in \cite{liu1997manin}, and \v{S}evera classification theorem of exact Courant algebroids \cite{vsevera2017letters}. This interpretation is natural since the maps $\mathcal{K}$ and $\tilde{\mathcal{K}}$ in the case of Lie bialgebroids are given by
\begin{equation}
    \mathcal{K}_V \omega = - \mathcal{L}_V \omega + d \iota_V \omega \, , \qquad \qquad \qquad \qquad \tilde{\mathcal{K}}_{\omega} V = - \mathcal{L}^*_{\omega} V + d^* \iota_V \omega \, ,
\end{equation}
where now the pair of maps $\mathcal{L}, d$ and $\mathcal{L}^*, d^*$ on the right-hand sides are the ones that are constructed as in (\ref{liederivative}, \ref{exteriorderivative}) from the brackets on $A$ and $A^*$. 

With the assumption that both $A$ and $Z$ have brackets whose symmetric part can be decomposed as in Equation (\ref{symmetricpart}) where their metrics $g_A$ and $g_Z$ take values in some arbitrary vector bundles $\mathbb{A}$ and $\mathbb{Z}$, we see that the bracket (\ref{thebracketold}) has a symmetric part given by the metric 
\begin{equation}
    g_E(U + \omega, V + \eta) = g_A(U, V) \oplus \left( \tilde{\iota}_{\omega} V + \tilde{\iota}_{\eta} U \right) \oplus g_Z(\omega, \eta) \oplus \left( \iota_U \eta + \iota_V \omega \right) \, , 
\label{metricdecomp}
\end{equation}
which takes values in the vector bundle
\begin{equation}
    \mathbb{E} = \mathbb{A} \oplus \mathcal{A} \oplus \mathbb{Z} \oplus \mathcal{Z} \, ,
\end{equation}
and the first-order differential operator
\begin{equation}
    \mathbb{D}_E = \mathbb{D}_A \oplus \tilde{d} \oplus \mathbb{D}_Z \oplus d 
\end{equation}
is acting on this metric. 

Next, we check the left-Leibniz rule and its consistency with the symmetric part. An algebroid~$E$ is said to satisfy the left-Leibniz rule \cite{grabowski2011supergeometry} if it is equipped with a \textit{locality operator} $L_E: C^{\infty} M \times E \times E \to E$\footnote{This map is $C^{\infty}M$-linear in the second and third entries and satisfies the Leibniz rule for the first entry. Yet, the map $L_E$ is usually defined as $C^{\infty}M$-multilinear map of the form $E^* \times E \times E \to E,$ with $L_E(D_E f, u, v)$ \cite{jurvco2015leibniz}.}, such that
\begin{equation}
    [f u, v]_E = f [u, v]_E - \rho_E(v)(f) u + L_E(f, u, v) \, .
    \label{left-leibniz}
\end{equation}
Algebroids satisfying both right- and left-Leibniz rules are sometimes referred as \textit{local} \cite{baraglia2012leibniz}, and it is possible to construct metric-connection geometries on them \cite{dereli2021metric}. Since the bracket of an arbitrary algebroid is not necessarily anti-symmetric as in the Lie algebroid case, one cannot directly induce the left-Leibniz rule from the other one. But the symmetric part has still an intricate relation with the locality operator, since the decomposition (\ref{symmetricpart}) forces one to have the locality operator \cite{ccatal2022pre}
\begin{equation}
    L_E(f, u, v) = \Delta_{\mathbb{D}_E}(g_E(u, v)) \, .
\label{loplop}
\end{equation}
Here, we denote the non-tensorial part, \textit{the symbol map},  of a first-order differential operator $\delta$ by~$\Delta_{\delta}$ which is defined as\footnote{For a first-order differential operator $\delta: E \to E'$, the map $\Delta_{\delta}$ is usually defined as a map $T^*M \times E \to E'$ of the form $\Delta_{\delta}(d f, u)$, which is $C^{\infty}M$-bilinear, where $d$ is the usual exterior derivative \cite{mackenzie2005general}. }
\begin{equation}
    \Delta_{\delta}(f, u) := \delta(f u) - f \delta(u) \, ,
\label{asadasda}
\end{equation}
for $f \in C^{\infty}M$. If the domain of the differential operator is the form of a product as for the maps~$\mathcal{L}$ and $\mathcal{K}$, then we use the notation $\Delta_{\delta}^{(i)}$ for the symbol map in the $i$-th entry.  

By straightforward calculations, we observe that the bracket (\ref{thebracketold}) on $E = A \oplus Z$ satisfies the left-Leibniz rule if and only if both $A$ and $Z$ satisfy it for some $L_A$ and $L_Z$ and 
\begin{align}
    \Delta_{\mathcal{L}}^{(1)}(f, U, \eta) &= \text{pr}_Z L_E(f, U, \eta) \, , \qquad  \Delta_{\mathcal{L}}^{(2)}(f, V, \omega) = \Delta_d(f, \iota_V \omega) + \rho_A(V)(f) \omega - \text{pr}_A L_E(f, \omega, V) \, , \nonumber\\
    \Delta_{\tilde{\mathcal{L}}}^{(1)}(f, \omega, V) &= \text{pr}_Z L_E(f, \omega, V) \, , \quad \ \ \, \Delta_{\tilde{\mathcal{L}}}^{(2)}(f, \eta, U) = \Delta_{\tilde{d}}(f, \tilde{\iota}_{\eta} V) + \rho_Z(\eta)(f) U - \text{pr}_A L_E(f, U, \eta) \, . 
\label{symbolsleft}
\end{align}
On the other hand, the $C^{\infty}M$-bilinearity assumption for $\iota$ together with the $C^{\infty}M$-linearity properties (\ref{leibnizlinearity}) of $\mathcal{L}$ and $\mathcal{K}$ force us to have
\begin{align}
    \Delta^{(1)}_{\mathcal{L}}(f, V, \omega) &=  \Delta_d(f, \iota_V \omega) \, , \nonumber\\
    \Delta^{(2)}_{\mathcal{L}}(f, V, \omega) &= \rho_A(V)(f) \omega \, .
\label{symbols}
\end{align}
The conditions (\ref{symbolsleft}) are consistent with these conditions (\ref{symbols}), since Equation (\ref{loplop}) implies that the locality operator $L_E$ can be decomposed as
\begin{equation}
    L_E(f, U + \omega, V + \eta) = L_A(f, U, V) + \Delta_{\tilde{d}} \left( f, \tilde{\iota}_{\omega} V + \tilde{\iota}_{\eta} U \right) \oplus L_Z(f, \omega, \eta) + \Delta_d \left( f, \iota_U \eta + \iota_V \omega \right) \, .
\label{locality}
\end{equation}

Before we analyze the remaining properties, we further investigate Leibniz representation definition, in particular the conditions coming from the Jacobi identity of the matched pairs of Leibniz algebroids \cite{ibanez2001matched}. We first observe that with Equation (\ref{decomp}), the third condition in (\ref{leibnizrepresentation}) takes the following form
\begin{equation}
    \mathcal{L}_W d \iota_V \omega - d \iota_W d \iota_V \omega = 0 \, .
\label{cuba}
\end{equation}
Moreover, the second condition in (\ref{leibnizrepresentation}) can be written as
\begin{equation}
    \mathcal{L}_U \mathcal{L}_W \omega - \mathcal{L}_W \mathcal{L}_U \omega  - \mathcal{L}_{[U, W]_A} \omega = - d \iota_{[U, W]_A} \omega - \mathcal{L}_W d \iota_U \omega - d \iota_W \mathcal{L}_U \omega + d \iota_W d \iota_U \omega + \mathcal{L}_U d \iota_W \omega \, ,
\end{equation}
after we use Equation (\ref{decomp}). Using the first equation of (\ref{leibnizrepresentation}), we see that the left-hand side vanishes, so that we get the following requirement
\begin{equation}
    - d \iota_{[U, W]_A} \omega - \mathcal{L}_W d \iota_U \omega - d \iota_W \mathcal{L}_U \omega + d \iota_W d \iota_U \omega + \mathcal{L}_U d \iota_W \omega = 0 \, .
\end{equation}
Now using Equation (\ref{cuba}) in this, we see that this becomes
\begin{equation}
    - d \iota_{[U, W]_A} \omega - d \iota_W \mathcal{L}_U \omega + \mathcal{L}_U d \iota_W \omega = 0 \, .
\end{equation}
Combining all of these results from Leibniz rules and Jacobi identity, we are led to the main definition of this paper: Given three vector bundles $A, Z$ and $\mathcal{Z}$, a triplet of maps $(\mathcal{L}, \iota, d)$ is called a \textit{calculus on $Z$ induced by $A$} if 
\begin{equation}
    \mathcal{L}: A \times Z \to Z \, , \qquad \qquad \iota: A \times Z \to \mathcal{Z} \, , \qquad \qquad d: \mathcal{Z} \to Z
\end{equation}
are first-order differential operators satisfying the following \textit{linearity conditions}
\begin{align}
    \Delta^{(1)}_{\mathcal{L}}(f, V, \omega) &= \Delta_d(f, \iota_V \omega) \, , \qquad \qquad \qquad \Delta^{(2)}_{\mathcal{L}}(f, V, \omega) = \rho_A(V)(f) \omega \, , \nonumber\\
    \Delta^{(1)}_{\iota}(f, V, \omega) &= 0 \, , \qquad \qquad \qquad \qquad \qquad \ \Delta^{(2)}_{\iota}(f, V, \omega) = 0 \, ,
\label{linearityconditions}
\end{align}
provided that the following constraints, \textit{calculus conditions,}
\begin{align}
    \mathcal{L}_U \mathcal{L}_V \mu - \mathcal{L}_V \mathcal{L}_U \mu - \mathcal{L}_{[U, V]_A} \mu &= 0 \, , \nonumber\\
    \mathcal{L}_U d \iota_W \eta - d \iota_{[U, W]_A} \eta - d \iota_W \mathcal{L}_U \eta &= 0 \, , \nonumber\\
        \mathcal{L}_W d \iota_V \omega - d \iota_W d \iota_V \omega &= 0 \, 
\label{calculusconditions}
\end{align}
are satisfied for all $f \in C^{\infty}M, U, V, W \in A, \omega, \eta, \mu \in Z$. We sometimes refer to a second set of calculus elements $(\tilde{\mathcal{L}}, \tilde{\iota}, \tilde{d})$ as \textit{tilde-calculus}. We present some details on the derivation of these linearity (\ref{linearityconditions}) and calculus (\ref{calculusconditions}) conditions in Appendix C. We also note that the first equation yields an interesting condition upon symmetrization
\begin{equation}
    \mathcal{L}_{\mathbb{D}_A g_A(U, V)} \mu = 0 \, . 
\end{equation}

First of the calculus condition is of course valid for the usual Lie derivative due to the Jacobi identity of the Lie bracket. The second one follows from the commutativity of the exterior and Lie derivatives, and the commutation relation of the latter with the interior product. Moreover, the last one holds for the usual operators since Cartan magic formula and the fact that exterior derivative squares to zero. Hence, our notion of calculus is a generalization of the usual Cartan calculus.

The notion of calculus introduced here is intended to be considered as a reinterpretation of algebroid representations with small additional requirements in the linearity conditions coming from left-Leibniz rule. In particular, given a calculus $(\mathcal{L}, \iota, d)$, the pair $\mathcal{L}$ and $\mathcal{K}$ becomes a Leibniz representation with the identification $\mathcal{K}_V = - \mathcal{L}_V + d \iota_V$. Of course not all algebroid representations can be decomposed as in Equation~(\ref{decomp}) into maps $\iota$ and~$d$. However when this is the case, we can view the algebroid representations as calculus elements subject to a more ``general'' set of Cartan calculus relations. We believe this perspective is especially propitious for understanding/extending algebroids appearing in physics literature for the simple reason that one typically works with usual Cartan calculus of exterior forms on manifolds. Moreover this practical point of view also holds for algebroids that appear in mathematics literature such as generalizations of Lie bialgebroids; see Section \ref{s5} for explicit demonstrations of this fact. We also note that there are other approaches involving generalizations of Cartan calculus. For instance in \cite{hohm2015tensor,wang2015generalized}, different generalizations of the Cartan calculus for exceptional field theories is studied. In another context, a graded generalization of calculus is introduced in \cite{dolgushev2008formality}. Moreover a notion of calculus is introduced on multigraded vector spaces and its relation with derived brackets is investigated in \cite{Kosmann_Schwarzbach_2004}. Derived bracket construction is also useful in physical context, for instance exceptional Courant brackets can be expressed neatly using them \cite{arvanitakis2018brane}.

Note that we use all the conditions, linear or quadratic, that do not mix the two calculus elements together in order to define a calculus. Yet, we see from the main theorem of the matched pair results \cite{ibanez2001matched}, there are some conditions (\ref{compatibility1old}) together with their duals (\ref{compatibility2old}) that do mix them in order to be able to satisfy the Jacobi identity. By using the decomposition of $\mathcal{K}, \tilde{\mathcal{K}}$ and the symmetric part (\ref{symmetricpart}) of the bracket $[\cdot,\cdot]_Z$, we observe that after some manipulation, the first three of the conditions (\ref{compatibility1old}) become
\begin{align}
    \mathcal{D}^Z_{\mathcal{L}_U}(\eta, \mu) &= \mathcal{L}_{\tilde{\mathcal{K}}_{\eta} U} \mu + \mathcal{K}_{\tilde{\mathcal{K}}_{\mu} U} \eta \, , \label{comp1} \\
    \mathcal{L}_{\tilde{d} \tilde{\iota}_{\eta} U} \mu &= - [d \iota_U \eta, \mu]_Z \, , \label{comp2} \\
    d \iota_{\tilde{\mathcal{L}}_{\omega} W} \eta - d \iota_{\tilde{d} \tilde{\iota}_{\eta} W} \omega + d \iota_W [\omega, \eta]_Z &= \mathbb{D}_Z g_Z(d \iota_W \eta, \omega) - \mathbb{D}_Z g_Z(\mathcal{K}_W \omega, \eta) \, .  \label{comp3} 
\end{align}
Here, we define the \textit{derivator} of a map $\Phi: E \to E$ as
\begin{equation}
    \mathcal{D}^E_{\Phi}(u, v) := \Phi [u, v]_E - [\Phi u, v]_E - [u, \Phi v]_E \, ,
\label{derivator}
\end{equation}
which is not a tensorial quantity in general. Equation (\ref{comp1}) is identical to the first equation in~(\ref{compatibility1old}). Equation (\ref{comp2}) follows from the third equation in (\ref{compatibility1old}) after taking the difference with Equation (\ref{comp1}). The last remaining one (\ref{comp3}) similarly follows from the second equation in (\ref{compatibility1old}) by using the other two. Completely analogously, from (\ref{compatibility2old}), we get the dual conditions as follows:
\begin{align}
    \mathcal{D}^A_{\tilde{\mathcal{L}}_{\omega}}(V, W) &= \tilde{\mathcal{L}}_{\mathcal{K}_V \omega} W + \tilde{\mathcal{K}}_{\mathcal{K}_W \omega} V \, , \nonumber\\
    \tilde{\mathcal{L}}_{d \iota_V \omega} W &= - [\tilde{d} \tilde{\iota}_{\omega} V, W]_A \, , \nonumber\\
    \tilde{d} \tilde{\iota}_{\mathcal{L}_U \mu} V - \tilde{d} \tilde{\iota}_{d \iota_V \mu} U + \tilde{d} \tilde{\iota}_{\mu} [U, V]_A &=  \mathbb{D}_A g_A(\tilde{d} \tilde{\iota}_{\mu} V, U) - \mathbb{D}_A g_A(\tilde{K}_{\mu} U, V)  \, . 
\label{compdual}
\end{align}

First of these conditions (\ref{comp1}), implies that the map $\tilde{\mathcal{K}}$ somewhat measures the deviation of~$\mathcal{L}$ to be a derivation of the bracket on $Z$. The second condition (\ref{comp2}) is identical to a crucial property of Lie bialgebroids as proven in Proposition 3.4 of \cite{mackenzie1994lie}, and it holds for Dirac structures as explained in Lemma 5.2 of \cite{liu1997manin}. For the third condition (\ref{comp3}), we observe that the symmetric part of the bracket $[\cdot,\cdot]_Z$ appear as a deformation to a condition which solely is given in terms of calculus elements. We refer to these six conditions (\ref{comp1}, \ref{comp2}, \ref{comp3}) together with their duals~(\ref{compdual}) as \textit{Jacobi compatibility conditions} between two calculi, and we call two calculi as each others' \textit{duals} when these conditions are satisfied.

As we will see, our calculus elements serve as good book-keeping devices. For instance in Section \ref{s6}, we will construct certain examples of tilde-calculus of the form
\begin{equation}
    \tilde{\mathcal{L}}_{\omega} V = [\Pi \omega, V]_{\text{Lie}} + \Pi \iota_V d \omega \, , 
\end{equation}
for the usual Cartan calculus elements, where $\Pi$ is a Nambu-Poisson structure. This map defines a Lie derivative-like object which takes the derivative of a vector with respect to a $p$-form, as this specific combination $[\Pi \omega, V]_{\text{Lie}} + \Pi \iota_V d \omega$ satisfies the properties of calculus. Our constructions might be also useful to present some of the results, including certain compatibility conditions, from the exceptional Drinfel'd algebras in a frame independent formalism. For instance, Bianchi identities for fluxes can be obtained from the Jacobi identity of the related bracket \cite{blumenhagen2012bianchi}, and our formalism make it possible to interpret these Bianchi identities as generalizations of Cartan calculus relations together with Jacobi compatibility conditions between two dual calculi. Compatibility conditions for pairs of Lie algebroids constitute the algebraic origin \cite{mori2020doubled} of the strong constraint of DFT related to metric algebroids of Vaisman \cite{vaisman2012geometry}. Our second Jacobi compatibility condition~(\ref{comp1}) is in particular interesting due to its relation to Dirac structures which have ubiquitous applications in both physical and mathematical contexts. As we will be interested in higher Courant algebroids \cite{bi2011higher} and higher Roytenberg bracket \cite{jurvco2013p}, we hope to relate the condition with higher Dirac structures~\cite{zambon2012l_, bursztyn2019higher}, which is out of scope of this paper. A similar analysis to our axiomatic approach is done in \cite{mori2020more}, where they closely investigate suitable subsets of Courant algebroid axioms, and write down certain compatibility conditions. 

Since Lie algebroids are Leibniz algebroids equipped with an anti-symmetric bracket, let us finish the section by commenting on the relation of these conditions to the ones that appear in the matched pair of Lie algebroids \cite{mokri1997matched}. The single relevant condition that we have for matched pairs of Lie algebroids is (\ref{comp1}) and its dual. This is because there is only a single map for a Lie algebroid representation with $\mathcal{K} = - \mathcal{L}$ and $\tilde{\mathcal{K}} = - \tilde{\mathcal{L}}$, so that both $d \iota$ and $\tilde{d} \tilde{\iota}$ vanish. Hence, the second~(\ref{comp2}), and the third (\ref{comp3}) Jacobi compatibility conditions together with their duals are trivially satisfied because one also has $g_A = g_Z = 0$. Moreover, the last remaining Jacobi compatibility condition~(\ref{comp1}) becomes
\begin{equation}
    \mathcal{D}^Z_{\mathcal{L}_U}(\eta, \mu) + \mathcal{L}_{\tilde{\mathcal{L}}_{\eta} U} \mu - \mathcal{L}_{\tilde{\mathcal{L}}_{\mu} U} \eta = 0 \, ,     
\end{equation}
which is exactly the same as Condition $(i)$ of Theorem 4.3 of \cite{mokri1997matched}, whereas its dual yields Condition~$(ii)$. Condition $(iii)$ is again unnecessary as the Jacobi identity together with the right-Leibniz rule implies that the anchor is a morphism of brackets.


\subsection{Metric Invariance Property and Bracket Morphisms}
\label{s4c}

\noindent In this subsection, we consider the last remaining algebroid axioms, namely the metric invariance property in its rather general form and certain bracket morphisms. The symmetric part decomposition (\ref{symmetricpart}) into a vector bundle valued metric and a first-order differential operator will be useful to choose certain metrics whose invariance can be checked with respect to another first-order differential operator.

Recall that for a Courant algebroid $(E, \rho_E, [\cdot,\cdot]_E, g_E)$, the following metric invariance property is satisfied
\begin{equation}
    \rho_E(u)(g_E(v, w)) = g_E([u, v]_E, w) + g_E(v, [u, w]_E) \, .
\end{equation}
For higher Courant algebroids \cite{bi2011higher}, which are of the form $TM \oplus \Lambda^p T^*M$, the metric takes values in $(p-1)$-forms, and the left-hand side of the metric invariance is given by the usual Lie derivative~$\mathcal{L}_{\rho_E(u)}(g_E(v, w))$ where $\rho_E(u)=\text{proj}_{TM}(u)$. When there is a metric structure in the picture taking values in some arbitrary vector bundle, it is natural to ask whether a generalization of such an invariance of the metric can be made meaningful. This seems to require an additional ingredient, a map $\mathbb{L}^E$ that is a generalization of the Lie derivative of forms, but it must be distinct from our calculus elements $\mathcal{L}$ or $\tilde{\mathcal{L}}$ as they simply do not have the same domain and ranges\footnote{For Courant and higher Courant algebroids, two Lie derivatives are acting on forms with different degrees. The one in the bracket is acting on $p$-forms, whereas the one in the metric invariance is acting on $(p-1)$-forms.}. 

In general, we study a metric invariance property for a metric $g_E: E \times E \to \mathbb{E}$ of the form
\begin{equation}
    \mathbb{L}^E_u g_E(v, w) = g_E([u, v]_E, w) + g_E(v, [u, w]_E) \, ,
\label{metricinv}
\end{equation}
with a first-order differential operator $\mathbb{L}^E: E \times \mathbb{E} \to \mathbb{E}$, which we term as the \textit{metric invariance operator}. When the vector bundle $E$ has the decomposition $E = A \oplus Z$, we want to consider this unique $\mathbb{L}^E$ as a combination of two maps where each is associated with one of the calculi. Therefore in this case, we decompose $\mathbb{L}^E$ as
\begin{equation}
    \mathbb{L}^E_{U + \omega} \xi = \pounds_U \xi + \tilde{\pounds}_\omega \xi \, ,
\end{equation}
where the maps $\pounds$ and $\tilde{\pounds}$ are of the form $\pounds: A \times \mathbb{E} \to \mathbb{E}$ and $\tilde{\pounds}: Z \times \mathbb{E} \to \mathbb{E}$. With this decomposition, we observe that the bracket (\ref{thebracketold}) on $E = A \oplus Z$ satisfies the metric invariance property with the metric (\ref{metricdecomp}) 
\begin{equation}
    g_E(U + \omega, V + \eta) = g_A(U, V) \oplus \left( \tilde{\iota}_{\omega} V + \tilde{\iota}_{\eta} U \right) \oplus g_Z(\omega, \eta) \oplus \left( \iota_U \eta + \iota_V \omega \right) \, , 
\end{equation}
if and only if both $A$ and $Z$ themselves satisfy the metric invariance property with
\begin{align}
    \pounds_U g_A(U, V) &= g_A([U, V]_A, W) + g_A(V, [U, W]_A) \nonumber \, , \\
    \tilde{\pounds}_\omega g_Z(\eta, \mu) &= g_Z([\omega, \eta]_Z, \mu) + g_Z(\eta, [\omega, \mu]_Z) \, , \label{metricinvAZ}
\end{align}
and the following relations are satisfied
\begin{align}
    \pounds_U (\tilde{\iota}_\mu V \oplus \iota_V \mu) &= g_A(V, \tilde{\mathcal{K}}_\mu U) \oplus \tilde{\iota}_\mu [U, V]_A +\tilde{\iota}_{\mathcal{L}_U \mu} V \oplus 0 \oplus \iota_{[U, V]_A} \mu + \iota_V \mathcal{L}_U \mu \, ,  \label{metricinvcond1} \\
    \pounds_U g_Z(\eta, \mu)  &= 0 \oplus \tilde{\iota}_\mu \tilde{\mathcal{K}}_\eta U + \tilde{\iota}_\eta \tilde{\mathcal{K}}_\mu U \oplus g_Z(\mathcal{L}_U \eta, \mu) +g_Z(\eta, \mathcal{L}_U \mu) \oplus \iota_{\tilde{\mathcal{K}}_\eta U} \mu + \iota_{\tilde{\mathcal{K}}_\mu U} \eta \, , \label{metricinvcond2}
\end{align}
together with their duals
\begin{align}
    &\tilde{\pounds}_\omega (\tilde{\iota}_\mu V \oplus \iota_V \mu) = 0 \oplus \tilde{\iota}_\mu \tilde{\mathcal{L}}_\omega V + \tilde{\iota}_{[\omega, \mu]_Z} V \oplus g_Z(\mathcal{K}_V \omega, \mu) \oplus \iota_{\tilde{\mathcal{L}}_\omega V} \mu + \iota_V [\omega, \mu]_Z \, , \nonumber\\ 
    &\tilde{\pounds}_\omega g_A(V, W) = g_A(\tilde{\mathcal{L}}_\omega V, W) + g_A(V, \tilde{\mathcal{L}}_\omega W) \oplus \tilde{\iota}_{\mathcal{K}_V \omega} W + \tilde{\iota}_{\mathcal{K}_W \omega} V \oplus 0 \oplus \iota_W \mathcal{K}_V \omega + \iota_V \mathcal{K}_W \omega \, , \label{metricinvtildeconds}
\end{align}
where we explicitly put zeroes on the right-hand sides when there is no corresponding part in the direct sum decomposition. We will refer to these four conditions, (\ref{metricinvcond1}, \ref{metricinvcond2}) together with their duals (\ref{metricinvtildeconds}), as \textit{metric invariance compatibility conditions}. They can be considered as compatibility conditions as they all mix calculus and tilde-calculus elements together.

Lastly, we consider morphisms of brackets, and comment on their effects on the calculus elements. As we have discussed, the right-Leibniz rule and the Jacobi identity imply that the anchor is a morphism of brackets, \textit{i.e.}, $\rho_E([u, v]_E) = [\rho_E(u), \rho_E(v)]_{\text{Lie}}$. Yet, for some examples like $E$-Courant algebroids \cite{Chen_2010}, there are certain maps $\phi: E \to E'$ which are required to be a morphism of brackets in the sense that 
\begin{equation}
    \phi([u, v]_E) = [\phi(u), \phi(v)]_{E'} \, ,
\end{equation}
holds for all $u, v \in E$\footnote{This tempts us to define the ``predator'' of $\Phi$, as a map $\mathcal{P}^{E E'}_{\Phi}(u, v) := \Phi([u, v]_E) - [\Phi(u), \Phi(v)]_{E'}$, which is not a tensorial quantity in general.}.

When the vector bundle $E$ is of the form $E = A \oplus Z$, decomposing $\phi = \phi_A \oplus \phi_Z$ where $\phi_A: A \to E'$ and $\phi_Z:Z\to E'$, the requirement that $\phi$ is a bracket morphism is identical to the condition that $\phi_A$ and $\phi_Z$ are bracket morphisms themselves together with the following \textit{bracket morphism compatibility conditions}
\begin{align}
    \phi_A(\tilde{K}_{\eta} U) + \phi_Z(\mathcal{L}_U \eta) &= [\phi_A(U), \phi_Z(\eta)]_{E'} \, , \nonumber\\
    \phi_A(\tilde{\mathcal{L}}_{\omega} V) + \phi_Z(\mathcal{K}_V \omega) &= [\phi_Z(\omega), \phi_A(V)]_{E'} \, ,
    \label{ggaha}
\end{align}
which mix calculus and tilde-calculus elements together. Note that if we choose $E'$ as the tangent Lie algebroid $TM$, and $\phi$ as the anchor $\rho_E$, these conditions are identical to the last equations of~(\ref{compatibility2old}) and (\ref{compatibility1old}), respectively. Combining above conditions (\ref{ggaha}), we get
\begin{equation}
    \phi_A(\tilde{d} \tilde{\iota}_{\omega} V) + \phi_Z(d \iota_V \omega) = - \mathbb{D}_{E'} g_{E'}(\phi_A(V), \phi_Z(\omega)) \, ,
    \label{yaha}
\end{equation}
which vanishes if $E'$ has an anti-symmetric bracket.


\subsection{Bialgebroids and Metric-Bourbaki Algebroids}
\label{s4d}

\noindent In this subsection, we collect all of our observations about algebroid axioms in order to formalize the notion of a bialgebroid. In the previous subsections, we explicitly worked on the conditions coming from right- and left-Leibniz rules, Jacobi identity, the decomposition of the symmetric part given by (\ref{symmetricpart}), metric invariance and certain bracket morphisms. In a sense, the extra conditions that we worked on extend the notions of matched pairs of Leibniz algebroids \cite{ibanez2001matched} and Lie bialgebroids \cite{mackenzie1994lie}, where the latter's relation to our framework is explained explicitly in this subsection. By using the compatibility conditions we derived, we now define the Drinfel'd doubles, and state our main theorems. We also comment on the relation between the formalism in this paper with our earlier work on metric-Bourbaki algebroids \cite{ccatal2022pre}.

In our calculations from the previous subsections, we observe that when the doubled bracket satisfies an particular property, both of the initial subalgebroids have to satisfy the same property. As we analyze each algebroid axiom separately, we have the right to use the term bialgebroid freely without any adjectives. We term the pair $(A, Z)$ a \textit{bialgebroid} if both $A$ and $Z$ are algebroids with common desired properties, which are equipped with two calculi satisfying compatibility conditions corresponding to each property. For any bialgebroid $(A, Z)$, we call the algebroid structure on the direct sum $E = A \oplus Z$ induced by the bracket (\ref{thebracket}) as the bialgebroid's \textit{Drinfel'd double}. With these definitions our main theorem can be stated as follows: Given a bialgebroid $(A, Z)$ where the pair shares the same subset of properties we analyzed, then the Drinfel'd double $E = A \oplus Z$ satisfies the same subset of properties. As we have included the dual versions of all compatibility conditions for a bialgebroid $(A, Z)$, by definition we automatically have that $(Z, A)$ is also a bialgebroid.

Following our earlier work \cite{ccatal2022pre}, we refer to algebroids that satisfy each of algebroid axioms that we have discussed as \textit{metric-Bourbaki algebroids}\footnote{We slightly modify the definition from \cite{ccatal2022pre} though.}. More concretely, a metric-Bourbaki algebroid is a septet $(E, \rho_E, [\cdot,\cdot]_E, \mathbb{E}, g_E, \mathbb{D}_E, \mathbb{L}^E)$ where $(E, \rho_E, [\cdot,\cdot]_E)$ is a Leibniz algebroid equipped with an $\mathbb{E}$-valued metric $g_E: E \times E \to \mathbb{E}$, and two first-order differential operators $\mathbb{D}_E: \mathbb{E} \to E$ and $\mathbb{L}^E: E \times \mathbb{E} \to \mathbb{E}$ satisfying 
\begin{align}
    [u, v]_E + [v, u]_E &= \mathbb{D}_E g_E(u, v) \, , \nonumber\\
    \mathbb{L}^E_u g_E(v, w) &= g_E([u, v]_E, w) + g_E(v, [u, w]_E) \, ,
\end{align}
for all $u, v, w \in E$. With this notion, we can introduce \textit{metric-Bourbaki bialgebroids} as a pair of metric-Bourbaki algebroids $A$ and $Z$ equipped with dual calculi which satisfy the metric invariance compatibility conditions (\ref{metricinvcond1}, \ref{metricinvcond2}, \ref{metricinvtildeconds}). In particular our observations from the previous subsections can be stated as the following theorem: The Drinfel'd double of a metric-Bourbaki bialgebroid is a metric-Bourbaki algebroid. 

Even though the metric-Bourbaki algebroids have almost each of the algebroid properties that we have discussed. They are still quite general, and cover many physically and mathematically motivated examples. In particular, we will see that this notion is related to generalized Lie \cite{iglesias2001generalized}, omni-Lie \cite{chen2010omni}, $E$-Courant algebroids \cite{Chen_2010}, exceptional Courant brackets \cite{pacheco2008m}, and the triangularity notion for Lie bialgebroids \cite{mackenzie1994lie} that we will extend in Section \ref{s6}. We will also see that many bialgebroid structures also fit into our framework, in particular we next focus on the prototypical example of Lie bialgebroids.

Metric-Bourbaki bialgebroid structures that we introduced can be considered as an extension of Lie bialgebroids, as we now outline the details. For Lie bialgebroids, one chooses $Z = A^*$. In this case, both interior products are the same, $\iota_V \omega = \tilde{\iota}_{\omega} V$, as they both coincide with the usual pairing between two dual bundles, and they take values in smooth functions, so one has $\mathcal{A}  =\mathcal{Z}= C^{\infty}M$. Moreover, both brackets on $A$ and $A^*$ are anti-symmetric, so that the metrics~$g_A$ and $g_{A^*}$ can be chosen as 0 which is degenerate. Hence, even though there is no metric for the set up of Lie bialgebroids, the symmetric part decompositionis trivially satisfied for any first-order differential operators $\mathbb{D}_A$ and $\mathbb{D}_{A^*}$. On the other hand, as they are Lie brackets, one can induce their corresponding Lie derivatives $\mathcal{L}$ and $\mathcal{L}^*$ by Equation (\ref{liederivative}), which would coincide with our~$\mathcal{L}$ and~$\tilde{\mathcal{L}}$. Similarly, one can induce two corresponding  exterior derivatives $d$ and~$d^*$ by Equation~(\ref{exteriorderivative}), which would coincide with our $d$ and $\tilde{d}$. Note that in the case of Lie algebroids, both of the maps~$\mathcal{L}$ and $\mathcal{L}^*$ act on the whole tensor algebra on $A$. Similarly, the exterior derivatives $d$ and~$d^*$ act on the associated exterior bundles of~$A^*$ and $A$, respectively. Crucially, they satisfy all of the relations of the usual Cartan calculus. Consequently, they also satisfy calculus (\ref{calculusconditions}) and linearity~(\ref{linearityconditions}) conditions, so that they form a calculus in the sense that we have introduced. We see that the Jacobi compatibility condition (\ref{comp2}) and its dual are satisfied directly as it is proven in Proposition~3.4 and 3.8 of \cite{mackenzie1994lie} and Proposition 3.1 of \cite{kosmann1995exact}. Using the anti-symmetry of the brackets, the Jacobi compatibility condition (\ref{comp3}) becomes
\begin{equation}
    d \iota_{\tilde{\mathcal{L}}_{\omega} W} \eta - d \iota_{\tilde{d} \tilde{\iota}_{\eta} W} \omega + d \iota_W [\omega, \eta]_{A^*} = 0 \, ,
\end{equation}
for all $W \in A, \omega, \eta \in A^*$, which together with its dual is satisfied trivially by $\tilde{d} f(\omega) = \rho_{A^*}(\omega)(f)$. The last remaining Jacobi compatibility condition (\ref{comp1}) is equivalent to the usual compatibility condition for Lie bialgebroids given by Equation (\ref{compatibilityLie}) as they both read Equation (\ref{complieframe})
on a frame. As one has the fact that when $(A, A^*)$ is a Lie bialgebroid, so is $(A^*, A)$, this completes the proof that our Jacobi compatibility conditions reduce to the usual \textit{single} compatibility condition~(\ref{compatibilityLie}). For the metric invariance property, we see that metric invariance properties (\ref{metricinvAZ}) for individual $A$ and $A^*$ are trivially satisfied for any metric invariance operator because $g_A = g_{A^*} = 0$. Moreover, the metric invariance compatibility conditions (\ref{metricinvcond1}, \ref{metricinvcond2}, \ref{metricinvtildeconds}) also hold trivially with the choice $\pounds = \mathcal{L}$ and $\tilde{\pounds} = \tilde{\mathcal{L}}$ since one has $\iota_U \eta = \tilde{\iota}_{\eta} U$ and the commutation relation analogous to the one for the Lie derivative and the interior product holds. Hence, we conclude that any Lie bialgebroid is an example of metric-Bourbaki bialgebroid. The Drinfel'd double of a Lie bialgebroid is a Courant algebroid \cite{liu1997manin}, and they are of course metric-Bourbaki algebroids. 

We finish this section with some further remarks on our previous work on calculus on algebroids. In \cite{ccatal2022pre}, we also included a second Lie derivative-like map analogous to $\pounds$ of this paper acting on the sections of $\mathcal{Z}$ and defined the notion of \textit{Bourbaki calculus}. The other three maps $(\mathcal{L}, \iota, d)$ had the same $C^{\infty}M$-linearity properties, but the constraints were chosen in order to be able to prove an extension of \v{S}evera classification theorem. There were some extra conditions which read in this paper's notation\footnote{In \cite{ccatal2022pre} a Bourbaki calculus, or rather a Bourbaki $A$-calculus, is defined as a quadruplet $(\iota, d, \mathcal{L}^R, \mathcal{L}^Z)$, where the maps $\mathcal{L}^R$ and $\mathcal{L}^Z$ correspond to $\pounds$ and $\mathcal{L}$, whereas $\iota$ and $d$ are the same.}
\begin{align}
    \pounds_U \iota_V \omega &= \iota_{[U, V]_A} \omega + \iota_V \mathcal{L}_U \omega \, , \nonumber\\
    - \iota_U \mathcal{K}_V \omega &= \iota_V \mathcal{K}_U  \omega \, .
\label{bourbaki}
\end{align}
Hence, we can say that every Bourbaki calculus is a calculus in the sense of this paper, equipped with the extra map $\pounds$ which is related to metric invariance operator, satisfying these additional constraints. The first one here is identical to the usual Cartan relation between the Lie derivative and the interior product, whereas the second one is valid for the usual Cartan calculus because the interior product squares to zero. Hence, the usual Cartan calculus is a Bourbaki calculus. As we will see in Section \ref{s6}, these additional constraints do not necessarily hold for crucial examples, so an extension of \v{S}evera's results in the sense of \cite{ccatal2022pre} is not directly applicable. We note that, although the second condition of (\ref{bourbaki}) does not affect neither the calculus conditions (\ref{calculusconditions}) nor the Jacobi compatibility conditions (\ref{comp1} - \ref{comp3}), it simplifies some of the terms in the metric invariance compatibility conditions. For example, Equation (\ref{metricinvcond2}) becomes
\begin{equation}
    \pounds_U g_Z(\eta, \mu) = 0 \oplus 0 \oplus g_Z(\mathcal{L}_U \eta, \mu) +g_Z(\eta, \mathcal{L}_U \mu) \oplus \iota_{\tilde{\mathcal{K}}_\eta U} \mu + \iota_{\tilde{\mathcal{K}}_\mu U} \eta \, ,
\end{equation}
which further simplifies to 
\begin{equation}
    \pounds_U g_Z(\eta, \mu)  = 0 \oplus 0 \oplus g_Z(\mathcal{L}_U \eta, \mu) +g_Z(\eta, \mathcal{L}_U \mu) \oplus 0 \, ,
\end{equation}
with the additional assumption $\iota_U \omega = \tilde{\iota}_{\omega} U$, which holds for many important examples as we will discuss. The last equation resembles the vanishing of the Lie derivative of the metric $g_Z$.


\section{Examples from the Literature}
\label{s5}

One can find many examples of bialgebroids both from the physics and mathematics literatures. Yet, most of these examples, which we will cover in this section, are either the case where $A$ and~$Z$ are dual in some sense, or the Leibniz algebroid structure on $Z$ is trivial, that is its bracket and anchor are given by the zero map. Physically motivated vector bundles, for instance the ones in (\ref{physicsbundles}), are certainly not dual in the usual sense. Moreover, for exceptional field theoretical purposes, one needs non-trivial algebroid structures in all of the components, since the full bracket should be analogous to the D-bracket of DFT. We believe that our dual calculus framework can be useful for such constructions, as we will build up an explicit example in the consequent section, which is directly relevant for physics including exceptional Drinfel'd algebras. 

In the following, we will refer the reader to the main references for the details. To keep the discussion focused, we will only note pertinent formulas. For an extensive list of examples of metric-Bourbaki algebroids, we refer to our previous work \cite{ccatal2022pre}.


\subsection{Generalized Lie Bialgebroids}
\label{s5a}

As we will investigate further in the following section, it is well-known that there is a close relation between Lie bialgebroids and Poisson structures. Moreover, this relation is intimately tied with the Poisson-Lie T-duality as we discussed in Section \ref{s3}. However, if one considers a generalization of Poisson structures, namely Jacobi structures, then Lie bialgebroids should be also ``generalized''~\cite{iglesias2001generalized}. Moreover, generalized Lie bialgebroid structures seem to be related to a further extension of T-duality, namely Jacobi-Lie T-plurality \cite{fernandez2021jacobi}. 

In order to define generalized Lie bialgebroids, we need a Lie algebroid $A$ equipped with a 1-cocycle $\phi \in A^*$ in the Lie algebroid cohomology with trivial coefficients. One can then twist the exterior derivative operator $d$ of the Lie algebroid $A$ defined via Equation (\ref{exteriorderivative}) by the 1-cocycle $\phi$:
\begin{equation}
    d^{\phi} \xi := d \xi + \phi \wedge \xi \, ,
\end{equation}
for $\xi \in \Lambda^p A^*$. Then by Cartan magic formula, one can define the twisted Lie derivative $\mathcal{L}^{\phi}: A \times \Lambda^p A^* \to \Lambda^p A^*$
in terms of the twisted exterior derivative and the usual interior product operations. Subsequently, one can extend this twisted Lie derivative to a twisted Schouten-Nijenhuis bracket~$[\cdot,\cdot]_{\text{SN}, A}^{\phi}: \Lambda^p A \times \Lambda^q A \to \Lambda^{p + q - 1} A$ defined by Equation (\ref{schouten}). 
With this setup, let $A$ and $A^*$ be a pair of Lie algebroids whose calculus operations are twisted by 1-cocycles $\phi \in A^*$ and $X \in A$, respectively. Then the pair $(A,A^*)$ is a generalized Lie bialgebroid if
\begin{align}
    \tilde{d}^X[U, V]_A^\phi &=  [U, \tilde{d}^X V]_{\text{SN},A}^\phi - [\tilde{d}^X U, V]_{\text{SN},A}^\phi \, , \label{generalizedlie1} \\
    (\tilde{\mathcal{L}}^X)_\phi P &= -  (\mathcal{L}^\phi)_X P \, , \label{generalizedlie2}
\end{align}
for any $U, V \in A$ and $P \in \Lambda^p A$.

To see that generalized Lie bialgebroids are indeed metric-Bourbaki bialgebroids in the sense that we defined, let us first start by noting that the triplet of operations $(\mathcal{L}^\phi,\iota, d^\phi)$ and $(\tilde{\mathcal{L}}^X, \tilde{\iota}, \tilde{d}^X)$ both are calculi, since by construction they satisfy (\ref{calculusconditions}). Considering the Jacobi compatibility conditions: the first one (\ref{comp1}) agrees with the first condition (\ref{generalizedlie1}) given above. The second Jacobi compatibility condition (\ref{comp2})'s equivalence to the second condition above (\ref{generalizedlie2}) follows from Proposition 4.3 of \cite{iglesias2001generalized} due to the graded anti-symmetry of the Schouten-Nijenhuis bracket. The last one (\ref{comp3}) is trivially satisfied by the anti-symmetry of the Lie bracket similarly to the usual Lie bialgebroid case. Finally we check the metric invariance compatibility conditions. For this, we note that $g_A = g_{A^*} = 0$, and hence metric invariance properties (\ref{metricinvAZ}) of the algebroids $A$ and $A^*$ trivially hold. The metric invariance compatibility conditions (\ref{metricinvcond1}) and (\ref{metricinvcond2}) reduce to a couple of (fourth and second, respectively) Cartan calculus relations (\ref{cartancalculusrelations}) by noting that $\iota_V \omega = \tilde{\iota}_{\omega} V$ and making the identification $\pounds = \mathcal{L}$. The dual metric invariance compatibility conditions (\ref{metricinvtildeconds}) also follow similarly with the identification $\tilde{\pounds} = \tilde{\mathcal{L}}$. Therefore we see that generalized Lie bialgebroids are examples of metric-Bourbaki bialgebroids. Their Drinfel'd double is a generalized Courant algebroid \cite{da2004dirac}, and it is a metric-Bourbaki algebroid. Moreover, the calculus on generalized Lie algebroids is an example of Bourbaki calculus, so an extension of \v{S}evera classification theorem should hold for the exact cases as it can be inferred from \cite{ccatal2022pre}.


\subsection{Omni-Lie Algebroids}
\label{s5b}

Omni-Lie algebras are introduced by Weinstein \cite{weinstein2000omni} in order to obtain a linearization of the Courant bracket at a point. These algebras are defined on $\mathfrak{gl}(V) \oplus V$ for a vector space $V$, and they are not Lie algebras. Yet, all possible Lie algebra structures on $V$ are in one-to-one correspondence with their Dirac structures. This situation can be generalized to a vector bundle~$\mathbb{E}$, and its omni-Lie algebroid $E=\mathfrak{D} \mathbb{E} \oplus \mathfrak{J} \mathbb{E}$, which classifies all possible Lie algebroid structures on $\mathbb{E}$ \cite{chen2010omni}. Here~$\mathfrak{D} \mathbb{E}$ and $\mathfrak{J} \mathbb{E}$ denote the gauge or covariant differential operator bundle \cite{mackenzie1987lie}, and the first jet bundle of $\mathbb{E}$, respectively. These two vector bundles are $\mathbb{E}$-dual, that is there is a non-degenerate $\mathbb{E}$-valued pairing between them \cite{chen2010omni}. As the gauge bundle \cite{kosmann2002differential} and the jet bundle \cite{crainic2005secondary} are both Lie algebroids, this $\mathbb{E}$-duality can be used to generate analogous operators to usual Cartan elements, which also yield a calculus in the sense that we introduced. For details of this, we refer to \cite{chen2010omni}. The omni-Lie algebroid is then defined as the vector bundle $E = \mathfrak{D} \mathbb{E} \oplus \mathfrak{J} \mathbb{E}$ equipped with the Dorfman-like bracket, so that the tilde-calculus elements are taken to be vanishing, that is $\tilde{\mathcal{L}}=\tilde{\iota} = \tilde{d} = 0$.

An omni-Lie algebroid $E$ satisfies many Courant-like properties by Theorem 3.1 of \cite{chen2010omni}, and consequently it is a metric-Bourbaki algebroid. In particular, since $E$ is a Leibniz algebroid, the pair $(\mathfrak{D} \mathbb{E}, \mathfrak{J} \mathbb{E})$ is a matched pair. Therefore all Jacobi compatibility conditions (\ref{comp1} - \ref{comp3}) together with their duals (\ref{compdual}) are satisfied. Furthermore, $E$ satisfies the metric invariance property (\ref{metricinv}) where the metric invariance operator is given by the map
\begin{equation}
    \mathbb{L}^E_u = \text{proj}_{\mathfrak{D} \mathbb{E}} u \, .
\end{equation}
Since $\mathfrak{D} \mathbb{E}$ and $\mathfrak{J} \mathbb{E}$ are both Lie algebroids, we have $g_{\mathfrak{D} \mathbb{E}} = g_{\mathfrak{J} \mathbb{E}} = 0$ and their metric invariance properties (\ref{metricinvAZ}) are trivially satisfied. Similarly metric invariance compatibility condition (\ref{metricinvcond2}) is trivially satisfied because the tilde-calculus elements are trivial. The condition (\ref{metricinvcond1}) is satisfied as a Cartan-like calculus identity (\ref{cartancalculusrelations}) with the identification $\pounds=\mathcal{L}$. Finally the first of the dual metric compatibility conditions (\ref{metricinvtildeconds}) imposes $\tilde{\pounds}=0$ at least on the image of $\iota$ and the second condition of (\ref{metricinvtildeconds}) does not impose any conditions on $\tilde{\pounds}$. Finally, the second dual condition follows from the Cartan-like calculus relation $\iota_U \iota_V = - \iota_V \iota_U$. Therefore the pair $(\mathfrak{D} \mathbb{E}, \mathfrak{J} \mathbb{E})$ can be considered as a metric-Bourbaki bialgebroid where the jet bundle $\mathfrak{J} \mathbb{E}$ is equipped with the trivial structures, and its Drinfel'd double is the omni-Lie algebroid, which is of course a metric-Bourbaki algebroid.

A special case for omni-Lie algebroids of the form $\left( TM \times C^{\infty}M \right) \oplus \left( T^*M \times C^{\infty}M \right)$ was first considered in \cite{aissa101181conformal} about conformal extensions of Dirac structures. Moreover, the omni-Lie algebroids are further generalized to higher omni-Lie algebroids of the form $E = \mathfrak{D} \mathbb{E} \oplus \mathfrak{J}^p \mathbb{E}$. Analogously to the higher Courant case, these algebroids satisfy similar properties \cite{bi2018higher}; see Theorem 2.11, and they are also metric-Bourbaki algebroids.


\subsection{$\mathbb{E}$-Lie Bialgebroids}
\label{s5c}

In the following subsection, the omni-Lie algebroids of the form $E = \mathfrak{D} \mathbb{E} \oplus \mathfrak{J} \mathbb{E}$ are discussed. These algebroids can be seen as generalizations of the standard Courant algebroid where one replaces the tangent bundle $TM$ with the gauge bundle $\mathfrak{D} \mathbb{E}$ and $T^*M$ with the fet bundle $\mathfrak{J} \mathbb{E}$. $E$-Courant algebroids, or for notational issues $\mathbb{E}$-Courant algebroids in this discussion, are defined as axiomatizations of omni-Lie algebroids \cite{Chen_2010}. An $\mathbb{E}$-Courant algebroid is a quadruplet $(E, \phi_E, [\cdot,\cdot]_E, g_E)$, where the metric $g_E$ takes values in a vector bundle $\mathbb{E}$, and the anchor-like map now is of the form $\phi_E: E \to \mathfrak{D} \mathbb{E}$ taking values in the gauge bundle instead of $TM = \mathfrak{D} C^{\infty}M$. They satisfy~$\mathbb{E}$-valued generalizations of the usual properties of Courant algebroids together with some technical additional ones, see \cite{Chen_2010} for details. All of these properties hold for generalized Lie and omni-Lie algebroids.

One can extend the notion of Lie bialgeorids to $\mathbb{E}$-Lie bialgebroids whose Drinfel'd double yields an $\mathbb{E}$-Courant algebroid. They are defined as $\mathbb{E}$-dual pair of Lie algebroids $A$ and $Z$ such that 
\begin{align}
    \tilde{d} [U, V]_A &= \mathcal{L}_U (\tilde{d} V) - \mathcal{L}_V (\tilde{d} U) \, , \label{eliecomp1} \\
    \tilde{\mathcal{L}}_{d \xi} U &= - \mathcal{L}_{\tilde{d} \xi} U \, , \label{eliecomp2} \\
    g_E(\tilde{d} \xi_1, d \xi_2) &= 0 \label{eliecomp3} \, , 
\end{align}
for all $U, V \in A, \xi, \xi_i \in \mathbb{E}$, where these calculus maps are constructed with the help of the $\mathbb{E}$-duality.

The first and second conditions above,~(\ref{eliecomp1}) and (\ref{eliecomp2}), agree with our first and second Jacobi compatibility conditions (\ref{comp1}) and (\ref{comp2}), respectively. The third one is related to the morphism property of the anchor-like map $\phi_E$:
\begin{equation}
    \phi_E([u, v]_E) = [\phi_E(u), \phi_E(v)]_{\mathfrak{D} \mathbb{E}} \, ,
\end{equation}
which is included in the definition of an $\mathbb{E}$-Courant algebroid \cite{Chen_2010}. By Remark 6.2 of \cite{Chen_2010}, our bracket morphism compatibility conditions (\ref{ggaha}) imply the third condition (\ref{eliecomp3}). This can be seen from Equation (\ref{yaha}) whose right-hand side vanishes since the differential operator bundle~$\mathfrak{D} \mathbb{E}$ is a Lie algebroid. The metric invariance compatibility conditions (\ref{metricinvcond1}, \ref{metricinvcond2}, \ref{metricinvtildeconds}) also hold with~$\pounds = \phi_A$ and $\tilde{\pounds} = \phi_Z$ for the maps coming from the decomposition $\phi_E = \phi_A \oplus \phi_Z$. This is due to~$g_A = g_Z = 0$, together with $\iota_U \omega = \tilde{\iota}_{\omega} U$ and $\iota_U \iota_V = - \iota_V \iota_U$. Consequently, $\mathbb{E}$-Lie bialgebroids are special cases for metric-Bourbaki bialgebroids, and their Drinfel doubles, $\mathbb{E}$-Courant algebroids, are for metric-Bourbaki algebroids. Moreover, since the calculus in this case is also a Bourbaki calculus \cite{ccatal2022pre}, exact $\mathbb{E}$-Courant algebroids satisfy an extension of \v{S}evera classification theorem which is proven in Theorem 5.4 of \cite{Chen_2010}.


\section{Nambu-Poisson Structures and Higher Courant Algebroids}
\label{s6}

\noindent In this section, we construct a specific example, which is important from both physical and mathematical perspectives. We set $A = TM$ as the tangent Lie algebroid and $Z = \Lambda^p T^*M$ for some positive integer $p$, and take the calculus elements as the usual Cartan calculus elements. We construct an algebroid structure on $\Lambda^p T^*M$ by using a Nambu-Poisson structure \cite{nambu1973generalized, takhtajan1994foundation}, which coincides with the one induced by the Koszul bracket \cite{bi2011higher}. Moreover, with the help of the Nambu-Poisson structure, we construct a tilde-calculus which will be dual to the usual Cartan calculus, so that we get a matched pair of Leibniz algebroids. Furthermore, we explicitly prove that the metric invariance compatibility conditions also hold for a certain metric that we will construct. Hence, we get a metric-Bourbaki algebroid structure on the Drinfel'd double $TM \oplus \Lambda^p T^*M$. Our construction in this section is a direct extension of \cite{halmagyi2009non} in which the Roytenberg algebra is obtained in terms of certain twists of the Dorfman bracket via a bivector. As will explicitly discuss, this can be considered as an alternative way to construct triangular Lie bialgebroids \cite{mackenzie1994lie}, when the bivector is a Poisson structure. We generalize these constructions for Nambu-Poisson structures, and extend the notion of triangularity in the realm of higher Courant algebroids  \cite{bi2011higher}, whose results will be used heavily. See also the works of \cite{hagiwara2002nambu, zambon2012l_}. 

Nambu-Poisson structures are natural generalizations of Poisson structures, where the latter has many interesting connections to sigma models \cite{bojowald2005lie, alekseev2005current, kotov411112dirac, kotov2010generalizing, chatzistavrakidis2014dirac}. Furthermore, D-branes can be described by using Dirac structures \cite{asakawa2012d}, whose alternative descriptions in Poisson generalized geometry are related by Seiberg-Witten maps \cite{asakawa2014d}. The non-geometric $Q$- and $R$-fluxes, and topological T-duality that exchanges them have been also explained in the realm of Poisson generalized geometry \cite{asakawa2015poisson, asakawa2015topological}, which will be the main topic of Section \ref{s8a}. Halmagyi studied Polyakov models from a worldsheet charge algebra perspective \cite{halmagyi2008non}, and by twisting the Dorfman bracket by a bivector he constructed the Roytenberg bracket under which the algebra closes. This Hamiltonian approach was based on earlier work of Alekseev and Strobl \cite{alekseev2005current}, and he later continued in the Lagrangian setting, and by lifting the action to 3-dimensions he realized the same Roytenberg bracket as generalized Wess-Zumino terms \cite{halmagyi2009non}. 

As M theory requires generalizations of the ideas that we outlined in the previous paragraph to the context of branes, the study of them in a more abstract setting is of paramount importance. Nambu-Poisson structures in sigma models appear after a search of such generalizations, and it gained attention after seminal works of Bagger, Lambert and Gustavson \cite{bagger2007modeling,  bagger2008gauge, bagger2008comments, gustavsson2009algebraic} on membranes and fivebranes, see also \cite{basu2005m2, ho2008m5, ho2010nambu}. An AKSZ type construction \cite{alexandrov1997geometry} for the extension of Poisson sigma models to Nambu-Poisson framework is discussed related to open $p$-branes in \cite{bouwknegt2013aksz}. Analysis of this Nambu-Poisson sigma models with a similar approach to Halmagyi yields the higher Roytenberg bracket \cite{jurvco2013p}, which again appears as generalized Wess-Zumino terms. This higher Roytenberg bracket in the absence of twists will be one of the main subjects of this section. We will show that Halmagyi's methods can be used for the construction of triangular Lie bialgebroids \cite{mackenzie1994lie}. Then by building upon these ideas, we will extend the triangularity in the realm of higher Courant algebroids and Nambu-Poisson structures. This will establish a formal framework for such higher Roytenberg brackets.

We saw that the usual Cartan calculus can be considered as a calculus on $\Lambda^p T^*M$ induced by the tangent bundle $TM$. In order to have a bialgebroid structure $(TM, \Lambda^p T^*M)$, we also need a tilde-calculus on the tangent bundle $TM$ induced by $\Lambda^p T^*M$. Unfortunately, without an additional structure on $M$, there exists no such calculus in general. For the case $p = 1$, when~$T^*M$ is also equipped with a Lie algebroid structure, one could induce such a tilde-calculus. An important special case for this situation is when the base manifold $M$ is a Poisson manifold \cite{lichnerowicz1977varietes}. More generally, a Poisson structure on a Lie algebroid $A$ induces a Lie algebroid structure on $A^*$ by the Koszul bracket. This automatically yields a Lie bialgebroid $(A, A^*)$, since the compatibility condition (\ref{compatibilityLie}) is automatically satisfied. Such Lie bialgebroids are known as triangular \cite{mackenzie1994lie}, and they have close relations with Poisson-Lie T-duality \cite{klimvcik1995dual}. Triangular Lie bialgebroids (with the absence of $H$- and $R$-twists for the purposes of this paper) are also constructed with physical motivations \cite{halmagyi2009non}. The constructions are done by a twisting of the Dorfman bracket on the standard Courant algebroid $TM \oplus T^*M$ with a bivector in order to get the full Roytenberg bracket~\cite{roytenberg2002quasi}. We will extend this procedure for $TM \oplus \Lambda^p T^*M$, and show that for $p = 1$ it yields the same tilde-calculus as the one for triangular Lie bialgebroids.

For the construction of triangular bialgebroids, we need a specific vector bundle automorphism 
\begin{equation}
    \Psi_{\Pi}: TM \oplus \Lambda^p T^*M \to TM \oplus \Lambda^p T^*M \, ,
\end{equation}
induced by a $(p+1)$-vector $\Pi$, which will be taken as a Nambu-Poisson structure so that it will satisfy each of the necessary conditions that we have discussed. Such automorphisms are intimately related to twist matrices frequently used in physics literature, and in particular to beta-transformations. We start the procedure by considering the higher Dorfman bracket on $TM \oplus \Lambda^p T^*M$ \cite{bi2011higher}
\begin{equation}
    [U + \omega, V + \eta]_{\text{Dor}} = [U, V]_{\text{Lie}} \oplus \mathcal{L}_U \eta + \mathcal{K}_V \omega \, ,
\end{equation}
for $U, V \in TM, \omega, \eta \in \Lambda^p T^*M$. Here, $\mathcal{L}$ is the usual Lie derivative acting on $p$-forms and $\mathcal{K}_V = - \mathcal{L}_V + d \iota_V$ where $d$ and $\iota$ are the usual exterior derivative and interior product acting on $(p-1)$- and $p$-forms, respectively\footnote{As the calculus that we consider is the usual Cartan calculus, of course we have $\mathcal{K}_V = - \iota_V d$, by the Cartan magic formula.}. Following an analogous approach to Halmagyi \cite{halmagyi2009non}, we continue with a \textit{twist} induced by the automorphism~$\Psi_{\Pi}$ for a $(p+1)$-vector $\Pi$, which we symbolically write as
\begin{equation}
    \Psi_\Pi = \begin{pmatrix}
            1 & \Pi \\
            0 & 1
            \end{pmatrix} \, , \qquad \qquad \qquad 
    \Psi_\Pi^{-1} = \begin{pmatrix}
            1 & - \Pi \\
            0 & 1
            \end{pmatrix} \, ,
\label{psitwist}
\end{equation}
in the sense that we define a new Dorfman-like bracket of the form
\begin{equation}
    [U + \omega, V + \eta]_{\tilde{\mathcal{D}}} = \Psi_{\Pi}^{-1} \left( [\Psi_{\Pi}(U + \omega), \Psi_{\Pi}(V + \eta)]_{\text{Dor}} \right) - [U + \omega, V + \eta]_{\text{Dor}} \, . 
\end{equation}
Here, we consider the $(p+1)$-vector $\Pi$ as a (musical sharp) map $\Pi: \Lambda^p T^*M \to TM$. Note that when the new bracket is zero, it means that the automorphism is a morphism of the higher Dorfman bracket. Next, we observe that due to the upper triangular form of $\Psi_{\Pi}$, the new bracket $[\cdot,\cdot]_{\tilde{\mathcal{D}}}$ does not have any common parts with the initial bracket $[\cdot,\cdot]_{\text{Dor}}$, so that we decompose it as
\begin{equation}
    [U + \omega, V + \eta]_{\tilde{\mathcal{D}}} = \tilde{\mathcal{L}}_{\omega} V + \tilde{\mathcal{K}}_{\eta} U + R(\omega, \eta) \oplus [\omega, \eta]_Z + H(U, V) \, ,
\end{equation}
for some appropriate maps. With this decomposition, we can directly read off the tilde-calculus elements as
\begin{align}
    \tilde{\mathcal{L}}_\omega V &= [\Pi \omega, V]_{\text{Lie}} - \Pi \mathcal{K}_V \omega \, , \nonumber\\ 
    \tilde{\mathcal{K}}_\eta U &= [U, \Pi \eta]_{\text{Lie}} - \Pi  \mathcal{L}_U \eta \, . \label{tildecalculuspi}
\end{align}
Combining these two operations via (\ref{calculusdecompose}) we obtain the composite operator $\tilde{d}\tilde{\iota}$ which we decompose as
\begin{equation}
    \tilde{\iota}_{\omega} V = \iota_V \omega\, , \qquad \qquad \qquad \qquad \qquad \qquad \tilde{d} = - \Pi d\, , \label{poissonexterior}
\end{equation}
where $\tilde{\iota}$ takes values in $\Lambda^{p-1} T^*M$ and is $C^{\infty}M$-bilinear, so that $\tilde{d}$ is a first-order differential operator of the form $\Lambda^{p-1} T^*M \to TM$. Moreover, the bracket on $Z$ can be identified as
\begin{equation}
    [\omega, \eta]_Z = \mathcal{L}_{\Pi \omega} \eta + \mathcal{K}_{\Pi \eta} \omega =: [\omega, \eta]_{\text{Kos}} \, , \label{koszulbracket}
\end{equation}
which coincides with the Koszul bracket $[\cdot,\cdot]_{\text{Kos}}$ of $p$-forms that yields a Leibniz algebroid structure on $\Lambda^p T^*M$ \cite{koszul1985crochet, bi2011higher}\footnote{There is another Leibniz algebroid structure on $\Lambda^p T^*M$ which is closely related but different to the one that we consider here \cite{ibanez1999leibniz}.}. We can observe that the procedure cannot induce an $H$-twist due to upper triangular form of $\Psi_{\Pi}$, so that $H = 0$. On the other hand, we see that the following $R$-twist is introduced
\begin{align}
    R(\omega, \eta) &= [\Pi \omega, \Pi \eta]_{\text{Lie}} - \Pi(\mathcal{L}_{\Pi \omega} \eta + \mathcal{K}_{\Pi \eta} \omega) \nonumber\\
    &= [\Pi \omega, \Pi \eta]_{\text{Lie}} - \Pi [\omega, \eta]_{\text{Kos}} \, .
\end{align}
Hence, this procedure induces the following bracket on $TM \oplus \Lambda^p T^*M$:
\begin{equation}
    [U + \omega, V + \eta]_{\tilde{\mathcal{D}}} = [\Pi \omega, V]_{\text{Lie}} - \Pi \mathcal{K}_V \omega + [U, \Pi \eta]_{\text{Lie}} - \Pi \mathcal{L}_U \eta + [\Pi \omega, \Pi \eta]_{\text{Lie}} - \Pi [\omega, \eta]_{\text{Kos}} \oplus [\omega, \eta]_{\text{Kos}} \, . \nonumber
\end{equation}
With all of these structures in our hand, we claim that with the only assumption of $\Pi$ being a Nambu-Poisson structure on $M$, we can prove the following claims:
\begin{enumerate}
    \item The induced $R$-twist vanishes, so that $\Lambda^p T^*M$ is a subalgebroid.
    \item The Koszul bracket $[\cdot,\cdot]_{\text{Kos}}$ satisfies the Jacobi identity, so that $\Lambda^p T^*M$ is a Leibniz algebroid.
    \item The tilde-calculus elements $(\tilde{\mathcal{L}}, \tilde{\iota}, \tilde{d})$ indeed form a calculus.
    \item The pair of calculi $(\mathcal{L}, \iota, d)$ and $(\tilde{\mathcal{L}}, \tilde{\iota}, \tilde{d})$ are \textit{dual} in the sense that we introduced. 
    \item The full bracket $[\cdot,\cdot]_{\text{Dor}} + [\cdot,\cdot]_{\tilde{\mathcal{D}}}$ yields a Leibniz algebroid structure on $TM \oplus \Lambda^p T^*M$, \textit{i.e.}, we get a matched pair $(TM, \Lambda^p T^*M)$ of Leibniz algebroids.
    \item The full bracket and the metric read from its symmetric part satisfy a metric invariance property with the maps
    \begin{equation}
        \pounds_U \xi = \mathcal{L}_U \xi \, , \qquad \qquad \qquad \tilde{\pounds}_\omega \xi = \mathcal{L}_{\Pi \omega} \xi \, ,
    \end{equation}
    where $\mathcal{L}$ on the right-hand side of both expressions is the usual Lie derivative acting on $(p-1)$-forms instead of $p$-forms. 
    \item Consequently, the pair $(TM, \Lambda^p T^*M)$ is a metric-Bourbaki bialgebroid, and its Drinfel'd double $TM \oplus \Lambda^p T^*M$ is a metric-Bourbaki algebroid.
\end{enumerate}
Before continuing, let us recall the usual definition of the Nambu-Poisson structures on manifolds. An anti-symmetric $\mathbb{R}$-multilinear map $\{ \cdot, \ldots, \cdot \}: \otimes_{i = 1}^{p+1} C^{\infty} M \to C^{\infty} M$ is called a Nambu-Poisson structure \cite{nambu1973generalized, takhtajan1994foundation} of order $p$ on $M$ if the following conditions are satisfied
\begin{align} 
    \left\{ f_1, \ldots, h f_i, \ldots f_{p+1} \right\} &= h \left\{ f_1, \ldots, f_i, \ldots f_{p+1} \right\} + f_i \left\{ f_1, \ldots, h, \ldots f_{p+1} \right\} \, , \nonumber\\
    \left\{ f_1, \ldots, f_p, \left\{h_1, \ldots, h_{p+1} \right\} \right\} &= \sum_{i = 1}^{p+1} \left\{ h_1, \ldots, \left\{ f_1, \ldots, f_p, h_i \right\}, \ldots, h_{p+1} \right\} \, ,
\end{align}
for all $f_i, h_i, h \in C^{\infty} M$. For $p = 1$, the second one, called the fundamental identity, reduces to the Jacobi identity for a Poisson structure. For every Nambu-Poisson structure of order $p$, there is a corresponding $(p+1)$-vector field $\Pi \in \Lambda^{p+1} TM$ such that 
\begin{equation}
    \left\{ f_1, \ldots, f_{p+1} \right\} = \Pi(d f_1, \ldots, d f_{p+1}) \, ,
\end{equation}
where $d$ is the usual exterior derivative. The fundamental identity in terms of $\Pi$ can be expressed as
\begin{equation}
    \mathcal{L}_{\Pi(d f_1 \wedge \ldots \wedge d f_{p+1})} \Pi = 0 \, ,
\label{NPdef}
\end{equation}
for all $f_i \in C^{\infty} M$, where $\mathcal{L}$ is the usual Lie derivative acting on multivectors. 

First two of our claims are proven in \cite{bi2011higher}, and the fact that $\Pi$ is a bracket morphism is actually equivalent to the fact that it is a Nambu-Poisson structure, see also Proposition 3.2 of \cite{bouwknegt2013aksz}. Hence, we get a vanishing $R$-twist, $R = 0$. For the second one, we observe that the Jacobiator of the Koszul bracket can be expressed in terms of the $R$-twist:
\begin{equation}
    \mathcal{J}_Z(\omega, \eta, \mu) = \mathcal{L}_{R(\omega, \eta)} \mu + \iota_{R(\omega, \eta)} d \eta + \iota_{R(\eta, \mu)} d \omega \, ,
\end{equation}
which vanishes since $R = 0$. Moreover, the anchor of the Koszul bracket is given by $\Pi$ due to the right-Leibniz rule
\begin{equation}
    [\omega, f \eta]_Z = f [\omega, \eta]_Z + \Pi(\omega)(f) \eta\, , 
\end{equation}
Hence, we automatically have the other implication since the vanishing of the Jacobiator implies that the anchor, $\Pi$, is a bracket morphism. An important difference with the Lie bialgebroid ($p=1$) case is that, the Koszul bracket now ($p>1$) is not anti-symmetric. Its symmetric part is given by
\begin{equation}
    [\omega, \eta]_Z + [\eta, \omega]_Z = \mathbb{D}_Z g_Z(\omega, \eta) \, ,
\label{metricnp}
\end{equation}
with the identifications
\begin{equation}
    \mathbb{D}_Z = \frac{1}{2} d \, , \qquad \qquad \qquad \qquad g_Z(\omega, \eta) = 2 \left( \iota_{\Pi \omega} \eta + \iota_{\Pi \eta}\omega \right) \, ,
\label{zazan}
\end{equation}
where the metric $g_Z$, defined in terms of the usual interior product acting on $p$-forms, takes values in $\mathcal{Z} =\Lambda^{p-1} T^*M$, and $d$ is the usual exterior derivative acting on $(p-1)$-forms. Consequently, the left-Leibniz rule is satisfied with the locality operator
\begin{equation}
    L_Z(f, \omega, \eta) = d f \wedge \left( \iota_{\Pi\omega} \eta + \iota_{\Pi \eta} \omega \right) \, . 
\end{equation}
The factors $\frac{1}{2}$ and $2$ in the identification (\ref{zazan}) are crucial for the metric invariance property.

We leave the proof of the remaining claims to Appendix D. With these proofs, we conclude that for a Nambu-Poisson structure, our procedure induces a matched pair of Leibniz algebroids $(TM, \Lambda^p T^*M)$. Moreover, the full bracket on $E = TM \oplus \Lambda^p T^*M$ satisfies the metric invariance property (\ref{metricinv}) with 
\begin{equation}
    \mathbb{L}^E_{U + \omega} = \mathcal{L}_{U + \Pi \omega} \, ,
\end{equation}
where $\mathcal{L}$ is the usual Lie derivative acting on $(p - 1)$-forms, since the symmetric part of the bracket induces the following metric
\begin{equation}
    g_E(U + \omega, V + \eta) = 0 \oplus \left( \tilde{\iota}_{\omega} V + \tilde{\iota}_{\eta} U \right) \oplus 2 \left( \iota_{\Pi \omega} \eta + \iota_{\Pi \eta} \omega \right) \oplus \left( \iota_U \eta + \iota_V \omega \right)\, . 
\label{nambumetric}
\end{equation} 
This metric takes values in $\mathbb{E} = \mathbb{A} \oplus \mathcal{A} \oplus \mathbb{Z} \oplus \mathcal{Z} \cong \Lambda^{p-1} T^*M$, since the first one does not contribute and the last three of them are identical to $(p-1)$-forms, it is valued in $\Lambda^{p-1} T^*M$.

The vector bundle $E = TM \oplus \Lambda^p T^*M$ equipped with the full bracket constructed by the tilde-calculus elements then satisfies each property (except bracket morphisms) that we have discussed since every compatibility condition is satisfied. Hence, the pair $(TM, \Lambda^p T^*M)$ is a metric-Bourbaki bialgebroid, and consequently $E = TM \oplus \Lambda^p T^*M$ is a metric-Bourbaki algebroid. We refer to these bialgebroids constructed from a Nambu-Poisson structure as \textit{triangular metric-Bourbaki bialgebroids}.

It seems possible to extend this definition for an arbitrary Lie algebroid $A$ and its $p$-forms~$\Lambda^p A^*$ by using Nambu-Poisson structures on Lie algebroids \cite{wade2002nambu}. Moreover, it is interesting to extend this triangularity condition for ``exact'' bialgebroids in the sense of \cite{liu1996exact}, yet these will be out of scope of this paper. This (extended) definition of triangularity naturally includes\footnote{At least the non-extended notion recovers the triangular Lie algebroid structures related to the tangent and cotangent bundles.} triangular Lie bialgebroids which are defined in terms of a Poisson structure. In this case, the Koszul bracket~$[\cdot,\cdot]_{\text{Kos}}$ (\ref{koszulbracket}) is anti-symmetric since
\begin{equation}
    g_{T^*M}(\omega, \eta) = 2 \left( \iota_{\Pi \omega} \eta + \iota_{\Pi \eta} \omega \right) = 2 \left[ \Pi(\omega, \eta) + \Pi(\eta, \omega) \right] = 0 \, ,
\end{equation}
so that $(T^*M, \Pi, [\cdot,\cdot]_{\text{Kos}})$ is a Lie algebroid \cite{mackenzie1994lie}. Since it is a Lie algebroid, one can induce a Lie derivative $\mathcal{L}^*$ from the Koszul bracket $[\cdot,\cdot]_{\text{Kos}}$ by Equation (\ref{liederivative}) acting on all tensors. In particular for sections of $TM$, it reads 
\begin{align}
    (\mathcal{L}^*_{\omega} V) (\eta) &= (\Pi \omega)(\eta(V)) - V([\omega, \eta]_{\text{Kos}}) \nonumber\\
    &= \mathcal{L}_{\Pi \omega } \iota_V \eta - \iota_V([\omega, \eta]_{\text{Kos}}) \nonumber\\
    &= \iota_{[\Pi \omega , V]_{\text{Lie}}} \eta + \iota_V \mathcal{L}_{\Pi \omega} \eta - \iota_V (\mathcal{L}_{\Pi \omega} \eta - \iota_{\Pi \eta} d \omega) \nonumber\\
    &= [\Pi \omega, V]_{\text{Lie}} (\eta) - \iota_{\Pi \eta} \iota_V d \omega \nonumber\\
    &= [\Pi \omega, V]_{\text{Lie}}(\eta) + \Pi \iota_V d \omega (\eta) \nonumber\\
    &= [\Pi \omega, V]_{\text{Lie}}(\eta) - \Pi \mathcal{K}_V \omega (\eta) = (\tilde{\mathcal{L}}_{\omega} V)(\eta) \, ,
\end{align}
where we made use of the commutation relation between the Lie derivative and interior product and the following small identity after we set $\iota_V d \omega= \mu$
\begin{equation}
    -\iota_{\Pi \eta} \mu  = - (\Pi \eta)(\mu) = - \Pi(\eta, \mu) = \Pi(\mu, \eta) = (\Pi \mu) (\eta) \, .
\end{equation}
As both $TM$ and $T^*M$ are Lie algebroids, we can choose $g_{TM} = g_{T^*M} = 0$ and they trivially satisfy a metric invariance property (\ref{metricinvAZ}). Moreover, the metric invariance compatibility conditions (\ref{metricinvcond1}, \ref{metricinvcond2}) together with their duals (\ref{metricinvtildeconds}) are all satisfied with $\pounds = \mathcal{L}$ and $\tilde{\pounds} = \mathcal{L}^*$. The usual compatibility condition (\ref{compatibilityLie}) for Lie bialgebroids is trivially satisfied since the Koszul bracket is constructed in terms of a Poisson structure. In this sense, our triangular metric-Bourbaki bialgebroids generalize triangular Lie bialgebroids since every necessary condition is automatically satisfied due to the Nambu-Poisson condition. 

Above discussion on $\mathcal{L}^*$ and $\tilde{\mathcal{L}}$ also shows that Halmagyi's procedure \cite{halmagyi2009non} in the case that the bivector is chosen to be a Poisson structure reproduces the Drinfel'd double of the triangular Lie bialgebroid structure $(TM, T^*M)$. For $p > 1$ cases, the construction of a Lie derivative from the Koszul bracket does not work since it induces a map acting on only tensor fields whose type is a multiple of $p$. In this case it may yield a tilde-calculus on $\Lambda^{k p} TM$ induced by $\Lambda^p T^*M$ for some natural number $k$, but it cannot yield a map acting on the tangent bundle itself. Yet, the Halmagyi's approach about the automorphism procedure which we present can circumvent this problem, since it directly yields a tilde-calculus on $TM$ induced by $\Lambda^p T^*M$, as we have explicitly constructed. 

We finish this section with some comments on the properties of the full bracket that we have constructed on $E = TM \oplus \Lambda^p T^*M$:
\begin{align}
    [U + \omega, V + \eta]_E &= [U + \omega, V +  \eta]_{\text{Dor}} + [U + \omega, V + \eta]_{\tilde{\mathcal{D}}} \nonumber\\
    & \ = [U, V]_{\text{Lie}} + [\Pi \omega, V]_{\text{Lie}} + \Pi \iota_V d \omega + [U, \Pi \eta]_{\text{Lie}} - \Pi \mathcal{L}_U \eta \nonumber\\
    & \quad \ \oplus \mathcal{L}_{\Pi \omega} \eta - \iota_{\Pi \eta} d \omega + \mathcal{L}_U \eta - \iota_V d \omega \, ,
\label{papa}
\end{align}
in terms of the usual Cartan calculus elements. This \textit{higher Roytenberg bracket} in the presence of an $H$-twist was first observed in \cite{bouwknegt2013aksz, jurvco2013p}. Its anchor explicitly reads
\begin{equation}
    \rho_E(U + \omega) = \text{id}_{TM} U + \Pi \omega \, ,
\end{equation}
whereas its symmetric part is given by the metric $g_E$ in Equation (\ref{nambumetric}) and the following first-order differential operator acting on $g_E$
\begin{equation}
    \mathbb{D}_E = 0 \oplus - \Pi d \oplus \frac{1}{2} d \oplus d \, .
\end{equation}
Consequently, its locality operator is given by
\begin{equation}
    L_E(f, U + \omega, V + \eta) = 0 + \Pi \left(df \wedge (\iota_V \omega + \iota_U \eta) \right) \oplus d f \wedge \left( \iota_{\Pi \omega} \eta + \iota_{\Pi \eta} \omega \right) \oplus d f \wedge \left( \iota_U \eta + \iota_V \omega \right) \, .
\end{equation}
It is worth noting that the tilde-calculus elements that we have constructed satisfy an interesting property:
\begin{equation}
    \tilde{\mathcal{L}}_{\omega} \Pi \eta + \tilde{\mathcal{K}}_{\eta} \Pi \omega = \Pi \mathcal{L}_{\Pi \omega} \eta + \Pi \mathcal{K}_{\Pi \eta} \omega = \Pi [\omega, \eta]_{\text{Kos}} = [\Pi \omega, \Pi \eta]_{\text{Lie}} \, , \nonumber
\end{equation}
which immediately follows from the definitions of the terms. We also note that the tilde-calculus constructed from a Nambu-Poisson structure does not satisfy Equation (\ref{bourbaki}), so that it is not a Bourbaki calculus in the sense of \cite{ccatal2022pre}.


\section{Exceptional Courant Brackets}
\label{s7}

\noindent In this section, we make some observations on exceptional Courant brackets of various dimensions, and comment on their relations to our calculus framework.

Compactification of 11-dimensional supergravity over a manifold $M$ with $\text{dim} \hh M = 4$ results in a geometry where the relevant vector bundle is given by \cite{hull2007generalised, pacheco2008m}
\begin{equation}
    E = T M \oplus \Lambda^2 T^*M \, ,
\end{equation}
equipped with the following exceptional Courant bracket in terms of the usual Cartan calculus operators
\begin{equation}
    [U + \omega, V + \eta]_E = [U, V]_{\text{Lie}} \oplus \mathcal{L}_U \eta - \iota_V d \omega = [U + \omega, V + \eta]_{\text{Dorf}} \, .
\end{equation}
This bracket coincides with the higher Dorfman bracket, and thus it yields a higher Courant algebroid structure \cite{bi2011higher}. We can also construct the ``dual'' of such a bracket if the manifold $M$ is endowed with a Nambu-Poisson structure. Both of these constructions follow from the discussion of the previous section about triangular metric-Bourbaki bialgebroids for the case $p = 2$. The full bracket can be constructed as in Equation (\ref{papa}). The bracket is directly related to $SL(5)$ exceptional Drinfel'd algebra defined by relations (\ref{SL5EDA}), where the tilde-calculus elements on a frame yield non-constant fluxes $X_{A B}{}^C$ of \cite{sakatani2020u}. This relation is out of scope of this paper, and we plan to investigate it further in an upcoming paper.

When the dimension of the manifold $M$ exceeds four, the higher form degrees start to contribute \cite{hull2007generalised}, and if $\text{dim} \hh M = 5, 6$ the relevant vector bundle becomes \cite{pacheco2008m, arvanitakis2018brane}
\begin{equation}
    E = T M \oplus \Lambda^2 T^*M \oplus \Lambda^5 T^*M \, , 
\label{fiveformbundle}
\end{equation}
which is equipped with the exceptional Courant bracket  in terms of the usual Cartan calculus operators
\begin{equation}
    \left[U + \omega_2 + \omega_5, V + \eta_2 + \eta_5 \right]_E = [U, V]_{\text{Lie}} \oplus \left( \mathcal{L}_U \eta_2 - \iota_V d \omega_2 \right) \oplus \left( \mathcal{L}_U \eta_5 - \iota_V d \omega_5 - \eta_2 \wedge d \omega_2 \right) \, ,
\end{equation}
where we explicitly display the form degree for convenience. In the decomposition of the vector bundle $E$ in (\ref{fiveformbundle}) as $A \oplus Z$, choosing $A = TM, Z = \Lambda^2 T^*M \oplus \Lambda^5 T^*M$ together with $\mathcal{Z} = \Lambda^1 T^*M \oplus \Lambda^4 T^*M, \mathbb{Z} = \Lambda^4 T^*M$, we identify
\begin{align}
    [U, V]_A &= [U, V]_{\text{Lie}} \, , \nonumber\\
    \mathcal{L}_U(\eta_2 + \eta_5) &= \mathcal{L}_U \eta_2 + \mathcal{L}_U \eta_5 \, , \nonumber\\
    \mathcal{K}_V (\omega_2 + \omega_5) &= - \iota_V d \omega_2 - \iota_V d \omega_5 \, , \nonumber\\
    \iota_V (\omega_2 + \omega_5) &= \iota_V \omega_2 + \iota_V \omega_5 \, , \nonumber\\
    d (\xi_1 + \xi_4) &= d \xi_1 + d \xi_4 \, , \nonumber\\ 
    [\omega_2 + \omega_5, \eta_2 + \eta_5]_Z &= - \eta_2 \wedge d \omega_2 \, , \nonumber\\
    g_Z(\omega_2 + \omega_5, \eta_2 + \eta_5) &= - \eta_2 \wedge \omega_2 \, , \nonumber\\
    \mathbb{D}_Z \xi_4 &= d \xi_4 \, , 
\end{align}
where the operators on the right-hand side are the usual Cartan calculus operations. Here in our decomposition we choose the metric $g_Z$ with minus sign in order to be able to prove the metric invariance property depending on the graded-distributivity of the exterior derivative on the wedge product. As there is no tilde-calculus elements in this bracket, we set $\tilde{\mathcal{L}}, \tilde{\mathcal{K}}, \tilde{\iota}$ and $\tilde{d}$ to zero, so that $\mathcal{A}$ is irrelevant. Moreover, because the bracket $[\cdot,\cdot]_A$ has no symmetric part, $g_A$ is set to zero, so that $\mathbb{A}$ is also arbitrary. Since the calculus elements are just combinations of the usual Cartan calculus operations acting on various degrees, all calculus conditions (\ref{calculusconditions}) are trivially satisfied. Similarly, all six of the Jacobi compatibility conditions (\ref{comp1}, \ref{comp2}, \ref{comp3}, \ref{compdual}) can be directly proven. The metric invariance property also holds with the following map
\begin{equation}
    \mathbb{L}^E_{U + \omega_2 + \omega_5}(\xi_1 + \xi_4) = \mathcal{L}_U \xi_1 \oplus \mathcal{L}_U \xi_4 + d \omega_2 \wedge \xi_1 \, ,
\label{asdasd}
\end{equation}
since $Z$ satisfies the metric invariance property with the map $\tilde{\pounds} = 0$, and all the metric invariance compatibility conditions (\ref{metricinvcond1}, \ref{metricinvcond2}) and (\ref{metricinvtildeconds}) are satisfied.

Choosing a different decomposition, one ends up with some twists. For example, in the decomposition $A = TM \oplus \Lambda^2 T^* M, Z = \Lambda^5 T^*M$, there is an $H$-twist of the form $H: A \times A \to Z$ given by
\begin{equation}
    H(U + \omega_2, V + \eta_2) = - \eta_2 \wedge d \omega_2 \, .
\end{equation}
On the other hand, the decomposition $A = TM \oplus \Lambda^5 T^*M, Z = \Lambda^2 T^*M$ yields the following $R$-twist:
\begin{equation}
    R(\omega_2, \eta_2) = 0 \oplus - \eta_2 \wedge d \omega_2 \, .
\end{equation}
We plan to investigate brackets and their corresponding dual calculus elements in the presence of~$H$- and $R$-twists in a consecutive paper. 
Our primary results indicate that in the presence of such twists, the notion of two separate calculi somewhat dissolves as every condition coming from the Jacobi identity mixes both calculi together. Yet, the twisted exceptional Courant bracket of course still satisfies much more complicated compatibility conditions that we will explicitly present. We also note that the map (\ref{asdasd}) coming from the metric invariance property resembles the map
\begin{equation}
    \mathcal{L}_{U + \omega} \eta := \mathcal{L}_U \eta + d \omega \wedge \eta \, ,
\end{equation}
which yields a calculus (when $\omega$ is an odd degree form) with the usual exterior derivative and the modified ``interior product'' operations 
\begin{equation}
    \iota_{U + \omega} \eta := \iota_U + \omega \wedge \eta \, . 
\end{equation}
We also aim to focus on such ``twisted''\footnote{The exterior derivative operator can be also twisted by an odd degree form.} calculus examples \cite{witten1982supersymmetry}, where we also plan to investigate extensions of our constructions more explicitly for the vector bundles with three or more direct sum components.

When we have $\text{dim} \hh M =7$, the relevant vector bundle reads
\begin{equation}
    E = TM \oplus \Lambda^2 T^*M \oplus \Lambda^5 T^*M \oplus \left( T^*M \otimes \Lambda^7 T^*M \right) \, ,
\end{equation}
and the exceptional Courant bracket is given in terms of the usual Cartan calculus operators
\begin{align}
    \left[U + \omega_2 + \omega_5 + \omega_{1,7}, V + \eta_2 + \eta_5 + \eta_{1,7} \right]_E &= [U, V]_{\text{Lie}} \oplus \left( \mathcal{L}_U \eta_2 - \iota_V d \omega_2 \right) \oplus \left( \mathcal{L}_U \eta_5 - \iota_V d \omega_5 - \eta_2 \wedge d \omega_2 \right) \nonumber\\
    &\quad \oplus \left( \mathcal{L}_U \omega_{1,7} - j \eta_5 \wedge d \omega_2 - j \eta_2 \wedge d \omega_5 \right) \, ,
\label{obarey}
\end{align}
where the map $j: \Lambda^p T^*M \otimes \Lambda^{8-p} T^*M \to  T^*M \otimes \Lambda^7 T^*M$ is defined as
\begin{equation}
    (j \omega_p \wedge \omega_{8-p})(U) := (\iota_U \omega_p) \wedge \omega_{8-p} \, ,
\end{equation}
which is considered as a map of the form $TM \to \Lambda^7 T^*M$ \cite{pacheco2008m, baraglia2012leibniz, hulik2023algebroids}. Even though one can read off the calculus elements from this bracket without any twist for the decomposition $A = TM$, and~$Z = \Lambda^2 T^*M \oplus \Lambda^5 T^*M \oplus \left( T^*M \otimes \Lambda^7 T^*M \right)$, we observe that linearity condition (\ref{linearityconditions}) about the first symbol of the calculus element $\mathcal{L}$ does not hold, so that they do not form an actual calculus. Moreover, the bracket on $Z$ reads
\begin{equation}
    [\omega_2 + \omega_5 + \omega_{1,7}, \eta_2 + \eta_5 + \eta_{1, 7}]_Z = 0 \oplus - \eta_2 \wedge d \omega_2 \oplus - j \eta_5 \wedge d \omega_2 - j \eta_2 \wedge d \omega_5 \, ,
\end{equation}
which has a symmetric part that cannot be decomposed as in Equation (\ref{symmetricpart}) like the full bracket (\ref{obarey}). 

In order to take into account of such brackets, recently $Y$-algebroids have been introduced \cite{hulik2023algebroids} as generalizations of $G$-algebroids \cite{bugden2021g} where the latter's symmetric part can be decomposed as in this paper. These $Y$-algebroids are special cases of anti-commutable Leibniz algebroids of \cite{dereli2021anti}, which is introduced by one of us. An anti-commutable Leibniz algebroid $E$ is defined as a Leibniz algebroid on which there exists special, ``admissible'', (linear) $E$-connections $\nabla$ such that the symmetric part of the bracket is given by
\begin{equation}
    [u, v]_E + [v, u]_E = L_E(e^a, \nabla_{e_a} u, v) + L_E(e^a, \nabla_{e_a} v, u) \, ,
\end{equation}
where $(e_a)$ is a frame of $E$ whose dual is $(e^a)$. Here, $L_E$ is the locality operator related to the left-Leibniz rule, but it is seen as a map of the form $E^* \times E \times E \to E$ as mentioned in Footnote 2, which is denoted by $Y$ in $Y$-algebroids. On such algebroids if the $E$-connection $\nabla$ further satisfies some properties, one can construct certain algebroid representations which can be referred from Propositions 3.16 and Corollary 3.4 of \cite{dereli2021anti}. Given an $E$-connection $\nabla$, one can induce the following modified bracket
\begin{equation}
    [u, v]_E^{\nabla} := [u, v]_E - L(e^a, \nabla_{e_a} u, v) \, ,
\end{equation}
which is anti-symmetric for admissible $\nabla$. When this modified bracket also satisfies the Jacobi identity, then one can prove that the corresponding Lie and exterior derivatives with the interior product indeed form a calculus on $\Lambda^p E^*$ induced by $E$ \cite{dereli2021anti}. The additional required properties of such admissible $E$-connections for individual algebroid axioms, with their possible relations to non-uniqueness of Levi-Civita $E$-connections, will be the subject of another future paper. It is also interesting to analyze various properties of more complicated algebroid structures such as the ones considered in \cite{bugden2022exceptional, hulik2022exceptional} in our calculus framework or its possible extensions.


\section{Further Topics}
\label{s8}

In this section, we concentrate on two further topics: Nambu-Poisson exceptional generalized geometry and formal bundle rackoids. In the first subsection, we summarize Muraki's Poisson generalized geometry \cite{muraki2015new, asakawa2015poisson} and extend it to Nambu-Poisson exceptional generalized geometry in light of our constructions from Section \ref{s6}. In the second part, we focus on formal bundle rackoids introduced in \cite{ikeda2021global}. We slightly extend the notion of metric rackoids for algebroids equipped with a vector bundle valued metric. We only take the first preliminary steps in these directions, which we plan to further investigate in the near future.


\subsection{Nambu-Poisson Exceptional Generalized Geometry}
\label{s8a}

\noindent In this subsection, we summarize the Poisson generalized geometry introduced in \cite{asakawa2015poisson}. We explicitly show that our calculus elements constructed by using a Nambu-Poisson structure reproduces the Poisson calculus \cite{vaisman2012lectures}. Hence, we are able to construct an extension which we call as Nambu-Poisson exceptional generalized geometry, and prove some preliminary results on the similarities and differences between Poisson and Nambu-Poisson cases. 

Conventionally, in generalized geometry, one constructs an exact Courant algebroid starting from the tangent Lie algebroid $(TM, \text{id}_{TM}, [,]_{\text{Lie}})$, whereas the cotangent bundle has vanishing algebroid structures. The Drinfel'd double of these algebroids is the standard Courant algebroid, which in this section we denote by $(TM)_{\text{Lie}} \oplus (T^*M)_0$. In Poisson generalized geometry \cite{asakawa2015poisson, muraki2015new}, one takes the ``dual'' approach where the cotangent and tangent bundles' roles are interchanged, and a Courant algebroid structure on the generalized tangent bundle is constructed starting from the cotangent bundle. Of course for this construction to work, the base manifold should be equipped with a Poisson structure $\Pi$. As noted earlier, by virtue of the Poisson bivector, one can induce a Lie algebroid structure $(T^*M, \Pi, [\cdot,\cdot]_{\text{Kos}})$ on the cotangent bundle. Here, the musical map $\Pi:T^*M \to TM$ is the anchor, and sections of $T^*M$ are equipped with the Koszul bracket (\ref{koszulbracket}). One can furnish the tangent bundle with vanishing algebroid structures, and the Drinfel'd double of these two algebroids is the Courant algebroid denoted by $(TM)_0 \oplus (T^*M)_{\Pi}$ \cite{asakawa2015poisson}, which fits into the following exact sequence
\begin{equation} 
    0 \rightarrow (TM)_0 \rightarrow E \rightarrow \left( T^*M \right)_{\Pi} \rightarrow 0\, .
\end{equation}  

As demonstrated in Section \ref{s3}, by using the Lie algebroid structure on $T^*M$, one can induce the corresponding Lie derivative and exterior derivative operations on its dual bundle $TM$ by Equations (\ref{liederivative}) and (\ref{exteriorderivative}), respectively. We also discussed the equivalence of this approach to the one of Halmagyi's \cite{halmagyi2008non} in Section \ref{s6}. As we also have shown, together with the interior product $\tilde{\iota}$ coming from canonical pairing, they define a calculus $(\tilde{\mathcal{L}}, \tilde{\iota}, \tilde{d})$ on $TM$ induced by $T^*M$, known as Poisson calculus in the literature \cite{vaisman2012lectures}. The operator $\tilde{d}$ corresponds to the differential operator $\hat{d}$ of the Poisson complex \cite{lichnerowicz1977varietes}, which reads specifically on smooth functions 
\begin{equation}
    \tilde{d} f = - \Pi d f = [\Pi, f]_{\text{SN}} = \hat{d} f \, .
\end{equation}
This map $\hat{d} P := [\Pi, P]_{\text{SN}}$ defined via the usual Schouten-Nijenhuis bracket constructed from the Lie bracket can act on any multivector $P$, but since the canonical pairing takes values in smooth functions, $\tilde{d}$ as a calculus element in our sense only acts on smooth functions, and it appeared in Equation (2.2) of \cite{asakawa2015poisson}. Moreover, the operator $\tilde{\mathcal{L}}$ is constructed by the Cartan magic formula, and its equivalence to our formulation is proven in Equation (2.13) of \cite{asakawa2015poisson}, where similar formulas also appeared in \cite{blumenhagen2013non}. With these objects, the full bracket on the Courant algebroid~$E = (TM)_0 \oplus (T^*M)_{\Pi}$ is given by the Dorfman-like bracket (\ref{dtildebracket}):
\begin{equation}
    [U + \omega, V + \eta]_E = \tilde{\mathcal{L}}_\omega V + \tilde{\mathcal{L}}_\eta U \oplus [\omega, \eta]_{\text{Kos}} \, , 
\end{equation}
for $U, V \in TM, \omega, \eta \in T^*M$ together with the anchor $\rho_E = 0 \oplus \Pi$ and the metric 
\begin{equation}
    g_E(U + \omega, V + \eta) = \tilde{\iota}_\omega V + \tilde{\iota}_\eta U = \iota_V \omega + \iota_U \eta \, .     
\end{equation}

As proven in \cite{asakawa2015poisson}, the symmetries of $E = (TM)_0 \oplus (T^*M)_{\Pi}$ are given by $\beta$-transformations and $\beta$-diffeomorphisms, where the latter is first introduced in \cite{blumenhagen2013non}. The  $\beta$-diffeomorphism is defined by the diagonal action of a 1-form $\xi$ as
\begin{equation}
    \xi \blacktriangleright (U + \omega) := \tilde{\mathcal{L}}_{\xi} U \oplus [\xi, \omega]_{\text{Kos}}\, .
\end{equation}
It is a symmetry of the Courant algebroid $E$; it preserves all the structures on the Courant algebroid in the sense:
\begin{align}
    \rho_E(\xi \blacktriangleright (U + \omega)) &= \xi \blacktriangleright \rho_E(U + \omega) \, , \label{betadiffanch} \\
    [\xi \blacktriangleright (U + \omega), V + \eta]_E + [U + \omega, \xi \blacktriangleright (V + \eta)]_E &= \xi \blacktriangleright [U + \omega, V + \eta]_E \, , \label{betadiffbrac} \\
    g_E(\xi \blacktriangleright (U + \omega), V + \eta) + g_E(U + \omega, \xi \blacktriangleright (V + \eta)) &= \xi \blacktriangleright g_E(U + \omega, V + \eta) \, . \label{betadiffmetr}
\end{align}
Last two equations hold for arbitrary $\xi$, whereas the first one is true only if $\Pi \xi$ is $\hat{d}$-closed, \textit{i.e.}, $\hat{d} (\Pi \xi) = 0$. Moreover, a $\beta$-transformation is defined as
\begin{equation}
    e^{\beta}(U + \omega) := U + \beta(\omega) \oplus \omega \, ,
\end{equation}
which is actually the action of $\Psi_{\Pi}$ of the previous section where we use $\beta$ in the place of $\Pi$. Similarly, a $\beta$-transformation is also a symmetry in the sense:
\begin{align}
    \rho_E(e^{\beta}(U + \omega)) &= \rho_E(U + \omega) \, , \nonumber\\
    [e^{\beta}(U + \omega), e^{\beta}(V + \eta)]_E &= e^{\beta}[U + \omega, V + \eta]_E \, , \nonumber\\
    g_E \left( e^{\beta}(U + \omega), e^{\beta}(V + \eta) \right) &= g_E(U + \omega, V + \eta) \, .
\label{betatransformsym} 
\end{align}
Now the first and third equations are valid for arbitrary $\beta$, but the second only is true only for $\hat{d}$-closed $\beta$. We note that when $\beta$ is chosen as $\Pi$, the second condition is satisfied because $\Pi$ is Poisson, so that its Schouten-Nijenhuis bracket with itself vanishes and hence it is $\hat{d}$-closed.

In this framework of Poisson generalized geometry, Poisson-analogues of constructions in standard generalized geometry can be studied. For instance, in \cite{asakawa2012d}, a geometrical treatment of D-branes as fluctuations of Dirac structures in generalized geometry is studied. Based on this work, an alternative description of D-branes in Poisson generalized geometry is studied, and its relation with the standard description is noted to be governed by a Seiberg-Witten map \cite{asakawa2014d}. Furthermore in \cite{asakawa2015poisson} $R$-flux is defined similarly to how $H$-flux is defined in generalized geometry: it is realized as twisting of the Courant algebroid $E = (TM)_0 \oplus (T^*M)_{\Pi}$ in terms of a triangular twist matrix associated with $\beta$-transformations. Finally, by exploiting topological T-duality as a symmetry which relates $R$- and $Q$-fluxes, a definition of $Q$-flux is proposed in \cite{asakawa2015topological}. Consequently, Poisson generalized geometry provides a geometrical framework for the investigation of non-geometric fluxes and string backgrounds \cite{hull2005geometry} such as T-folds \cite{hull2007doubled}. All of these indicate that the algebraic structure of spacetime might be non-commutative and non-associative \cite{blumenhagen2011non, jurvco2013generalized}, and the relation of non-geometric fluxes with the algebroid framework is further investigated in \cite{blumenhagen2013non, blumenhagen2013intriguing, chatzistavrakidis2014dirac}.

We believe generalization to Nambu-Poisson manifolds will be useful in similar dual constructions of higher Courant algebroids \cite{bi2011higher} used in exceptional generalized geometries \cite{pacheco2008m}, physics of branes \cite{bagger2007modeling,  bagger2008gauge, bagger2008comments, gustavsson2009algebraic, basu2005m2, ho2008m5} and U-duality \cite{sakatani2020u, malek2020poisson} all of which will be out of scope of this paper. We now describe Nambu-Poisson exceptional generalized geometry, and study symmetry properties of $\beta$-diffeomorphisms and $\beta$-transformations in this context.

We can extend Poisson generalized geometry to a \textit{Nambu-Poisson exceptional generalized geometry} where the relevant underlying vector bundle becomes $TM \oplus \Lambda^p T^*M$. In this case, the standard construction yields the higher Courant algebroid denoted by $(TM)_{\text{Lie}} \oplus \left( \Lambda^p T^*M \right)_0$, whose bracket is the higher Dorfmann bracket \cite{bi2011higher}. We are interested in the dual construction on $(TM)_0 \oplus \left( \Lambda^p T^*M \right)_{\Pi}$, where now $\Pi$ is a Nambu-Poisson structure of order $(p + 1)$. Notice that all the formulas that we wrote for the case $p = 1$ immediately follow since in Section \ref{s6} we derived them for an arbitrary $p$ including $p = 1$. Of course there are important differences with the $p = 1$ case, and now we highlight these. 

For the remainder of this section, we write $E$ for the vector bundle $TM \oplus \Lambda^p T^*M$ where $p$ is an arbitrary positive integer different than one, and equip it with the $\tilde{\mathcal{D}}$-bracket:
\begin{equation}
    [U + \omega, V + \eta]_{\tilde{\mathcal{D}}} = \tilde{\mathcal{L}}_\omega V - \tilde{\mathcal{L}}_\eta U + \tilde{d} \tilde{\iota}_{\eta} U \oplus [\omega, \eta]_{\text{Kos}} \, ,
\end{equation}
where now the tilde-calculus elements are taken to be the ones (\ref{tildecalculuspi}) and (\ref{poissonexterior}) from Section \ref{s6}. Observe that the bracket coming from tilde-calculus operations is in agreement with the case $p = 1$. However, in this case the symmetric part of the bracket does not coincide with that of the higher Dorfmann bracket. For the higher Dorfman bracket, the symmetric part is given in terms of the metric
\begin{equation}
    \iota_V \omega + \iota_U \eta \, ,
\end{equation}
whereas for the $\tilde{\mathcal{D}}$-bracket the symmetric part is given in terms of the metric
\begin{equation}
    \tilde{\iota}_{\omega} V + \tilde{\iota}_{\eta} U + 2 \left( \iota_{\Pi \omega} \eta + \iota_{\Pi \eta} \omega \right) = \iota_V \omega + \iota_U \eta + 2 \left( \iota_{\Pi \omega} \eta + \iota_{\Pi \eta} \omega \right) \, , 
\label{metricpi}
\end{equation}
for $U, V \in TM, \omega, \eta \in \Lambda^p T^*M$ by Equation (\ref{zazan}). This difference is caused by the fact that for~$p > 1$ the Koszul bracket does not define a Lie algebroid structure but instead a Leibniz algebroid structure with a non-vanishing symmetric part. 

Let us now study the symmetry properties of $\beta$-diffeomorphisms and $\beta$-transformations in the Nambu-Poisson context. We can directly extend their definitions for a $p$-form $\xi$, and a $(p+1)$-vector $\beta$. We first note that Condition (\ref{betadiffbrac}) about the brackets for a $\beta$-diffeomorphism induced by a $p$-form $\xi$ follows from the Jacobi identity of the $\tilde{\mathcal{D}}$-bracket with $\omega = 0$, that is, it is equivalent to:
\begin{equation}
    [U, [V + \eta, W + \xi]_E]_E - [[U, V + \eta]_E, W + \xi]_E - [V + \eta, [U, W + \xi]_E]_E = 0 \, . 
\end{equation}
Similarly, Condition (\ref{betadiffmetr}) is equivalent to the metric invariance of $\tilde{\mathcal{D}}$-bracket with $U = 0$, that is, 
\begin{equation}
   g_E([\omega, V + \eta]_E, W + \xi) + g_E(V + \eta, [\omega, W + \xi]_E) =  \mathcal{L}_{\Pi \omega} g_E(V + \eta, W + \xi) \, .
\end{equation}
Finally, let us look at both sides of anchor preservation condition (\ref{betadiffanch}):
\begin{align}
    \rho_E(\xi \blacktriangleright (U + \omega)) &= (0 \oplus \Pi)(\tilde{\mathcal{L}}_{\xi} U \oplus [\xi, \omega]_{\text{Kos}}) = \Pi [\xi, \omega]_{\text{Kos}} = [\Pi \xi, \Pi \omega]_{\text{Lie}} \, ,  \nonumber\\ 
    \xi \blacktriangleright \rho_E(U + \omega) &= \xi \blacktriangleright (0 \oplus \Pi)(U + \omega) = \xi \blacktriangleright (\Pi \omega) = \tilde{\mathcal{L}}_{\xi} (\Pi \omega) = [\Pi \xi, \Pi \omega]_{\text{Lie}} -\Pi \mathcal{K}_{\Pi \omega} \xi \, .
\end{align}
Here we see that, for a $\beta$-diffeomorphism to be a symmetry of the anchor, we require:
\begin{equation}
    -\Pi \mathcal{K}_{\Pi \omega} \xi = \Pi \iota_{\Pi \omega} d \xi = -(\mathcal{L}_{\Pi \xi} \Pi) \omega = - [\Pi \xi, \Pi]_{\text{SN}} \omega = 0 \, ,
\end{equation}
where we use different equivalent definitions of Nambu-Poisson structures; for instance see Proposition~3.2 of \cite{bouwknegt2013aksz} or the proof of Lemma 3.2 of \cite{bi2011higher}. This in general is not true, and a $\beta$-diffeomorphism does not preserve the anchor. For $p = 1$ case, this is equivalent to $\hat{d}$-closedness of $\Pi \xi$ due the last equality.

Now we come to the $\beta$-transformations on the algebroid $E$, and check the symmetry relations given by (\ref{betatransformsym}). One can easily prove that a $\beta$-transformation is a symmetry of the $\tilde{\mathcal{D}}$-bracket if
\begin{equation}
    [\Pi \omega, \beta \eta]_{\text{Lie}} + [\beta \omega, \Pi \eta]_{\text{Lie}} - \Pi(\mathcal{L}_{\beta \omega} \eta + \mathcal{K}_{\beta \eta} \omega) - \beta(\mathcal{L}_{\Pi \omega} \eta + \mathcal{K}_{\Pi \eta} \omega) = 0 \, .
\label{ananana}
\end{equation}
When the $(p+1)$-vector $\beta$ is taken as $\Pi$, the above equation becomes
\begin{equation}
    [\Pi \omega, \Pi \eta]_{\text{Lie}} - \Pi[\omega, \eta]_{\text{Kos}} = 0 \, , 
\end{equation}
which holds since $\Pi$ is Nambu-Poisson. For the case $p = 1$, this expression is equivalent to
\begin{equation}
    [\Pi, \beta]_{\text{SN}} = 0 \, , 
\end{equation}
which amounts to the $\hat{d}$-closedness of the bivector $\beta$. This is of course not valid for higher order vectors. Any $\beta$-transformation preserves the anchor, that is the second equation in (\ref{betatransformsym}) is trivially satisfied. Finally, a $\beta$-transformation respects the metric $g_E$ given by (\ref{metricpi}) if 
\begin{equation}
    \tilde{\iota}_\omega \beta(\eta) + \tilde{\iota}_\eta \beta(\omega) = \iota_{\beta(\eta)} \omega + \iota_{\beta(\omega)} \eta = 0 \, . 
\end{equation}
The above displayed equation is trivially satisfied when $\beta$ is a bivector, that is for the case $p = 1$. In general for $p > 1$, a $\beta$-transformation is not a symmetry of the metric $g_E$. Nevertheless this very condition seems to be a defining property of higher Poisson structures that are related to higher Dirac structures \cite{bursztyn2015multisymplectic, bursztyn2019higher}.


\subsection{Metric Bundle Rackoids}
\label{s8b}

We now summarize the basics of metric bundle rackoids, which are groupoid-like objects that can be thought as the exponentiation or integration of the \textit{infinitesimal} algebroids following \cite{ikeda2021global}. We then slightly extend the latter notion for vector bundle valued metrics, as it is needed for examples including the ones in the framework of higher Courant algebroids and Nambu-Poisson structures. 

The relation between a Lie group and its Lie algebra is ubiquitous for many areas of both mathematics and physics. Under certain circumstances the former can be determined by the latter; or at least the Lie algebra carries a lot of information about its corresponding Lie group. It is now well-known that a Lie bialgebra is an infinitesimal version of a Poisson-Lie group, which underlies the algebraic setup for Poisson-Lie T-duality \cite{klimvcik1995dual}, which can be extended to the algebroid level \cite{mackenzie1994lie} for Poisson groupoids. It is natural to expect a similar relation between Leibniz algebras and some group-like objects, or an analogous extension even at the Leibniz algebroid level. The procedure to integrate an algebra to a group-like object is known as the \textit{coquecigrue problem} \cite{loday1993version}. A solution to the coquecigrue problem for a Leibniz algebra yields a group-like object called rack \cite{kinyon2004leibniz}. These are equipped with a binary operation that mimics the adjoint action of a Lie group on its algebra. Racks are usually defined in a category theoretical language, and certain algebroid extensions like Lie rackoids \cite{laurent2020lie}, or integration of exact Courant algebroids \cite{li2011integration} have been proposed in the same categorical manner.

More in accord with the algebroid literature, Ikeda and Sasaki introduced formal rackoids~\cite{ikeda2021global} which are defined as natural axiomatizations. Their approach is to formalize the exponential map for metric algebroids, which include Courant algebroids, and define the notion of formal pre-rackoids for a better understanding of the global structure of the doubled spacetime of double field theories. This notion is a generalization of the cotangent path rackoids introduced in \cite{laurent2020lie} where Courant algebroids correspond to the infinitesimal versions. As it is shown in \cite{ikeda2021global}, these rackoid structures are directly related to topological sigma models for doubled spacetime \cite{kokenyesi2018double, chatzistavrakidis2018double}, which is an extension of the usual Courant sigma model \cite{ikeda2003chern}, and they are realized as Wilson lines in the topological sigma model. 

In rackoids, there are binary operations called rack product and rack action, where the former's self-distributivity is directly related to the strong constraint of DFT, and the latter has some close ties with quantum Yang-Baxter equation. Moreover, pre-rackoids are also related with the large gauge transformations of DFT \cite{hohm2013large} as the exponentiation of D-bracket of DFT has appeared as such a transformation. The importance of rackoids, even though not explored in details, is noticed in the exceptional Drinfel'd algebra literature \cite{fernandez2021jacobi}. As exceptional field theories are certain generalizations of DFT, the notion of rackoids should be extended to higher Courant algebroids whose metric takes values in $p$-forms. Our aim in this subsection is to take the first steps in this direction, and construct a rackoid for algebroids who is equipped with a vector bundle valued metric satisfying the metric invariance property. As certain examples of these algebroid structures are directly related to exceptional Drinfel'd algebras as we have discussed, we expect them to be useful for analogous results in exceptional field theories. We plan to investigate these topics and their possible relation to sigma models parallel to \cite{ikeda2021global} thoroughly in the near future. 

A pair $(E, \triangleright)$ is called a bundle (Lie) rackoid \cite{ikeda2021global}, if the vector bundle $E$ is equipped\footnote{Note that we ignore the other product $\cdot: E \times C^{\infty}M \to C^{\infty}M$, and take it directly as the usual product of a smooth function with a section of a vector bundle.} with a rack operation $\triangleright: E \times E \to E$ and a rack action $\triangleright: E \times C^{\infty}M \to C^{\infty}M$ satisfying
\begin{align}
    u \triangleright \left( v \triangleright w \right) &= \left( u \triangleright v \right) \triangleright \left( u \triangleright w \right), \label{jacobirackoid} \\
    u \triangleright \left( v \triangleright f \right) &= \left( u \triangleright v \right) \triangleright \left( u \triangleright f \right), \label{anchorrackoid}\\
    u \triangleright (f v) &= \left( u \triangleright f \right) \left( u \triangleright v \right) \, ,
\label{leibnizrackoid}
\end{align}
for all $u, v, w \in E, f \in C^{\infty}M$. These properties are related to the Jacobi identity, the anchor being a bracket morphism and the right-Leibniz rules, respectively. If we ignore the first condition, then we get a bundle pre-rackoid. 

In the first place, similarly to the algebroid literature, we can prove that the second one can be implied by the others. Consider the expression
\begin{align}
   u \triangleright (v \triangleright fw) &= (u \triangleright v ) \triangleright (u \triangleright fw) \nonumber\\
   &= (u \triangleright v ) \triangleright \left( (u\triangleright f) (u \triangleright w) \right) \nonumber\\
   &= \left((u\triangleright v) \triangleright (u \triangleright f) \right) \left( (u \triangleright v) \triangleright ( u\triangleright w) \right) \nonumber\\
   &= \left((u\triangleright v) \triangleright (u \triangleright f) \right) \left( u \triangleright (v \triangleright w) \right) \, , 
\end{align}
where in the first line we use (\ref{jacobirackoid}), in the second and third equalities we use (\ref{leibnizrackoid}) and in the last equality we again make use of (\ref{jacobirackoid}) property. Now let us evaluate the same expression differently:
\begin{align}
   u \triangleright (v \triangleright fw) &= (u \triangleright v ) \triangleright (u \triangleright fw) \nonumber\\
   &= u \triangleright \left( (v\triangleright f) (v \triangleright w) \right) \nonumber\\
   &= \left(u \triangleright (v \triangleright f) \right) \left( u \triangleright (v \triangleright w) \right) \, . 
\end{align}   
Here, we use (\ref{leibnizrackoid}) in the first equality and (\ref{jacobirackoid}) in the last one. Comparing the two results for the same expression, we see that we obtain the second property (\ref{anchorrackoid}), and hence we can actually eliminate it from the definition. 

A rack operation and a rack action can be constructed by the exponentiation of the bracket and anchor of a Leibniz algebroid. Indeed, given a Leibniz algebroid $(E, \rho_E, [\cdot,\cdot]_E)$, it can be equipped with a bundle rackoid structure via the maps defined by
\begin{equation}
    u \triangleright v = \text{exp} \ \text{ad}(u) (v), \qquad \qquad \qquad \qquad u \triangleright f = \text{exp} \ \rho_E(u) (f) \, , \label{rackaction}
\end{equation}
where we define $ad(u)(v) := [u, v]_E$, analogously to the adjoint action for a Lie algebra. As these operations are defined as \textit{formal} exponentiation of the structures of an algebroid, the rackoids are dubbed as formal rackoids in \cite{ikeda2021global}. Furthermore, if the vector bundle $E$ is equipped with a metric $g_E$ that takes values in $C^{\infty}M$ satisfying the \textit{rackoid metric condition}
\begin{equation}
    u \triangleright g_E(v, w) = g_E(u \triangleright v, u \triangleright w) \, ,
\label{metricrack}
\end{equation}
then it is called a metric bundle rackoid. In \cite{ikeda2021global}, it is proven that any Courant algebroid yields a metric bundle rackoid, since the metric invariance property
\begin{equation}
    \rho_E(u) g_E(v, w) = g_E([u, v]_E, w) + g_E(v, [u, w]_E) \, ,
\end{equation}
can be exponentiated to yield the metric condition (\ref{metricrack}). As the metric $g_E$ of a Courant algebroid takes values in $C^{\infty}M$, there is no need for an additional rack action, and the formal exponentiation of the anchor $\rho_E$ is enough to show the metric condition (\ref{metricrack}) for a bundle rackoid. 

For more general algebroids, the metric might take values in an arbitrary vector bundle as we have discussed. Hence, for a larger class of algebroids we need to extend the definition of the metric bundle rackoids via a third and independent rack action. For a metric $g_E$ that takes values in the vector bundle $\mathbb{E}$, we need a new rack action as a map of the form
\begin{equation}
    \triangleright: E \times \mathbb{E} \to \mathbb{E} \, .
\label{metricaction}
\end{equation}
We introduce the notion of $\mathbb{E}$-metric bundle rackoid as a triplet $(E, \mathbb{E}, \triangleright)$, where $(E,\triangleright)$ is a bundle rackoid equipped with the additional rack action (\ref{metricaction}), satisfying 
\begin{equation}
    u \triangleright g_E(v, w) = g_E(u \triangleright v, u \triangleright w) \, ,
\label{rackmetriccondition}
\end{equation}
where now the rack action on the left-hand side is given by the new action (\ref{metricaction}), and the other is the usual rack operation defined in (\ref{rackaction}). Trivially, the metric bundle rackoid of \cite{ikeda2021global} corresponding to a Courant algebroid is a $C^{\infty}M$-metric bundle rackoid. 

For an arbitrary algebroid satisfying the general form of the metric invariance property (\ref{metricinv}) with operator $\mathbb{L}^E: E \times \mathbb{E} \to \mathbb{E}$, the third rack action $\triangleright: E \times \mathbb{E} \to \mathbb{E}$ satisfying (\ref{rackmetriccondition}) can be defined by the formal exponentiation of the operator~$\mathbb{L}^E$:
\begin{equation}
    u \triangleright \xi = \text{exp}\,  \mathbb{L}^E_u \xi \, ,
\end{equation}
for $u \in E, \xi \in \mathbb{E}$. The details of this simple proof can be found in the Appendix E. Therefore the metric invariance property (\ref{metricinv}) also yields a rackoid metric condition (\ref{rackmetriccondition}) by formal exponentiation in parallel with the other algebroid properties. 

Two important examples of algebroids satisfying metric invariance property are higher Courant algebroids \cite{bi2011higher} and the Drinfel'd double of triangular metric-Bourbaki bialgebroids coming from Nambu-Poisson structures constructed in Section \ref{s6}. Their new rackoid actions explicitly read
\begin{equation}
    u \triangleright \xi = \text{exp} \, \mathcal{L}_{\text{proj}_{TM}(u)} \xi \, , \qquad \qquad \qquad \qquad u \triangleright \xi = \text{exp} \, \mathcal{L}_{(\text{id}_{TM} \oplus \Pi) (u)} \xi \, ,
\end{equation}
respectively. Here, the maps on the right-hand side are usual Lie derivatives composed with their respective anchors. As the metric in both examples take values in $(p-1)$-forms, both can be considered as $\Lambda^{p-1} T^*M$-metric bundle rackoids.

After introducing formal (pre-)rackoids, the discussion in \cite{ikeda2021global} continues with the construction of a Lie path rackoid in the framework of topological sigma models, in particular Courant sigma model \cite{ikeda2003chern, roytenberg2007aksz} and its extension for DFT \cite{kokenyesi2018double, chatzistavrakidis2018double}. Exponentiation of the fields in such models can be identified with the rackoid structures and these are physically interpreted as the Wilson lines corresponding to gauge fields \cite{ikeda2021global}. In \cite{bouwknegt2013aksz}, AKSZ construction is extended for higher Courant algebroids by using Nambu-Poisson structures. We should expect a close relation between this model and our triangular metric-Bourbaki bialgebroids analogous to the Poisson case \cite{cattaneo2001aksz}, which we plan to investigate in the near future.


\section{Concluding Remarks}
\label{s9}

\noindent In this paper, we have constructed doubles of a rather general class of bialgebroids by reinterpreting and extending the results known for matched pairs \cite{ibanez2001matched} of Leibniz algebroids. We have achieved this by introducing the notion of calculus and dual calculi on algebroids and by analyzing the compatibility conditions required for having a bialgebroid structure. This analysis has led us to various compatibility conditions, which we have obtained by studying a general set of algebroid axioms:  right- and left-Leibniz rules, Jacobi identity, a specific decomposition of the symmetric part \cite{ccatal2022pre, bugden2021g}, metric invariance property and certain bracket morphisms. In the case of Lie bialgebroids, all but one of the compatibility conditions are satisfied trivially, and the non-trivial condition to be satisfied is equivalent to the usual compatibility condition known for Lie bialgebroids \cite{mackenzie1994lie}. As in the case of Lie bialgebroids, compatibility of the calculi makes it possible to construct a new algebroid structure on the sum of the given algebroids structure which are not dual in the usual sense. The new algebroid structure on the direct sum can and has been dubbed the Drinfel'd double. A special class of Lie bialgebroids are triangular Lie bialgebroids, for which the algebroid structure on the dual  $A^*$ of the Lie algebroid A is constructed by utilizing a bivector $\Pi \in \Lambda^2 A$ satisfying $[\Pi, \Pi]_{\text{SN}, A} = 0$. The algebroid structure on $A^*$ induced by such $\Pi$ is automatically compatible with the  algebroid structure on $A$ so that $(A, A^*)$ is a Lie bialgebroid and there is a Courant algebroid structure on $A \oplus A^*$. Such bialgebroids are important for applications in physics. For example, for Poisson-Lie T-duality, the relevant bialgebroid is of this type, with $A = TM$, where $M$ is a Poisson-Lie group and $\Pi \in \Lambda^2 TM$ is the associated Poisson bivector. We discuss in Section \ref{s6} how the methods used in \cite{halmagyi2009non} are equivalent to but more practical then the usual methods used for constructing triangular Lie bialgebroids. In this section, extending the procedure of \cite{halmagyi2009non} to the vector bundle $TM \oplus \Lambda^p T^*M$, we construct the analogues of triangular bialgebroids. We utilize a $(p+1)$-vector $\Pi$ in order to extend the notion of triangularity. As has already been shown in \cite{bi2011higher}, when $\Pi$ is a Nambu-Poisson structure, it induces a Leibniz algebroid structure on $\Lambda^p T^*M$ via the Koszul bracket. Moreover, it has been discussed that the Dorfman bracket induces a higher Courant algebroid structure on $TM \oplus \Lambda^p T^*M$. We use these results and combine them with the procedure of \cite{halmagyi2009non} to construct a calculus dual to the usual Cartan calculus. The full bracket on $TM \oplus \Lambda^p T^*M$ then is realized as the higher Roytenberg bracket in the absence of twists \cite{jurvco2013p}. Our results also make it possible to construct the analogue of Poisson generalized geometry studied in \cite{asakawa2015poisson}. We discuss this in Section \ref{s8} and call the resulting geometric framework Nambu-Poisson exceptional generalized geometry. In the last part of the paper, we also work on some preliminary results on the coquecigrue problem and extend the notion of metric bundle rackoids \cite{ikeda2021global} for vector bundle valued metrics.

When $\Pi$ is not a Poisson structure in above constructions one ends up with protobialgebroids \cite{roytenberg2002quasi}, and the resulting bracket on the double is the full Roytenberg bracket. This bracket was studied in \cite{halmagyi2009non} in order to construct a charge algebra which also includes non-geometric fluxes. The $R$-flux, for example, measures the non-vanishing part of the Schouten-Nijenhuis bracket of the bivector $\Pi$ with itself. Then, a natural extension of our constructions would be the case where we relax the condition of $A$ and $Z$ being subalgebroids of their double $E = A \oplus Z$. This would yield the ultimate most general case where both $H$- and $R$-twists are turned on, which would be an extension of protobialgebroids of Roytenberg. We have some preliminary results in this direction that we plan to publish in a consecutive paper. For such constructions, we see that the notion of two dual individual calculi dissolves, since all of the Jacobi compatibility conditions mix the calculus and tilde-calculus elements together. Yet, our constructions might sustain a useful framework for working on the details of such generalizations in the realm of Nambu-Poisson structures, including twisted Poisson \cite{vsevera2001poisson}, twisted $R$-Poisson \cite{chatzistavrakidis2022bv} structures and pre-Roytenberg algebras \cite{blumenhagen2012bianchi}, where we observe that the latter's relation to Roytenberg algebras is still valid.

We have discussed the matched pairs of Lie algebroids and Lie bialgebroids, and extended these notions for other algebroid axioms. There is a closely related third construction which is the Manin triples of Lie algebroids crucial for Poisson-Lie T-duality \cite{klimvcik1995dual}. It is also natural to expect the notion of Manin triples to be useful for a better understanding of Nambu-Poisson U-duality in the algebroid settings that we analyzed. We believe our constructions sustain a convenient framework to work on this relation, which we plan to investigate in the near future. In particular, exceptional Drinfel'd algebras are directly relevant to our discussion. Indeed, it can be checked that, in a frame, the tilde-calculus elements that we have constructed in Section \ref{s6} for the triangular bialgebroids reproduce the non-constant fluxes $X_{A B}{}^C$ of Equations (3.10-3.17) of \cite{sakatani2020u}. These non-constant fluxes reduce to the structure functions of the $SL(5)$ exceptional Drinfel'd algebra \cite{sakatani2020u} at a certain point on the base manifold where the trivector is assumed to vanish, and the higher Roytenberg bracket~(\ref{papa}) we constructed induce the correct commutation relations (\ref{SL5EDA}), Equation 3.23 of \cite{sakatani2020u}. Moreover, the Jacobi compatibility conditions give the quadratic constraints (\ref{quadratic}). We plan to investigate the further details of exceptional Drinfel'd algebras in our dual calculus framework in a consecutive paper. The preliminary results indicate that our dual calculus formalism is useful for frame independent formulations, and they can serve as good book-keeping devices. For example, since Bianchi identities for fluxes are equivalent to the Jacobi identity of the relevant bracket \cite{blumenhagen2012bianchi}, our formalism make it possible to interpret these identities as generalizations of Cartan calculus relations and Jacobi compatibility conditions that we explicitly derived in Section \ref{s4}.

As there are more complicated exceptional Drinfel'd algebras related to an algebroid given in terms of the direct sum of three or more vector bundles \cite{malek2021e6, blair2022generalised}, we also plan to work on the details of the extensions of our constructions for such decompositions. Since the structures in this paper have been constructed in a way as broad as possible, our methods can be directly applied to such cases as well. This then means that the same compatibility conditions must be satisfied for any pair of vector bundles in the decomposition. This paves the way of constructing ``duals'' of exceptional Courant brackets of \cite{pacheco2008m}, which we commented on in Section \ref{s7}. These duals should be directly related to algebroid versions of more complicated exceptional Drinfel'd algebras \cite{malek2021e6, blair2022generalised, kumar202310}. Related to these complicated exceptional brackets, recently the notion of $Y$-algebroids is introduced \cite{hulik2023algebroids}, which are special cases of anti-commutable Leibniz algebroids \cite{dereli2021anti}. These algebroids include the cases where the decomposition property (\ref{symmetricpart}) of the bracket is not satisfied, so we need to further modify the formalism presented in this paper for such cases. We plan to investigate such constructions in a consecutive paper. 

It is also interesting to relate our work with derived brackets \cite{Kosmann_Schwarzbach_2004}, which would have some implications on the sigma model level \cite{arvanitakis2018brane}. We hope to be able to analyze AKSZ-type \cite{alexandrov1997geometry} constructions in our calculus framework without using a graded language. We expect that our results on triangular bialgebroids constructed in Section \ref{s6} to be useful for Nambu-Poisson sigma models \cite{bouwknegt2013aksz}. As it is possible to extend the triple point of Bianchi identities \cite{blumenhagen2012bianchi}, gauge closure conditions \cite{ikeda2003chern} for Courant sigma models and Courant algebroid axioms for $SL(5)$ M theory case based on Nambu-Poisson structures of order 3 \cite{chatzistavrakidis2019fluxes}, we hope that our dual calculus framework make it possible to generalize this for more complicated cases of algebroids. 

``Nature is known to have made ample use of Lie groups \textit{(Lie algebras)}. It seems unlikely that Nature has restricted itself to this relatively rigid notion, not also making use of the much more flexible Lie groupoids \textit{(Leibniz algebroids)}—and this also at the level of fundamental physics. To be unravelled.'' \cite{kotov2016gauging}, italic words in brackets added for emphasis.


\subsection*{Acknowledgements}

We thank Oğul Esen, Serkan Sütlü and their research group for their valuable contributions in the early stages of this work. In particular, we are grateful to Oğul Esen for introducing the matched pair literature to us. Authors are supported by The Scientific and Technological Research Council of Türkiye (TÜBİTAK) through the ARDEB 1001 project with grant number 121F123. K.D. is funded by İstanbul Technical University BAP Postdoctoral Research Fellowship (DOSAP) with project number TAB-2021-43202.


\newpage

\subsection*{Appendix A: Cartan calculus}
\label{appa}

 The tangent bundle carries a Lie algebra structure due to the usual \textit{Lie bracket} $[\cdot,\cdot]_{\text{Lie}}$, which is an anti-symmetric $\mathbb{R}$-bilinear map satisfying the Jacobi identity and the (right- and left-)Leibniz rule. The Lie bracket of vector fields can be extended to the \textit{Lie derivative} $\mathcal{L}$ of an arbitrary tensor, and in particular it can be considered as a map
\begin{equation}
    \mathcal{L}: T M \times \Lambda^p T^*M \to \Lambda^p T^*M \, .
\label{eb1}
\end{equation}
By using the Lie bracket, one can also construct the \textit{exterior derivative}
\begin{equation}
    d: \Lambda^{p-1} T^*M \to \Lambda^p T^* M \, .
\label{eb2}
\end{equation}
On the set of $p$-forms, there is another operation called the \textit{interior product}, which is a map of the form
\begin{equation}
    \iota: T M \times \Lambda^p T^*M \to \Lambda^{p-1} T^*M \, .
\label{eb3}
\end{equation}
These maps satisfy the following $C^{\infty}M$-linearity properties
\begin{align}
    \mathcal{L}_V (f \omega) &= f \mathcal{L}_V \omega + V(f) \omega \, , \nonumber\\
    \mathcal{L}_{f V} \omega &= f \mathcal{L}_V \omega + d f \wedge \iota_V \omega \, , \nonumber\\
    \iota_{f V} \omega &= \iota_V (f \omega) = f \iota_V \omega \, , \nonumber\\
    d(f \omega) &= f d \omega + d f \wedge \omega \, , \label{cartancalculuslinearity} 
\end{align}
for all $f \in C^{\infty} M, V \in T M, \omega \in \Lambda^p T^*M\, $. Moreover, there are important relations between these operations
\begin{align}
    \mathcal{L}_U \mathcal{L}_V - \mathcal{L}_V \mathcal{L}_U &= \mathcal{L}_{[U, V]_{\text{Lie}}} \, , 
    \nonumber\\
    \iota_U \iota_V + \iota_V \iota_U &= 0 \, ,
    \nonumber\\
    d^2 &= 0 \, , 
    \nonumber\\
    \mathcal{L}_U \iota_V  - \iota_V \mathcal{L}_U &= \iota_{[U, V]_{\text{Lie}}} \, ,
    \nonumber \\
    \mathcal{L}_V d - d \mathcal{L}_V &= 0 \, ,
    \nonumber \\
    d \iota_V + \iota_V d &= \mathcal{L}_V \, ,
\label{cartancalculusrelations} 
\end{align}
for all $U, V \in T M$. We refer to these maps $(\mathcal{L}, \iota, d)$ acting on forms as \textit{Cartan calculus}, and to the identities in (\ref{cartancalculusrelations}) as Cartan calculus relations.


\subsection*{Appendix B: Jacobi identity and right-Leibniz rule imply that the anchor is a morphism of brackets}
\label{appb}

Here, as the title indicates we prove that Jacobi identity 
\begin{equation} 
    [u, [v, w]_E]_E - [[u, v]_E, w]_E - [v, [u, w]_E]_E =0 \, ,
\label{jacobilodayapp}
\end{equation}
together with the right-Leibniz rule
\begin{equation}
    [u, f v]_E = f [u, v]_E + \rho_E(u)(f) v \, ,
\label{leibnizrightapp}
\end{equation}
yield that the anchor is a morphism of the brackets
\begin{equation}
    \rho_E ([u, v]_E) = [\rho_E(u), \rho_E(v)]_{\text{Lie}} \, . 
\label{anchormorphismapp}
\end{equation}
Let us start by considering the Jacobi identity with $w$ replaced by $f w$ for some $f \in C^\infty M$, that is 
\begin{align}
    0 &= [u, [v, f w]_E]_E - [[u, v]_E, f w]_E - [v, [u, f w]_E]_E \nonumber\\
    &= [u, f [v, w]_E + \rho_E(v)(f) w]_E - f [[u, v]_E, w]_E - \rho_E([u, v]_E)(f) w - [v, f [u, w]_E + \rho_E(u)(f) w]_E \nonumber\\
    &= f \left( [u, [v,  w]_E]_E - [[u, v]_E, w]_E - [v, [u, w]_E]_E \right) + \left[ \rho_E(u) \rho_E(v)(f) - \rho_E([u, v]_E)(f) \right. \nonumber\\
    & \left. \quad - \rho_E(v) \rho_E(u)(f) \right] w + \left\{ \rho_E(u)(f) [v, w]_E + \rho_E(v)(f)[u, w]_E - \rho_E(v)(f) [u, w]_E - \rho_E(u)(f) [v, w]_E \right\}\, .
\end{align}
Here, in the second and third equalities we made use of the right-Leibniz rule (\ref{leibnizrightapp}). The first term in the third line vanishes by Jacobi identity (\ref{jacobilodayapp}), and terms in the curly parenthesis trivially cancel out. The remaining terms in the last equality yields that the anchor is a morphism of brackets (\ref{anchormorphismapp}), which is the desired result.


\subsection*{Appendix C: Relation between calculus and Leibniz algebroid representations}
\label{appc}

From Jacobi identity, we directly observe, the necessary conditions that do not mix two calculi elements together:
\begin{align}
    \mathcal{L}_U \mathcal{L}_V \mu - \mathcal{L}_V \mathcal{L}_U \mu - \mathcal{L}_{[U, V]_A} \mu &= 0 \, , \label{tildecalculusbare1} \\ 
    \mathcal{L}_U \mathcal{K}_W \eta - \mathcal{K}_W \mathcal{L}_U \eta - \mathcal{K}_{[U, W]_A} \eta &= 0 \, , \label{tildecalculusbare2} \\
    \mathcal{L}_V \mathcal{K}_W \omega + \mathcal{K}_W \mathcal{K}_V \omega - \mathcal{K}_{[V, W]_A} \omega &= 0 \, . \label{tildecalculusbare3}
\end{align}
We prove that our calculus conditions (\ref{calculusconditions}) are equivalent to these equations as follows: The first condition written in the calculus definition agree with the first displayed Equation (\ref{tildecalculusbare1}). The second one can be obtained by taking the difference of first two Equations (\ref{tildecalculusbare1}) and (\ref{tildecalculusbare2}). Finally the third condition follows from taking the difference of (\ref{tildecalculusbare3}) with the first two Equations (\ref{tildecalculusbare1}), (\ref{tildecalculusbare2}). The relation between Equations (\ref{tildecalculusbare1}, \ref{tildecalculusbare2}, \ref{tildecalculusbare3}) with the algebroid representation definition (\ref{leibnizrepresentation}) given in matched pair paper \cite{ibanez2001matched} is as follows: The first one (\ref{tildecalculusbare1}) is the same their first one in Equation (\ref{leibnizrepresentation}). The third equation (\ref{tildecalculusbare3}) is the same as their second equation in (\ref{leibnizrepresentation}). Taking the difference between Equation (\ref{tildecalculusbare3}) and (\ref{tildecalculusbare2}) yields the last equation in (\ref{leibnizrepresentation}). We presented the relation between algebroid representations and calculus conditions in Section \ref{s4}. As we commented in the main text, our calculus definition includes also the linearity conditions coming from the left-Leibniz rule in addition the right-Leibniz rule where the latter is the only one in the Leibniz algebroid representation definition.


\subsection*{Appendix D: Proofs of Nambu-Poisson claims}
\label{appd}

We will prove linearity (\ref{linearityconditions}), calculus (\ref{linearityconditions}, \ref{calculusconditions}), Jacboi compatibility conditions (\ref{comp1}, \ref{comp2}, \ref{comp3}, \ref{compdual}) and metric invariance compatibility conditions (\ref{metricinvAZ}, \ref{metricinvcond1}, \ref{metricinvcond2}, \ref{metricinvtildeconds}) for the tilde-calculus elements (\ref{tildecalculuspi}, \ref{poissonexterior}) constructed from Nambu-Poisson structures. In this section, we set $A = TM$ and $Z = \Lambda^p T^*M$.


\subsubsection*{Linearity conditions}
\label{appd1}

We show that $(\tilde{\mathcal{L}}, \tilde{\iota},\tilde{d})$ defined in Equation (\ref{tildecalculuspi}, \ref{poissonexterior}) satisfies the linearity conditions (\ref{linearityconditions}). As one has $\tilde{\iota}_{\omega} U = \iota_U \omega$, the map $\tilde{\iota}$ is $C^{\infty}M$-bilinear as $\iota$ itself. In order to prove the others, we directly evaluate
\begin{align}
    \tilde{\mathcal{L}}_{f \omega} V = [\Pi(f \omega), V]_{\text{Lie}} - \Pi \mathcal{K}_V (f \omega) &= f [\Pi \omega, V]_{\text{Lie}} - V(f) \Pi \omega + \Pi \iota_V d (f \omega) \nonumber\\
    &= f \tilde{\mathcal{L}}_{\omega} V - V(f) \Pi \omega + \Pi \iota_V (d f \wedge \omega) \nonumber\\
    &= f \tilde{\mathcal{L}}_{\omega} V - V(f) \Pi \omega + \Pi \left( (\iota_V d f) \omega - d f \wedge \iota_V \omega \right) \nonumber\\
    &= f \tilde{\mathcal{L}}_{\omega} V - \Pi(df \wedge \iota_V \omega) \, ,
\end{align}
and 
\begin{align}
    \tilde{d} (f \tilde{\iota}_\omega V) &= - \Pi d (f \iota_V \omega) = - f \Pi d \iota_V \omega - \Pi (d f \wedge \iota_V \omega) \nonumber\\
    &= f \tilde{d} \tilde{\iota}_{\omega} V - \Pi (d f \wedge \iota_V \omega) \, ,
\end{align}
which together imply that 
\begin{equation}
    \Delta^{(1)}_{\tilde{\mathcal{L}}}(f, \omega, V) = \Delta_{\tilde{d}}(f, \tilde{\iota}_\omega V) \, ,
\end{equation}
as desired. For the last remaining one, we evaluate
\begin{align}
    \tilde{\mathcal{L}}_{\omega} (f V) = [\Pi \omega, f V]_{\text{Lie}} - \Pi \mathcal{K}_{f V} \omega &= f [\Pi \omega, V]_{\text{Lie}} + (\Pi \omega)(f) V - \Pi \iota_{f V} d \omega \nonumber\\
    &= f \tilde{\mathcal{L}}_{\omega} V + \rho_Z(\omega)(f) V \, ,
\end{align}
which is the desired result, where we use the fact that $\Pi$ seen as a map from $p$-forms to tangent bundle is the anchor.


\subsubsection*{Calculus conditions}
\label{appd2}

We show that $(\tilde{\mathcal{L}}, \tilde{\iota},\tilde{d})$ defined in Equation (\ref{tildecalculuspi}, \ref{poissonexterior}) indeed form a calculus by proving calculus conditions (\ref{calculusconditions}). However instead of showing the conditions presented in the definition of the calculus, we will show the conditions in their bare forms given by Equations (\ref{tildecalculusbare1}, \ref{tildecalculusbare2}, \ref{tildecalculusbare3}). We will express the bare forms of calculus conditions, and show that all of the expressions can be simplified in a similar manner. The first condition (\ref{tildecalculusbare1}) reads:
\begin{align}
    &\tilde{\mathcal{L}}_\omega \tilde{\mathcal{L}}_\eta W - \tilde{\mathcal{L}}_\eta \tilde{\mathcal{L}}_\omega W - \tilde{\mathcal{L}}_{[\omega, \eta]_{\text{Kos}}} W \nonumber\\
    & \quad = [\Pi \omega, [\Pi \eta, W]_{\text{Lie}} - \Pi \mathcal{K}_W \eta]_{\text{Lie}} - \Pi \mathcal{K}_{[\Pi \eta, W]_{\text{Lie}} - \Pi \mathcal{K}_W \eta} \omega - [\Pi \eta, [\Pi \omega, W]_{\text{Lie}} - \Pi \mathcal{K}_W \omega]_{\text{Lie}} \nonumber\\
    & \qquad + \Pi \mathcal{K}_{[\Pi \omega, W]_{\text{Lie}} - \Pi \mathcal{K}_W \omega} \eta - [\Pi [\omega, \eta]_{\text{Kos}}, W]_{\text{Lie}} + \Pi \left( \mathcal{K}_W \mathcal{L}_{\Pi \omega} \eta +\mathcal{K}_W \mathcal{K}_{\Pi \eta} \omega \right) \nonumber\\
    &\quad = -\Pi \left[ \left( \mathcal{L}_{\Pi \omega} \mathcal{K}_W - \mathcal{K}_{[\Pi \omega, W]_{\text{Lie}}} - \mathcal{K}_W \mathcal{L}_{\Pi \omega} \right) \eta + \left( \mathcal{K}_{[\Pi \eta , W]_{\text{Lie}}} - \mathcal{L}_{\Pi \eta} \mathcal{K}_W - \mathcal{K}_W \mathcal{K}_{\Pi \eta} \right) \omega \right] = 0 \, . 
\end{align}
The second condition (\ref{tildecalculusbare2}) reads 
\begin{align}
    &\tilde{\mathcal{L}}_\omega \tilde{\mathcal{K}}_\eta W - \tilde{\mathcal{K}}_\eta \tilde{\mathcal{L}}_\omega W - \tilde{\mathcal{K}}_{[\omega, \eta]_{\text{Kos}}} W \nonumber\\
    & \quad = [\Pi \omega, [W, \Pi \eta]_{\text{Lie}} - \Pi \mathcal{L}_W \eta]_{\text{Lie}} - \Pi \mathcal{K}_{[W, \Pi \eta]_{\text{Lie}} - \Pi \mathcal{L}_W \eta} \omega - [[\Pi \omega, W]_{\text{Lie}} - \Pi \mathcal{K}_W \omega, \Pi \eta]_{\text{Lie}} \nonumber\\
    & \qquad + \Pi \mathcal{L}_{[\Pi \omega, W]_{\text{Lie}} - \Pi \mathcal{K}_W \omega} \eta - [W, \Pi [\omega, \eta]_{\text{Kos}}]_{\text{Lie}} + \Pi(\mathcal{L}_W \mathcal{L}_{\Pi \omega} \eta + \mathcal{L}_W \mathcal{K}_{\Pi \eta} \omega) \nonumber\\
    &\quad = - \Pi\left[ \left( \mathcal{L}_{\Pi \omega} \mathcal{L}_{W} - \mathcal{L}_{[\Pi \omega, W]_{\text{Lie}}} - \mathcal{L}_{W} \mathcal{L}_{\Pi \omega} \right) \eta + \left( \mathcal{K}_{[W, \Pi \eta]_{\text{Lie}}} - \mathcal{K}_{\Pi \eta}  \mathcal{K}_{W} - \mathcal{L}_{W}\mathcal{K}_{\Pi\eta} \right) \omega \right] = 0 \, . 
\end{align}
Finally, the third condition (\ref{tildecalculusbare3}) reads
\begin{align}
    &\tilde{\mathcal{L}}_\omega \tilde{\mathcal{K}}_\eta W + \tilde{\mathcal{K}}_\eta \tilde{\mathcal{K}}_\omega W - \tilde{\mathcal{K}}_{[\omega, \eta]_{\text{Kos}}} W \nonumber\\
    & \quad = [\Pi \omega, [W, \Pi \eta]_{\text{Lie}} - \Pi \mathcal{L}_W \eta]_{\text{Lie}}  - \Pi \mathcal{K}_{[W, \Pi \eta]_{\text{Lie}} - \Pi \mathcal{L}_W \eta} \omega + [[W, \Pi \omega]_{\text{Lie}} - \Pi\mathcal{L}_W \omega, \Pi \eta]_{\text{Lie}} \nonumber\\
    & \qquad -\Pi \mathcal{L}_{[W, \Pi \omega]_{\text{Lie}} - \Pi \mathcal{L}_W \omega} \eta - [W, [\Pi [\omega, \eta]_{\text{Kos}}]_{\text{Lie}} + \Pi \left( \mathcal{L}_W \mathcal{L}_{\Pi \omega} \eta + \mathcal{L}_W \mathcal{K}_{\Pi \eta} \omega \right) \nonumber\\
    &\quad = - \Pi \left[ \left( \mathcal{L}_{\Pi \omega}\mathcal{L}_{W} + \mathcal{L}_{[W, \Pi \omega]_{\text{Lie}}} - \mathcal{L}_{W} \mathcal{L}_{\Pi \omega} \right) \eta + \left( \mathcal{K}_{[W, \Pi \eta]_{\text{Lie}}} + \mathcal{K}_{\Pi \eta}  \mathcal{L}_{W} - \mathcal{L}_{W} \mathcal{K}_{\Pi \eta} \right) \omega \right] = 0 \, . 
\end{align}
In all of the equations above, when moving from second to third equalities, in addition to trivial cancellations, we made use of the fact that $\Pi$ is a morphism of brackets (see for example \cite{bouwknegt2013aksz}), \textit{i.e.} $\Pi[\omega, \eta]_{\text{Kos}} = [\Pi \omega, \Pi \eta]_{\text{Lie}}$, and the fact that the Lie bracket satisfies the Jacobi identity. The last equalities follow from the fact that $\mathcal{K}_W = - \iota_W d$ (using Cartan magic formula) and the usual Cartan calculus relations (\ref{cartancalculusrelations}). The derivation of linearity and calculus conditions implies that $(\tilde{\mathcal{L}}, \tilde{\iota}, \tilde{d})$ is indeed a calculus, which proves our Claim 3.


\subsubsection*{Jacobi compatibility conditions}
\label{appd3}

We now show the Jacobi compatibility conditions. The first condition (\ref{comp1}) reads
\begin{align}
    &\mathcal{D}^Z_{\mathcal{L}_U}(\eta, \mu) - \mathcal{L}_{\tilde{\mathcal{K}}_{\eta} U} \mu - \mathcal{K}_{\tilde{\mathcal{K}}_{\mu} U} \eta \nonumber\\
    &\quad = \left( \mathcal{L}_U \mathcal{L}_{\Pi \eta} - \mathcal{L}_{\Pi \eta}\mathcal{L}_U - \mathcal{L}_{[U,\Pi \eta]_{\text{Lie}}} \right) \mu + \left( \mathcal{L}_U \mathcal{K}_{\Pi \mu} - \mathcal{K}_{\Pi \mu} \mathcal{L}_U - \mathcal{K}_{[U, \Pi \mu]_{\text{Lie}}} \right) \eta = 0 \, .
\end{align}
In the above we used the definition of the derivator (\ref{derivator}) and the Koszul bracket (\ref{koszulbracket}). After trivial cancellations, we are left with the expression in the second line which follows from the usual Cartan calculus relations (\ref{cartancalculusrelations}). The second Jacobi compatibility condition (\ref{comp2}) reads
\begin{align}
    \mathcal{L}_{\tilde{d} \tilde{\iota}_{\eta} U} \mu + [d \iota_U \eta, \mu]_{\text{Kos}} = - \mathcal{K}_{\Pi\mu}d\iota_U \eta = 0 \, , 
\end{align}
which follows immediately from the fact that the usual exterior derivative squares to zero since $\mathcal{K}_V = - \iota_V d$. The third condition (\ref{comp3}) reads
\begin{align}
    &d \iota_{\tilde{\mathcal{L}}_{\omega} W} \eta - d \iota_{\tilde{d} \tilde{\iota}_{\eta} W} \omega + d \iota_W [\omega, \eta]_{\text{Kos}} + \mathbb{D}_Z g_Z(\mathcal{K}_W \omega, \eta) - \mathbb{D}_Z g_Z(d \iota_W \eta, \omega) \nonumber\\
    &\qquad = \left( d \iota_{[\Pi \omega, W]_{\text{Lie}}} + d \iota_W\mathcal{L}_{\Pi \omega} - d \iota_{\Pi \omega} d \iota_W  \right) \eta + \left( d \iota_W \mathcal{K}_{\Pi \eta} + d \iota_{\Pi \eta} \mathcal{K}_W \right) \omega = 0 \, . 
\end{align}
Here, we used the definition of metric $g_Z$ and the first-order differential operator $\mathbb{D}_Z$ given in (\ref{zazan}). The final equality again follows from the usual Cartan calculus relations (\ref{cartancalculusrelations}). Now we move to the dual Jacobi compatibility conditions (\ref{compdual}), the first of which reads
\begin{align}
    \mathcal{D}^A_{\tilde{\mathcal{L}}_{\omega}}(V, W) - \tilde{\mathcal{L}}_{\mathcal{K}_V \omega} W - \tilde{\mathcal{K}}_{\mathcal{K}_W \omega} V &= [\Pi \omega, [V, W]_{\text{Lie}}]_{\text{Lie}} - \Pi \mathcal{K}_{[V, W]_{\text{Lie}}} \omega - [[\Pi \omega, V]_{\text{Lie}}, W]_{\text{Lie}} \nonumber\\
    &\quad - [V, [\Pi \omega, W]_{\text{Lie}}]_{\text{Lie}} + \Pi \left( \mathcal{K}_W \mathcal{K}_V \omega + \mathcal{L}_V \mathcal{K}_W \omega \right) = 0 \, . 
\end{align}
Aside from the trivial cancellations, the equality follows from the fact that Lie bracket satisfies the Jacobi identity and the Cartan calculus relations (\ref{cartancalculusrelations}). The second condition in (\ref{compdual}) reads:
\begin{equation}
    \tilde{\mathcal{L}}_{d \iota_V \omega} W + [\tilde{d} \tilde{\iota}_{\omega} V, W]_{\text{Lie}} = [\Pi d \iota_V \omega, W]_{\text{Lie}} - \Pi \mathcal{K}_W d \iota_V \omega + [- \Pi d \iota_V \omega, W]_{\text{Lie}} = 0 \, , 
\end{equation}
which follows from $d^2 = 0$ since $\mathcal{K}_W = - \iota_W d$. The third and final dual Jacobi compatibility condition (\ref{compdual}) simplifies since $g_A = 0$, and becomes
\begin{align}
    &\tilde{d} \tilde{\iota}_{\mathcal{L}_U \mu} V - \tilde{d} \tilde{\iota}_{d \iota_V \mu} U + \tilde{d} \tilde{\iota}_{\mu} [U, V]_{\text{Lie}} = - \Pi d (\iota_V\mathcal{L}_U - \iota_U d \iota_V + \iota_{[U, V]_{\text{Lie}}}) \mu = 0 \, , 
\end{align}
which follows from another combination of Cartan calculus identities (\ref{cartancalculusrelations}). As Jacobi compatibility conditions between $(\mathcal{L}, \iota, d)$ and $(\tilde{\mathcal{L}}, \tilde{\iota}, \tilde{d})$ are all satisfied, they form a pair of dual calculi, which proves our Claim 4. Claim 5 directly follows from the combination of Claims 1 to 4, as we have explicitly showed in Section \ref{s4}.


\subsubsection*{Metric invariance compatibility conditions}
\label{appd4}

We now prove the metric invariance compatibility conditions. Let us start with metric invariance property (\ref{metricinvAZ}) for $A$ and $Z$ algebroids. Note that since $g_A = 0$, the first one is trivially satisfied. To identify the correct operator coming from the second one, let us write the right-hand side:
\begin{align}
    &g_Z([\omega, \eta]_{\text{Kos}}, \mu) + g_Z(\eta, [\omega, \mu]_{\text{Kos}}) \nonumber\\
    &\quad = \left( \iota_{[\Pi \omega, \Pi \eta]_{\text{Lie}}} + \iota_{\Pi \eta}\mathcal{L}_{\Pi \omega} \right) \mu + (\iota_{[\Pi \omega, \Pi \mu]} + \iota_{\Pi \mu} \mathcal{L}_{\Pi \omega}) \eta + \left( \iota_{\Pi \mu} \mathcal{K}_{\Pi \eta} + \iota_{\Pi \eta} \mathcal{K}_{\Pi \mu} \right) \omega \nonumber\\
    &\quad = \mathcal{L}_{\Pi \omega} \left( \iota_{\Pi \mu} \eta + \iota_{\Pi \eta} \mu \right) \nonumber\\
    &\quad = \mathcal{L}_{\Pi \omega} g_Z(\eta, \mu) \, . 
\end{align}
From this, we see that the metric invariance operator of $Z = \Lambda^p T^*M$ is given by 
\begin{equation}
    \tilde{\pounds}_{\omega} = \mathcal{L}_{\rho_Z(\omega)} = \mathcal{L}_{\Pi \omega}\, . 
\end{equation}
We now move to the first set of conditions mixing the two calculi (\ref{metricinvcond1}). Noting that $g_A = 0$, and $g_Z$ identified via (\ref{zazan}), the right-hand side of these equations read:
\begin{equation}
    2 (\iota_{[U, V]_{\text{Lie}}} + \iota_V \mathcal{L}_U) \mu = 2 \mathcal{L}_U \iota_V \mu \, ,
\end{equation}
and
\begin{align}
   &2 \left( \iota_{[U, \Pi \eta]_{\text{Lie}} - \Pi \mathcal{L}_U \eta} \mu + \iota_{[U, \Pi \mu]_{\text{Lie}} - \Pi \mathcal{L}_U \mu} \eta \right) + 2 \left( \iota_{\Pi \mathcal{L}_U \mu} \eta + \iota_{\Pi \mu} \mathcal{L}_U \eta + \iota_{\Pi \mathcal{L}_U \eta} \mu + \iota_{\Pi \eta} \mathcal{L}_U \mu \right) \nonumber\\
   &\quad = 2 \mathcal{L}_U \left( \iota_{\Pi \omega} \eta + \iota_{\Pi \eta} \omega \right) \nonumber\\
   &\quad = \mathcal{L}_U g_Z(\omega, \eta)\, . 
\end{align}
In both equations, we made use of a Cartan calculus relation (\ref{cartancalculusrelations}). We then see that for the metric invariance to hold, the operators on the left-hand side of both of these equations can be identified with the usual Lie derivative, \textit{i.e.}, $\pounds_U=\mathcal{L}_U$. We now move to the dual conditions (\ref{metricinvtildeconds}). The second equation is trivially satisfied since $g_A = 0$ and the remaining terms on the right-hand side vanish by Cartan calculus identities (\ref{cartancalculusrelations}). The right-hand side of first equation reads:
\begin{equation}
    2 \left( \iota_V \mathcal{L}_{\Pi \omega} \eta + \iota_V \mathcal{K}_{\Pi \omega} \eta + \iota_{[\Pi \omega, V]_{\text{Lie}} - \Pi \mathcal{K}_V \omega} \eta \right) + 2 \left( \iota_{\Pi \eta} \mathcal{K}_W \omega + \iota_{\Pi \mathcal{K}_W \omega} \eta \right) = 2 \mathcal{L}_{\Pi \omega} \iota_W \eta \, ,  
\end{equation}
where we made use of Cartan calculus identities once again, and we see that the operator on the left-hand side can be identified with $\tilde{\pounds}_\omega = \mathcal{L}_{\Pi \omega} = \mathcal{L}_{\rho_E(\omega)}$. Combining all, we see that the metric invariance operator is just
\begin{equation}
    \mathbb{L}^E_{U + \omega} = \mathcal{L}_{\rho_E(U + \omega)} = \mathcal{L}_{U + \Pi \omega} \, ,
\end{equation}
which proves our Claim 6 about metric invariance. Claim 7 directly follows from the combination of Claims 5 \& 6, as we have explicitly showed in Section \ref{s4}.


\subsection*{Appendix E: Proof of rackoid metric condition}
\label{appe}

Here we prove that, on an algebroid $E$ which satisfies the metric invariance property with operator $\mathbb{L}^E: E \times \mathbb{E} \to \mathbb{E}$
\begin{equation}
    \mathbb{L}^E_u g_E(v, w) = g_E([u, v]_E, w) + g_E(v, [u, w]_E) \, ,
\label{metricinvapp}
\end{equation}
the formal exponentiation of the metric invariance operation
\begin{equation}
    u \triangleright \xi = \text{exp}\,  \mathbb{L}^E_u \xi \, ,
\end{equation}
yields the rackoid metric condition
\begin{equation}
    u \triangleright g_E(v, w) = g_E(u \triangleright v, u \triangleright w) \, .
\label{rackmetricconditionapp}
\end{equation}

The left-hand side of $(\ref{rackmetricconditionapp})$ can be calculated inductively in the orders of operator $\mathbb{L}^E$. For this we can use the following binomial-like expansion formula:
\begin{equation}
(\mathbb{L}^E_u)^k g_E(v, w) = \sum_{i = 0}^k \binom{k}{i} g_E \big( ad^i(u)(v), ad^{k - i}(u)(w) \big) \, . \label{inductionstep}
\end{equation}
Note that at induction step one, this expression agrees with the metric invariance of the algebroid (\ref{metricinvapp}) as expected. Now assuming the induction step at $k$ holds, let us calculate the next order:
\begin{align}
    (\mathbb{L}^E_u)^{k+1} &g_E(v, w) = \mathbb{L}^E_u \bigg( \sum_{i = 0}^k \binom{k}{i} g_E \big( ad^i(u)(v), ad^{k-i}(u)(w) \big) \bigg) \nonumber\\
    &= \sum_{i = 0}^k \binom{k}{i} \bigg( g_E \big(ad^{i+1}(u)(v), ad^{k-i}(u)(w) \big) + g_E \big(ad^i(u)(v), ad^{k - i + 1}(u)(w) \big)  \bigg) \nonumber\\
    &= \sum_{i=1}^k \bigg( \binom{k}{i-1} g_E \big(ad^{i}(u)(v), ad^{k-i+1}(u)(w) \big) + \binom{k}{i} g_E \big(ad^i(u)(v), ad^{k-i+1}(u)(w) \big) \bigg) \nonumber\\
    &\qquad + \binom{k}{0} g_E \big( ad^{k+1}(u)(v), w \big) + \binom{k}{k} g_E \big(v, ad^{k+1}(u)(w) \big) \nonumber\\
    &= \sum_{i=1}^{k+1} \binom{k+1}{i} g_E \big( ad^{i}(u)(v), ad^{k-i+1}(u)(w) \big) + g_E \big( ad^{k+1}(u)(v), w \big) + g_E \big(v, ad^{k+1}(u)(w) \big) \nonumber\\
    &= \sum_{i=0}^{k+1} \binom{k+1}{i} g_E \big( ad^i(u)(v), ad^{k+1-i}(u)(w) \big) \, . \label{lhsrackoidmetricinv}
\end{align}
In the second equality above, we use the metric invariance property (\ref{metricinvapp}) and $\mathbb{R}$-bilinearity of the first-order differential operator $\mathbb{L}^E$. Afterwards in the third equality, we separated the sums and redefined the dummy variables. In the fourth equality we used the small combinatorial identity
\begin{equation}
    \binom{k}{i-1} + \binom{k}{i} = \binom{k+1}{i} \, ,
\end{equation}
and in the final step we combined each term into a single sum which is the desired result. 

For the right-hand side of rackoid metric condition $(\ref{rackmetricconditionapp})$, we can similarly expand the rackoid action
\begin{equation}
    u \triangleright v = \sum_{i=0}^\infty \frac{1}{i!} ad^i(u)(v) \, ,
\end{equation}
and we see that the condition at order $k+1$ equals Equation (\ref{lhsrackoidmetricinv}). We directly evaluate the product and rearrange the summation to get
\begin{align}
    g_E(u \triangleright v, u \triangleright w) &= \sum_{i=0}^\infty \sum_{j=0}^\infty  \frac{1}{i! j!} g_E(ad^i(u)(v), ad^j(u)(w)) \\
    &= \sum_{k=0}^\infty \sum_{i=0}^k \frac{1}{i! (k-i)!} g_E(ad^i(u)(v), ad^{k-i}(u)(w)) \, . 
\end{align}
Finally by noting that $\frac{1}{k!} \binom{k}{i} = \frac{1}{i! (k-i)!}$, we see that left-hand side and right-hand side are equal. This whole proof is (of course) just an imitation of Cauchy product formula.


\newpage

\bibliographystyle{unsrt}
\bibliography{bibliography}

\end{document}